\documentclass[fleqn,usenatbib]{mnras}

\usepackage{newtxtext,newtxmath}

\usepackage{graphicx}	
\usepackage{amsmath}	
\usepackage{rotating}
\usepackage{adjustbox}
\usepackage[graphicx]{realboxes}
\newcommand{\comment}[1]{}

\usepackage{pdflscape}
\usepackage{scrextend}
\usepackage{tabu} 
\usepackage{multirow} 
\usepackage{import} 
\usepackage{gensymb} 
\title[Reddening and accretion properties of red QSOs]{Fundamental differences in the properties of red and blue quasars: \\measuring the reddening and accretion properties with \textit{X-shooter}}

\author[V. A. Fawcett et al.]
{V. A. Fawcett,$^{1}$\thanks{E-mail: victoria.fawcett@durham.ac.uk}
D. M. Alexander,$^{1}$
D. J. Rosario,$^{2,1}$
L. Klindt,$^{1}$
E. Lusso,$^{3,4}$
\newauthor
L. K. Morabito,$^{1,5}$
\& G. Calistro Rivera$^{6}$
\\
$^{1}$Centre for Extragalactic Astronomy, Department of Physics, Durham University, DH1 3LE, UK\\
$^{2}$ School of Mathematics, Statistics and Physics, Newcastle University, NE1 7RU, UK\\
$^{3}$Dipartimento di Fisica e Astronomia, Universit\`a di Firenze,  via G. Sansone 1, I-50019 Sesto Fiorentino, Firenze, Italy\\
$^{4}$Osservatorio Astrofisico di Arcetri, Largo Enrico Fermi 5, I-50125 Firenze, Italy \\
$^{5}$Institute for Computational Cosmology, Department of Physics, Durham University, DH1 3LE, UK\\
$^{6}$European Southern Observatory, Karl-Schwarzchild-Strasse 2, 85748, Garching bei Mnchen, Germany
}

\date{Accepted XXX. Received YYY; in original form ZZZ}
\pubyear{2022}

\begin{document}
\label{firstpage}
\pagerange{\pageref{firstpage}--\pageref{lastpage}}
\maketitle

\begin{abstract}
We have recently found fundamental differences in the radio properties of red quasars when compared to typical blue quasars. In this paper we use data from the \textit{X-shooter} spectrograph on the Very Large Telescope, providing spectral coverage from $\sim3000$--$25000\,$\AA, of a sample of 40 red and blue luminous quasars at $1.45<z<1.65$ to explore the connections between the radio, emission-line, and accretion-disc properties. We fit various dust-extinction curves to the data and find that dust reddening can fully explain the observed colours for the majority of the red quasars in our sample, with moderate extinctions ranging from $A_V\sim0.06$--$0.7$\,mags. We confront our spectra with a simple thin accretion-disc model and find this can describe the continua of both the blue and red quasars, once corrected for dust extinction; we also find no significant differences in the accretion properties. We detect ionized outflows in a number of red and blue quasars, but do not find any significant evidence that they are more prevalent in the red quasar population. Overall our findings imply that the radio emission is more closely connected to circumnuclear/ISM opacity rather than accretion disc or outflow differences. 
\end{abstract}

\begin{keywords}
galaxies: active -- galaxies: evolution -- quasars: general -- quasars: supermassive black-holes -- accretion, accretion discs -- quasars: emission lines
\end{keywords}

\section{Introduction}
Quasi-stellar objects (QSOs), also known as quasars, are the most powerful class of Active Galactic Nuclei (AGN), with extremely high bolometric luminosities (up to $10^{47-48}$~erg\,s$^{-1}$). These high luminosities are now known to be due to mass accretion onto a supermassive black-hole (SMBH; masses of $10^8$--$10^9$~M$_{\odot}$). The majority of quasars are blue due to an unobscured view of the SMBH accretion disc which peaks in the ultra-violet (UV). However, there is a significant subset of QSOs with redder optical-infrared colours (referred to as ``red QSOs''). 
Although red QSOs have been well studied in the literature \citep{Webster1995,Serjeant1996,kim98,richards,glik,georg,urrutia,ban12,glik12,kim18,klindt,fawcett,rosario,calistro,rosario_21}, the origin of the red colours is still debated. The most popular explanation is reddening due to dust extinction, which attenuates the blue light from the accretion disc (e.g., \citealt{Webster1995,glik,klindt}). However, a red synchrotron component or stellar contamination from the host galaxy have also been shown to contribute significantly to the colours of QSOs in strongly radio-loud/gamma-ray-loud and lower luminosity systems, respectively (e.g, \citealt{whiting,glik,Shen_2012,klindt,calistro}).

To study reddening due to dust in QSOs, the two standard methods used are through fitting the continuum with extinction curves and measuring the Balmer decrement (the H$\upalpha$ to H$\upbeta$ broad line flux ratio). For extinction-curve analyses, the extinction laws of the Small and Large Magellanic Clouds (SMC and LMC; e.g., \citealt{SMC,LMC}) are used preferentially over a Milky Way standard extinction curve (MW; e.g., \citealt{cardelli}). The SMC/LMC laws have a weak or absent 2175\,\AA~graphite feature which, while prominent in the MW, has only been detected in a minority of QSOs \citep{Jiang_2011,shi}. Past studies have found that an SMC-like extinction curve best describes the dust extinction in red QSOs \citep{richards,hopkins,young,urrutia}. However, these analyses are usually limited to the SMC, LMC and MW-like curves, and based only on optical photometry or spectroscopy over a narrow wavelength range. The Balmer decrement provides an alternative constraint on the extinction towards the QSO through measuring the strength of different broad hydrogen emission lines (i.e., the Balmer and Paschen series). In previous studies, clear evidence has been found for larger Balmer decrements, and therefore larger amounts of dust extinction, in red QSOs compared to typical QSOs \citep{glik,kim18}. 

However, even amongst the studies that favour dust as the cause of the reddening, there is further debate on the location and nature of this intervening dust. For example, some studies \citep{wilkes,rose} postulate that the dust is associated with a grazing view through the putative AGN ``dusty torus'' \citep{antonucci,urry,netzer,Almeida2017}. An alternative model (often referred to as the evolutionary scenario; \citealt{sanders}) identifies red QSOs as a short-lived phase in which an obscured QSO rapidly reveals itself by blowing away the surrounding shroud of dust in the circumnuclear regions and interstellar medium (ISM) through the action of powerful winds \citep{hop6,hop,Farrah_2012,glik12,alex_hick_12}.

Recently, our group have found fundamental differences in the radio properties of red QSOs when compared to typical QSOs that cannot be explained just by the orientation of the dusty torus \citep{klindt,rosario,fawcett,fawcett21}. The key results from these analyses is the identification of excess compact radio emission in red QSOs, which arises around the radio-quiet--radio-loud threshold (defined as the ratio of the 1.4\,GHz luminosity to 6$\upmu$m luminosity; see \citealt{klindt} for details). In \cite{fawcett}, we found that this radio excess decreases towards the extreme radio-quiet end and appears to be driven by AGN mechanisms rather than star-formation. In a more recent study, we have showed that this excess is compact even on extremely small scales (typically below 2--3\,kpc; \citealt{rosario_21}). These radio properties could be connected to the accretion engine, depending on parameters such as black-hole mass, accretion rate and black-hole spin, or connected to larger-scale processes such as winds or outflows (i.e., the commonly referred to ``AGN feedback process''). However, star-formation cannot be conclusively ruled out for the low redshift/lower radio luminosity sources.

To test whether differences in the accretion properties between red and typical QSOs are driving the enhanced radio emission, accurate black-hole masses are required which can be then used for a characterization of the pure accretion disc (AD) emission through continuum fits (e.g., \citealt{capellup}). There are many methods to constrain black-hole mass, the most commonly used being the ``single epoch virial mass determination'' method. These methods combine a broad line full width at half-maximum (FWHM) with either a continuum or broad line luminosity to obtain the black-hole mass, building on calibrations between AD luminosity and the broad line region (BLR) size, derived from reverberation mapping studies (e.g., \citealt{Kaspi_2000,Bentz_2013}). In order to self-consistently constrain key accretion parameters, single-epoch spectroscopy with wide wavelength coverage is required: \textit{X-shooter} \citep{vernet} at the Very Large Telescope (VLT) offers this capability, providing continuous spectral coverage across observed wavelengths $\sim3000$--$25000\,$\AA. In a series of papers (\citealt{capellup}; hereafter C15, \citealt{capb}; hereafter M16 and \citealt{capc}; hereafter C16), \textit{X-shooter} spectra were found to be very effective at constraining the accretion and emission-line properties in a sample of $z\sim1.5$ QSOs. For example in C15, a thin AD model was found to provide a good fit to the majority of their QSO \textit{X-shooter} spectra. The QSOs that could not be fit with an AD model were found to either have modest amounts of dust ($A_V\sim0.1$--0.45\,mags) or required an additional disc-wind component which, once implemented, provided a good fit to the majority of the remaining spectra.

In this work we use new and archival \textit{X-shooter} UV--NIR spectra for a sample of 40 red and typical QSOs at $1.45<z<1.65$, in order to determine whether the cause of the red colours is consistent with arising entirely from dust, or whether differences in the accretion properties could be driving both the colours and enhanced radio emission in red QSOs. In Section~\ref{sec:method} we define our sample, describe the observations, the reduction process, and the modelling of the telluric absorption. In Sections~\ref{sec:ext_method}, \ref{sec:fit}, and \ref{sec:AD} we describe our dust extinction, emission line, and AD model fitting procedures, respectively. In Section~\ref{sec:ext} we quantify the amount of reddening present in our red QSOs through an extinction-curve comparison and analysis of the Balmer decrements, in Section~\ref{sec:emission} we explore the emission-line properties of our sample, and in Section~\ref{sec:bh_prop} we explore whether there are any differences in the AD properties between red and typical QSOs. In Sections~\ref{sec:dust} and \ref{sec:bh_dis} we discuss the nature of dust in red QSOs and the presence of accretion-driven winds in red QSOs, respectively.
Throughout our work we adopt a flat $\Lambda$-cosmology with $H_0$\,$=$\,70~km\,s$^{-1}$Mpc$^{-1}$, $\Omega$\textsubscript{M}\,$=$\,0.3 and $\Omega_{\Lambda}$\,$=$\,0.7.
\section{Sample selection and observations}\label{sec:method}
\subsection{Sample selection}\label{sec:sample}
\begin{figure}
    \centering
    \includegraphics[width=3.3in]{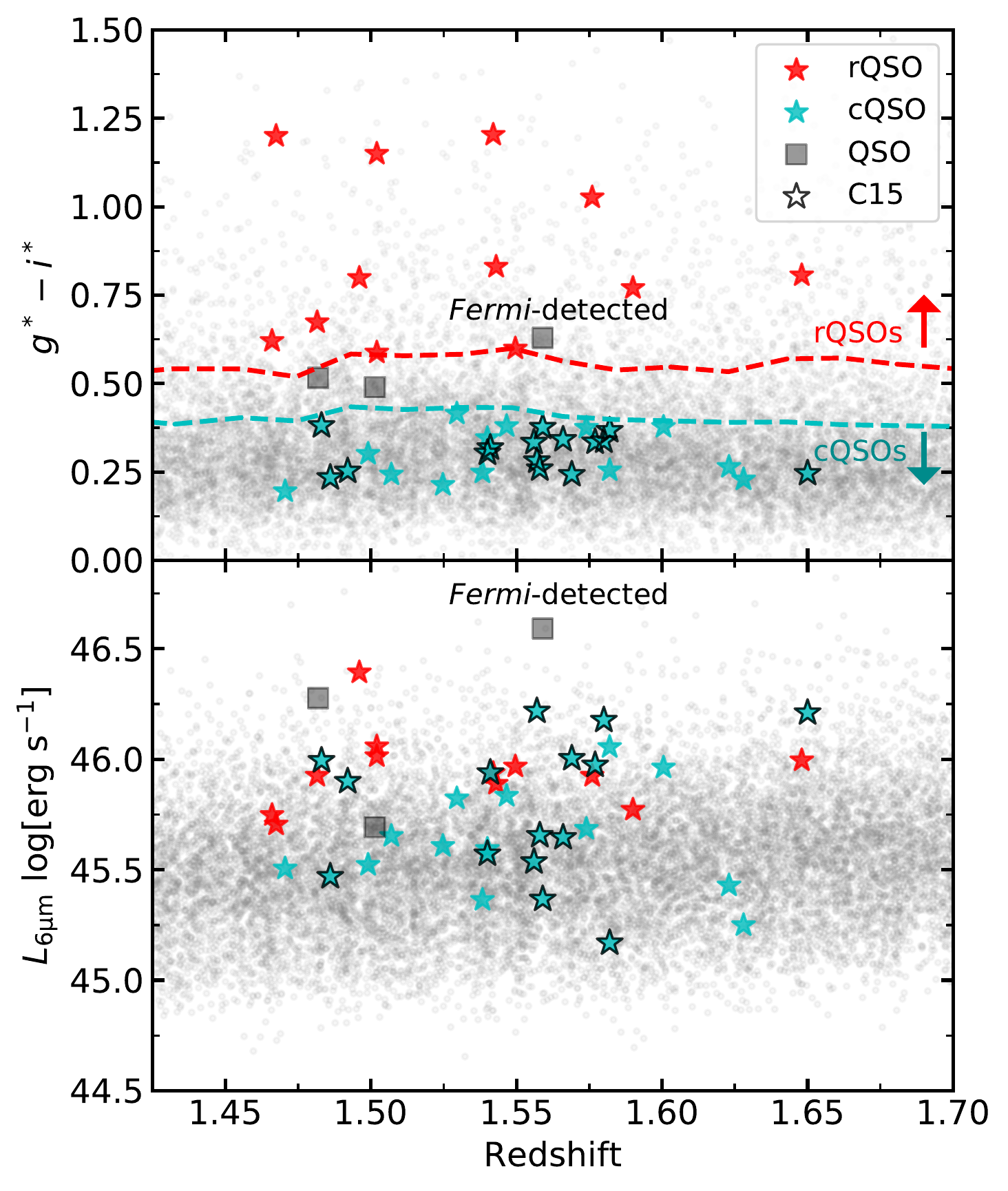}
    \caption{The optical $ g^*-i^*$ colour (top) and \textit{L}$_{\rm 6\,\upmu m}$ (bottom) versus redshift distributions for our full rQSO (red stars) and cQSO (cyan stars) samples; the cQSOs included from C15 are outlined in black. The three QSOs that are excluded from our main analysis are shown as grey squares and the \textit{Fermi}-detected source is annotated. The DR7 QSOs from our parent sample are shown in grey. In the top plot the lower and upper $g^*-i^*$ boundaries for our rQSO and cQSO selections, calculated from the DR7 QSO parent, sample are shown by the dashed red and cyan curves, respectively (see Section~\ref{sec:sample}).}
    \label{fig:z_gi}
\end{figure}

\begin{figure}
    \centering
    \includegraphics[width=3.3in]{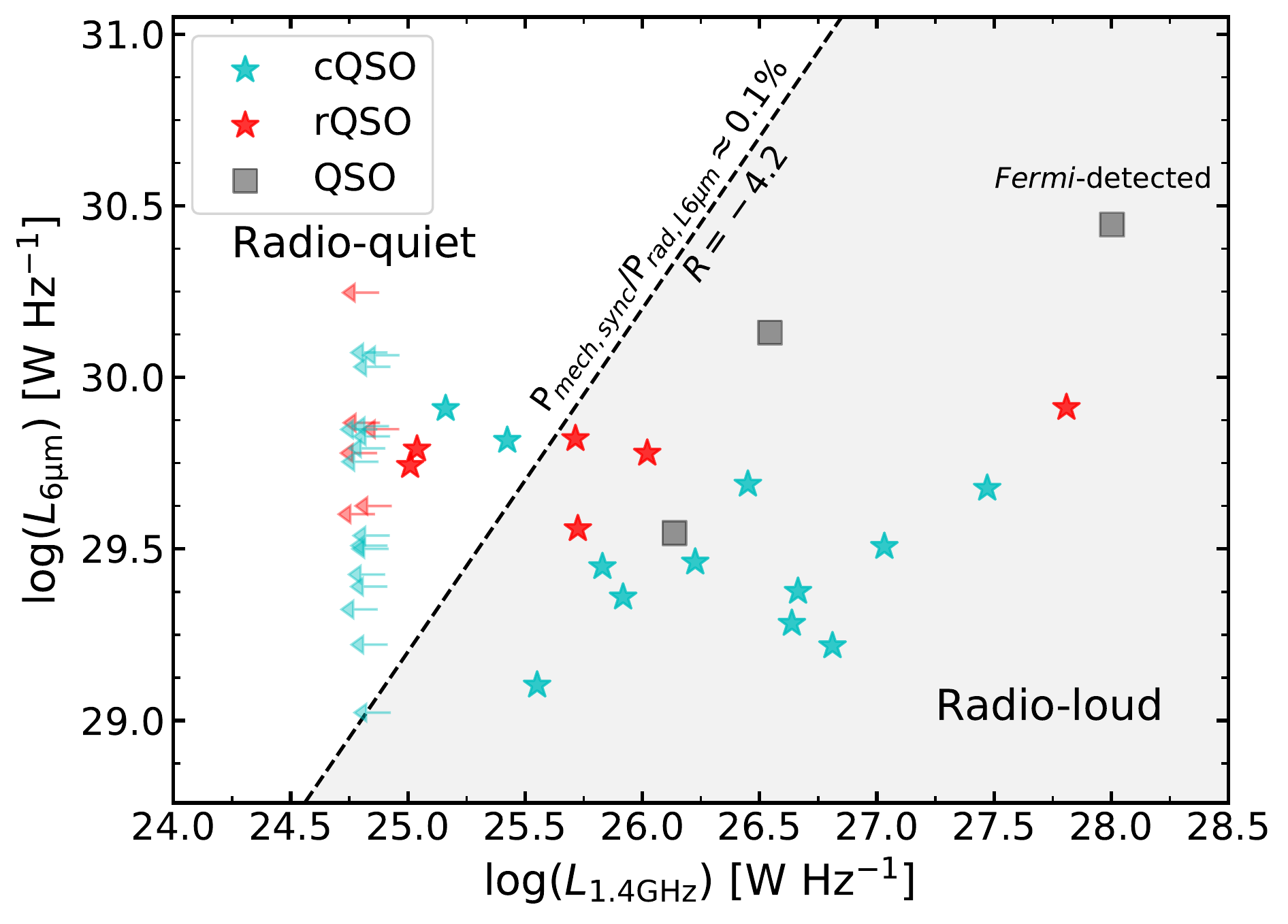}
    \caption{The \textit{L}$_{\rm 6\,\upmu m}$ versus \textit{L}$_{\rm 1.4 GHz}$ luminosity density distributions of our rQSOs and cQSOs. Upper limits for radio-undetected sources are displayed as arrows; see Fig.~\ref{fig:z_gi} for the colour and symbol definitions. The radio-quiet threshold is shown as a dashed line, defined as a mechanical-to-radiative power of 0.1 per~cent ($\mathcal{R}$\,$=$\,$-$4.2). None of the cQSOs included from C15 are radio-detected in FIRST.}
    \label{fig:radio_loud}
\end{figure}

Our sample is based on the Sloan Digital Sky Survey (SDSS) DR7 Quasar Catalogue \citep{dr7} which consists of 105783 spectroscopically confirmed quasars with luminosities of M$_{\rm I}<-22.0$ out to $z=5.48$. We used DR7 based on work by \cite{klindt}, who defined a uniformly-selected sample by requiring the uniform flag ({\sc UNIFORM\_TARGET = 1}; see \citealt{shen11}), which restricts the sample to the luminous end of the QSO population that satisfy the main QSO selection algorithm \citep{richardsb}, and removing the Faint Images of the Radio Sky at Twenty Centimeters (FIRST; \citealt{becker}) pre-selected sources. We will hereafter refer to sources using a shorthand version of the SDSS identifier; e.g., SDSSJ160808.12+120702.4 is shortened to 1608+1207. 

We require mid-infrared (MIR) counterparts from the Wide-Field Infrared Survey Explorer (\textit{WISE}; \citealt{wise}), an all-sky survey which provides photometry in four bands (3.4, 4.6, 12 and $22\,\upmu$m). It has been shown that MIR emission, arising from the hot dust heated by high-energy radiation from the accretion disc, is a useful, extinction-insensitive discriminant of QSOs (e.g., \citealt{stern05,lacy05,stern12,assef}), as well as a measure of their intrinsic luminosity; for example the commonly used rest-frame 6\,$\upmu$m luminosity (\textit{L}\textsubscript{6\,$\upmu$m}). We used the NASA/IPAC query engine to match SDSS DR7 QSOs to the All-Sky \textit{WISE} Source Catalogue (ALL-\textit{WISE}) adopting a $2\farcs7$ search radius, which ensures a 95.5 per~cent positional certainty \citep{lake11}, and required a detection with a signal-to-noise ratio (SNR) of greater than 2 in the \textit{WISE} \textit{W1}, \textit{W2} and \textit{W3} bands in order to derive an accurate estimation of \textit{L}\textsubscript{6\,$\upmu$m}. To compute the \textit{L}\textsubscript{6\,$\upmu$m} we used a log-linear interpolation or extrapolation of the fluxes in the \textit{W2} and \textit{W3} bands.

In this paper, we aim to extend our understanding of the fundamental differences between red and typical QSOs. To distinguish between these two populations, we follow the approach adopted in our previous papers (e.g., \citealt{klindt}). We select the red QSO sample as the top 10 per~cent of the redshift-dependent $g^*-i^*$ distribution. Importantly, we then select a ``control'' sample in a consistent way to the red QSOs, as the median 50 per~cent of the $g^*-i^*$ distribution, in order to identify a set of typical QSOs that we can reliably compare to within our analyses\footnote{To select the red and control samples, we first sorted the QSOs by redshift and then constructed $g^*-i^*$ distributions in contiguous bins of 1000 sources.} (see \citealt{klindt} for more details). From this colour-selected sample, we then targeted a sub-sample of red QSOs (rQSOs) and control QSOs (cQSOs), matched in redshift and $L_{\rm 6\,\upmu m}$ to be observed with VLT/\textit{X-shooter}. We chose our redshift range and observing strategy based on the approach taken in C15 (Program ID 088.B-1034). C15 used \textit{X-shooter} to observe 30 typical QSOs in a redshift range of $1.45<z<1.65$, which was chosen to ensure that the H\,$\upalpha$ and the H\,$\upbeta$--[\ion{O}{iii}] complex lay in the H-band and J-band, respectively. In addition to our own dedicated observations (see Section~\ref{sec:obs}), we included 15 QSOs from the C15 sample that were matched in $L_{\rm 6\,\upmu m}$ and $g^*-i^*$ colour to our cQSO sample, and also satisfy our uniform flag and \textit{WISE} constraints; these cQSOs from C15 are flagged in Table~\ref{tab:sample}. The cQSOs from the C15 sample are also indicated in Fig.~\ref{fig:z_gi}, but hereafter will be referred to as part of the full cQSO sample.

Due to scheduling restrictions, only 54 per~cent of our dedicated observations were completed: the observed targets were left with a \textit{L}\textsubscript{6\,$\upmu$m} bias (see Fig.~\ref{fig:z_gi}). In the relevant analyses (Section~\ref{sec:bh_prop}) we repeat our comparisons using a statistically-limited luminosity-matched sub-sample in order to account for this bias. For the matching, we used an \textit{L}\textsubscript{6\,$\upmu$m} tolerance of 0.2\,dex, following a similar approach to \cite{klindt} (see Section~2.3.2 therein). Within our sample we found that a mistake in the amount of Galactic extinction was originally applied to two rQSOs (2109+0045 and 2239+1222) which, when corrected, moved the sources out of our $g^*-i^*$ rQSO threshold. We do not consider these two sources in our main analyses and they will be referred to as ``QSO'' within the tables to distinguish them from the cQSOs and rQSOs.

The $g^*-i^*$ colours and \textit{L}\textsubscript{6\,$\upmu$m} versus redshift for our full sample are shown in Fig.~\ref{fig:z_gi} (top and bottom panel, respectively). The radio-detected sources, and upper limits for the radio-undetected sources, are displayed in Fig.~\ref{fig:radio_loud}; the dashed line indicates the radio-quiet threshold, defined here as the mechanical-to-radiative power of 0.1~per~cent. This corresponds to $\mathcal{R}$\,$=$\,$-$4.2, where $\mathcal{R}$ is the ``radio-loudness'' parameter defined as the dimensionless quantity:
\begin{equation}\label{eq:rl}
\mathcal{R}=\textrm{log\textsubscript{10}}\frac{\textit{L}\textsubscript{1.4\,GHz}}{\textit{L}\textsubscript{6\,$\upmu$m}} .\
\end{equation}
We calculated \textit{L}\textsubscript{1.4\,GHz} assuming a uniform radio spectral slope of $\upalpha=-0.7$, following a similar approach to \cite{Alexander_2003}.\footnote{We define the radio spectral index $\upalpha$ as $f_{\upnu}\propto \upnu^{-\upalpha}$.}
This is different from the canonical definition for radio-loudness, which is defined as a ratio of the 5\,GHz radio luminosity to 4400\,\AA~optical luminosity \citep{kell}; we do not use this definition since optical luminosities are heavily affected by dust extinction for the rQSOs. Our defined radio-quiet threshold of $\mathcal{R}=-4.2$ is roughly consistent to the classical definition (see \citealt{klindt} for more details). In both figures, the QSOs excluded from further analyses are displayed as grey squares.

Within our dedicated observed sample, we chose to include both FIRST radio-detected and undetected rQSOs and cQSOs in order to test whether the accretion and extinction properties are related to the radio properties. Our sample includes a flat-spectrum rQSO (2229-0832) which is also the only source in our sample that is detected in the fourth \textit{Fermi} Large Area Telescope catalog (4FGL; \citealt{Abdollahi_2020}). It is extremely radio-bright ($F_{\rm 1.4\,GHz,int}=1.03$\,Jy) and has the largest radio luminosity in our sample; see Fig.~\ref{fig:radio_loud} and Table~\ref{tab:sample}. The \textit{Fermi} detection suggests that the radio emission is relativistically beamed; also indicated by its flat spectrum. Our spectral analysis of 2229-0832 showed that the UV--near-IR emission is likely dominated by non-thermal processes (see Appendix~\ref{sec:intrinsic} for a more detailed discussion), which we also confirm from our radio--optical analysis, following the method from \citealt{klindt} (see Appendix~A therein). Due to the extreme nature of this source, we have excluded it from our analyses (indicated as `\textit{Fermi}-detected' in Figs.~\ref{fig:z_gi} and~\ref{fig:radio_loud}).

\subsection{Observations}\label{sec:obs}

\begin{table*}
    \centering
    \caption{Observation and sample details for our rQSO, cQSO and excluded QSO samples, including the 15 cQSOs from C15. The columns from left to right display the: (1) shortened source name utilized in this paper, (2) QSO sub-sample (rQSO, cQSO, QSO); the three sources labelled as QSO are excluded from our main analyses (see Section~\ref{sec:sample}), (3) date of observations (those prior to 2013 are from \protect\cite{capellup}, while those post 2017 are new dedicated observations presented here), (4)-(5) RA and Dec as defined in the optical SDSS DR7 catalogue, (6) redshift obtained from SDSS DR7, (7) redshift defined via visual inspection, based on the observed wavelength of H\,$\upalpha$, (8) the optical SDSS $g^*-i^*$ colour, (9) rest-frame 1.4~GHz FIRST radio luminosity (upper limits indicated by *) calculated assuming a radio spectral slope of $\upalpha=-0.7$, and (10) rest-frame 6\,$\upmu$m luminosity calculated using a log-linear interpolation or extrapolation of the fluxes in the \textit{W2} and \textit{W3} bands. \newline \textdagger QSOs from C15; \textsuperscript{$a$} \ion{C}{iv} absorption identified; \textsuperscript{$b$} Offset between VIS and NIR; \textsuperscript{$c$} \textit{Fermi}-detected; \textsuperscript{$d$} Poor seeing.}
    \begin{minipage}{\textwidth}
    \centering
    \begin{tabular}{cccccccccc} 
    \hline
    \hline
    Name & Sample & Date Observed  & RA & Dec & $z_{\rm SDSS}$ & $z_{\rm VI}$ & $( g^*-i^*)$ &  log $L_{\rm 1.4\,GHz}$ & log $L_{\rm 6\,\upmu m}$    \\ 
     & & & & & & &  [mag] & [WHz$^{-1}$]  & [erg\,s$^{-1}$] \\
    \hline
1020+1101 & rQSO & 16-Apr-2018 & 10 20 01.58 & +11 01 24.8 & 1.550 & 1.550 & 0.60 & 25.7 & 46.0 \\  
1049+1157\textsuperscript{$a$} & rQSO & 18-May-2018 & 10 49 44.52 & +11 57 58.9 & 1.544 & 1.542 & 1.20 & 25.0 & 45.9 \\  
1358+1145 & rQSO & 20-Jun-2018 & 13 58 09.49 & +11 45 57.6 & 1.484 & 1.482 & 0.67 & 24.9* & 45.9 \\  
1429-0112 & rQSO & 12-May-2018 & 14 29 48.67 & -01 12 52.2 & 1.499 & 1.502 & 1.15 & 27.8 & 46.1 \\  
1442+1426\textsuperscript{$d$} & rQSO & 16-Apr-2018 & 14 42 04.77 & +14 26 37.6 & 1.642 & 1.648 & 0.81 & 25.0* & 46.0 \\  
1523+0452\textsuperscript{$d$} & rQSO & 16-Apr-2018 & 15 23 35.99 & +04 52 35.5 & 1.587 & 1.590 & 0.77 & 24.9* & 45.8 \\  
1608+1207 & rQSO & 19-May-2018 & 16 08 08.13 & +12 07 02.4 & 1.464 & 1.466 & 0.62 & 24.9* & 45.7 \\  
1616+0931 & rQSO & 18-May-2018 & 16 16 39.57 & +09 31 17.1 & 1.466 & 1.467 & 1.20 & 25.7 & 45.7 \\  
1639+1135\textsuperscript{$a$} & rQSO & 16-Apr-2018 & 16 39 24.03 & +11 35 48.1 & 1.571 & 1.576 & 1.03 & 26.0 & 45.9 \\  
2133+1043 & rQSO & 15-Jun-2018 & 21 33 44.27 & +10 43 16.9 & 1.533 & 1.543 & 0.83 & 25.0 & 45.9 \\  
2223+1258 & rQSO & 15-Jun-2018 & 22 23 57.70 & +12 58 26.7 & 1.489 & 1.496 & 0.80 & 24.9* & 46.4 \\
2241-1006 & rQSO & 15-Jun-2018 & 22 41 01.44 & -10 06 50.4 & 1.498 & 1.502 & 0.59 & 24.9* & 46.0 \\  
\hline
0043+0114\textdagger & cQSO & 26-Nov-2011 & 00 43 15.08 & +01 14 45.6 & 1.563 & 1.569 & 0.24 & 24.9* & 46.0 \\
0152-0839\textdagger & cQSO & 10-Aug-2012 & 01 52 01.24 & -08 39 58.1 & 1.572 & 1.577 & 0.33 & 24.9* & 46.0 \\
0155-1023\textdagger & cQSO & 24-Oct-2011 & 01 55 04.73 & -10 23 28.3 & 1.546 & 1.577 & 0.28 & 24.9* & 46.2 \\
0213-0036\textdagger & cQSO & 12-Sep-2012 & 02 13 10.33 & -00 36 20.3 & 1.560 & 1.559 & 0.38 & 24.9* & 45.4 \\
0223-0007\textdagger & cQSO & 25-Nov-2011 & 02 23 21.38 & -00 07 33.8 & 1.534 & 1.540 & 0.38 & 24.9* & 45.6 \\
0303+0027\textdagger & cQSO & 21-Oct-2011 & 03 03 42.78 & +00 27 00.5 & 1.646 & 1.650 & 0.25 & 25.0* & 46.2 \\
0341-0037\textdagger\textsuperscript{$a$} & cQSO & 17-Dec-2011 & 03 41 56.07 & -00 37 06.3 & 1.553 & 1.556 & 0.33 & 24.9* & 45.5 \\
0404-0446\textdagger\textsuperscript{$a$} & cQSO & 23-Nov-2011 & 04 04 14.13 & -04 46 49.8 & 1.547 & 1.558 & 0.26 & 24.9* & 45.7  \\
0842+0151\textdagger & cQSO & 18-Dec-2011 & 08 42 40.63 & +01 51 34.1 & 1.493 & 1.492 & 0.25 & 24.9* & 45.9 \\
0927+0004\textdagger & cQSO & 23-Feb-2012 & 09 27 15.49 & +00 04 01.0 & 1.485 & 1.486 & 0.23 & 24.9* & 45.5 \\
0934+0005\textdagger & cQSO & 01-Mar-2012 & 09 34 11.14 & +00 05 19.7 & 1.534 & 1.541 & 0.32 & 24.9* & 45.9 \\
0941+0443\textdagger & cQSO & 19-Mar-2012 & 09 41 26.49 & +04 43 28.7 & 1.571 & 1.566 & 0.34 & 24.9* & 45.6 \\
1001+1015 & cQSO & 16-Apr-2018 & 10 01 57.73 & +10 15 49.7 & 1.532 & 1.530 & 0.41 & 27.5 & 45.8 \\  
1002+0331\textdagger\textsuperscript{$a$} & cQSO & 22-May-2012 & 10 02 48.15 & +03 31 55.9 & 1.481 & 1.483 & 0.38 & 24.9* & 46.0 \\
1013+0245\textdagger & cQSO & 16-May-2012 & 10 13 25.49 & +02 45 21.4 & 1.564 & 1.582 & 0.37 & 24.9* & 45.2 \\
1028+1456 & cQSO & 16-Apr-2018 & 10 28 23.58 & +14 56 48.9 & 1.633 & 1.628 & 0.23 & 25.6 & 45.2  \\  
1158-0322\textdagger\textsuperscript{$a$} & cQSO & 15-Apr-2012 & 11 58 41.37 & -03 22 39.9 & 1.573 & 1.580 & 0.34 & 24.9* & 46.2 \\
1352+1302 & cQSO & 20-Jun-2018 & 13 52 18.23 & +13 02 41.7 & 1.601 & 1.601 & 0.38 & 25.4 & 46.0 \\  
1357-0307 & cQSO & 12-May-2018 & 13 57 55.07 & -03 07 24.4 & 1.555 & 1.582 & 0.25 & 25.2 & 46.1 \\  
1358+1410\textsuperscript{$a$} & cQSO & 15-May-2018 & 13 58 04.10 & +14 10 59.7 & 1.543 & 1.540 & 0.35 & 25.8 & 45.6 \\  
1428+1001 & cQSO & 20-Jun-2018 & 14 28 53.08 & +10 01 17.8 & 1.499 & 1.599 & 0.30 & 26.7 & 45.6 \\  
1502+1016 & cQSO & 20-Jun-2018 & 15 02 26.47 & +10 16 11.0 & 1.629 & 1.623 & 0.26 & 26.6 & 45.4 \\  
1513+1011\textsuperscript{$b$} & cQSO & 11-Aug-2018 & 15 13 29.29 & +10 11 05.5 & 1.547 & 1.547 & 0.38 & 26.5 & 45.8 \\  
1521-0156 & cQSO & 12-May-2018 & 15 21 00.75 & -01 56 37.5 & 1.574 & 1.574 & 0.38 & 24.9* & 45.7 \\  
1539+0534 & cQSO & 10-Aug-2018 & 15 39 05.20 & +05 34 38.4 & 1.511 & 1.507 & 0.24 & 27.0 & 45.7 \\  
1540+1155 & cQSO & 10-Aug-2018 & 15 40 49.09 & +11 55 32.5 & 1.472 & 1.471 & 0.20 & 25.9 & 45.5 \\  
1552+0939 & cQSO & 19-May-2018 & 15 52 27.71 & +09 39 02.9 & 1.538 & 1.538 & 0.25 & 26.8 & 45.4 \\  
1618+1305 & cQSO & 19-May-2018 & 16 18 55.70 & +13 05 05.3 & 1.525 & 1.525 & 0.21 & 26.2 & 45.6 \\  
\hline
2109+0045\textsuperscript{$a$} & QSO & 13-May-2018 & 21 09 03.71 & +00 45 55.3 & 1.501 & 1.501 & 0.49 & 26.1 & 45.7 \\
2229-0832\textsuperscript{$c$} & QSO & 09-Jul-2018 & 22 29 40.09 & -08 32 54.5 & 1.560 & 1.559 & 0.63 & 28.0 & 46.6 \\  
2239+1222 & QSO & 16-Jun-2018 & 22 39 31.96  & +12 22 24.4 & 1.482 & 1.482 & 0.52 & 26.5 & 46.3 \\  
    \hline
    \hline
    \end{tabular} 
    \end{minipage}
    \label{tab:sample}
\end{table*} 


Our main analyses are based on observations obtained with the \textit{X-shooter} instrument on the VLT \citep{vernet}, which provides broad spectral coverage from $\sim$\,3000 to 25000\,\AA~by observing three wavelength ranges simultaneously: the UV-blue (UVB), visible (VIS), and near-infrared (NIR). At the redshift of our sample, this corresponds to rest wavelengths of $\sim1200-9800$\,\AA.

Dedicated observations for this project were carried out in nodding mode (nodding length $5''$) between April--August 2018 (Program ID 0101.B-0739(A); PI: Klindt). We adopted slit widths of $1\farcs6$, $1\farcs5$ and $1\farcs2$ in the UVB, VIS and NIR arms, respectively, corresponding to a resolving power of 3300, 5400 and 4300. For the brighter sources in our sample (J-band magnitude $<17.6$\,mags AB) we used 900\,sec exposures, and for the fainter targets we used 1200--1400\,sec exposures.
For the majority of the sources, observations were taken under conditions where the seeing was $\leq1\farcs6$ and airmass $\leq1.5$. However, for two objects the seeing was highly variable throughout the night, reaching $1\farcs8-1\farcs9$ at times: these QSOs are identified in Table~\ref{tab:sample}. From visually inspecting the data for these two sources, we found no issues with the spectra that could arise from poor seeing during observations. Telluric standard stars were not provided by the observatory following the specification change from P101: instead, the ESO software tool \texttt{molecfit} was used to fit and correct telluric lines (see Section~\ref{sec:data_red}). To obtain the flux-calibrated spectra, spectrophotometric standards were taken once a night. 

The full sample, including both our dedicated observations and the sub-sample from C15, consists of 12 rQSOs (6 radio-detected) and 28 cQSOs (12 radio-detected). We provide an overview of the observations for the full sample (including the three excluded QSOs) in Table \ref{tab:sample}.

\subsection{Data reduction}\label{sec:data_red}
We reduced the \textit{X-shooter} data from our dedicated program using version \texttt{2.9.1} of the ESO \textit{X-shooter} pipeline \citep{modigliani} within the ESO \texttt{Reflex} environment \citep{freud}. Data for each spectrographic arm were reduced individually, adopting the nodding mode setting in the pipeline. This provides an automated workflow, in which detector bias and dark current are subtracted followed by a wavelength and flux calibration, to produce the flux-calibrated one-dimensional spectra. In nodding mode, a sequence of science frames at different nodding positions are combined into an image sequence, and then the pairs of combined frames are subtracted. Sky subtraction is carried out by taking the difference of the images from nodding position A and B. Due to a gradient in one of the four raw VIS images for the cQSO 1552+0939, we were unable to effectively subtract the sky from the overall source spectrum. For this source we therefore carried out the reduction in the VIS arm excluding the pair of raw frames corresponding to this gradient.

In the majority of our sources there is good agreement in the flux for the overlapping regions of the spectra from the three arms. However, in one case (cQSO 1513+1011) there was a clear offset between the visible and NIR spectra output by the pipeline due to an issue with the flux calibration of the standard star. For this source we used the publicly available reduced data for the NIR arm from the ESO archive, which we found to be consistent with the overlapping region of our VIS reduced spectrum.

To further test the validity of our flux calibration, we calculated synthetic $g$, $r$, and $i$-band photometry by applying the SDSS photometric filters to our \textit{X-shooter} spectra, and compared these data to both the SDSS DR7 and Pan-STARRS1 (PS1; \citealt{PS1}) photometry (adopting a similar method to \citealt{selsing}). We found that there was a small amount of scatter between all three samples, with an average standard deviation in the photometric offsets of $\sim0.24$\,mag; consistent with the predicted variation due to QSO variability (0.26\,mags; \citealt{macleod}). We did find a small mean offset of $\sim-0.15$\,mag between our synthetic photometry compared to both SDSS and PS1 which is not wavelength dependent. This suggests that some flux is lost from the \textit{X-shooter} spectra which is likely due to extended emission beyond the QSO point source.

\begin{table}
    \centering
    \caption{Wavelength windows used in \texttt{molecfit} for telluric absorption corrections in the VIS and NIR arms. The main molecule responsible for the absorption features in each window is displayed. Windows were expanded slightly for low SNR sources in order to optimize the correction.} 
    \begin{tabular}{cccc} 
    \hline
    \hline
    Arm & $\uplambda_{\rm min}$ [$\upmu$m] & $\uplambda_{\rm max}$ [$\upmu$m] & Main molecule \\ 
    \hline
    VIS & 0.686 & 0.696 & {\sc O$_2$} \\
    VIS & 0.725 & 0.730 & {\sc O$_2$} \\
    VIS & 0.759 & 0.769 & {\sc O$_2$} \\
    VIS & 0.930 & 0.950 & {\sc H$_2$O} \\
    \hline
    NIR & 1.120 & 1.130 & {\sc H$_2$O} \\
    NIR & 1.460 & 1.470 & {\sc H$_2$O} \\
    NIR & 1.800 & 1.820 & {\sc H$_2$O} \\
    NIR & 1.960 & 1.970 & {\sc CO$_2$} \\
    NIR & 2.010 & 2.025 & {\sc CO$_2$} \\
    NIR & 2.350 & 2.360 & {\sc CH$_4$} \\
    \hline
    \hline
    \end{tabular}
    \label{tab:molec}
\end{table}

Ground-based observations suffer from atmospheric absorption (predominantly by {\sc H$_2$O}, {\sc O$_2$}, {\sc CO$_2$} and {\sc CH$_4$}), which is particularly prominent in the NIR waveband. After running the \textit{X-shooter} pipeline, we corrected the spectra for telluric absorption using a set of feature-free continuum windows (see Table~\ref{tab:molec}; some windows were expanded for a few low SNR sources). This was achieved through the ESO software tool \texttt{molecfit}\footnote{http://www.eso.org/sci/software/pipelines/skytools/} \citep{molecfit1,molecfit2}, which is based on synthetic modelling of the Earth's atmosphere. The model spectra are generated by the radiative transfer code Line-by-Line Radiative Transfer Model (LBLRTM), which takes an atmospheric profile created using local atmospheric information modelled from Global Data Assimilation System (GDAS\footnote{http://www.ready.noaa.gov/gdas1.php}) data and meteorological measurements from ESO Meteo Monitor (EMM\footnote{http://www.ls.eso.org/lasilla/dimm/}). For all sources, the correction in the NIR arm did not remove all features, and so the regions heavily affected by absorption were masked in our final analyses. 

For the C15 sources, we used reduced and telluric-corrected spectra obtained via private communication with the authors. We rebinned all the spectra to a consistent wavelength sampling and masked the same spectral regions in all of the QSOs. 

Finally, we corrected all of the spectra for Galactic extinction using the \cite{schlegel} map and the \cite{fitz} Milky Way extinction law. One cQSO (1521-0156) had an extreme level of Galactic extinction ($A_V=0.54$\,mags) and so for this source we masked the region around the 2175\,\AA~feature, in order to minimize the uncertainties this introduced to our extinction and spectral fitting analyses (see Sections~\ref{sec:ext_method} and \ref{sec:fit}, respectively). The final spectra for the rQSOs, cQSOs and QSOs are shown in Figs.~\ref{fig:spec_rQSO}, \ref{fig:all_spec}, and \ref{fig:ex_spec}, respectively.

\section{Dust-Extinction measurements}\label{sec:ext_method}

As previously mentioned, one rQSO (2229-0832) is \textit{Fermi}-detected and has the largest radio luminosity in our sample; due to the extreme nature of the source, we exclude it from our analyses. In Appendix~\ref{sec:intrinsic}, we conclude that this source is most likely reddened by a synchrotron component. Exploring whether the reddening for any of our other rQSOs can be explained with a synchrotron component, we find this to be the case only for 2229-0832. Moreover, in Fawcett et~al. (\textit{in prep}) we explore DR14 rQSO composites in bins of redshift and $L_{\rm 6\,\upmu m}$, and find that sources with $z>1$ and log\,$L_{\rm 6\,\upmu m}>45.0$\,erg\,s$^{-1}$ (consistent with our \textit{X-shooter} sample) display little to no host-galaxy absorption features (e.g., the Calcium H+K lines), and even the composite from the lowest redshift and luminosity bin, in which we see strong host-galaxy features, requires an additional dust-reddening component in order to explain the continuum emission (consistent with the broad-band spectral energy distribution (SED) analysis from \citealt{calistro}). Consequently, we expect the red colours of our \textit{X-shooter} rQSOs to be predominantly due to dust extinction, as suggested by previous literature (e.g., \citealt{glik,klindt}). Under this assumption, we predict the loss of flux at the rest-frame UV--VIS end of the spectrum to be frequency dependent, as seen in QSO dust-extinction curves. 
Modelling the dust extinction in QSOs via the fitting of extinction curves has been undertaken in many previous studies. However, these analyses are usually only based on optical photometry or spectroscopy over a narrow wavelength range, which will not necessarily provide reliable measurements; the wide wavelength coverage of \textit{X-shooter} can be used to robustly determine whether the red colours in the rQSO population are consistent with dust extinction. We adopt two methods widely utilized in the literature: fitting extinction curves to the continuum emission and measuring the Balmer decrement using the broad H\,$\upalpha$ and H\,$\upbeta$ emission-line fluxes.\footnote{The narrow lines are also sensitive to reddening but on larger scales than the nucleus. We use the broad lines to assess the reddening along the line-of-sight to the accretion disc and BLR.} 

\begin{figure}
    \centering
    \includegraphics[width=3.in]{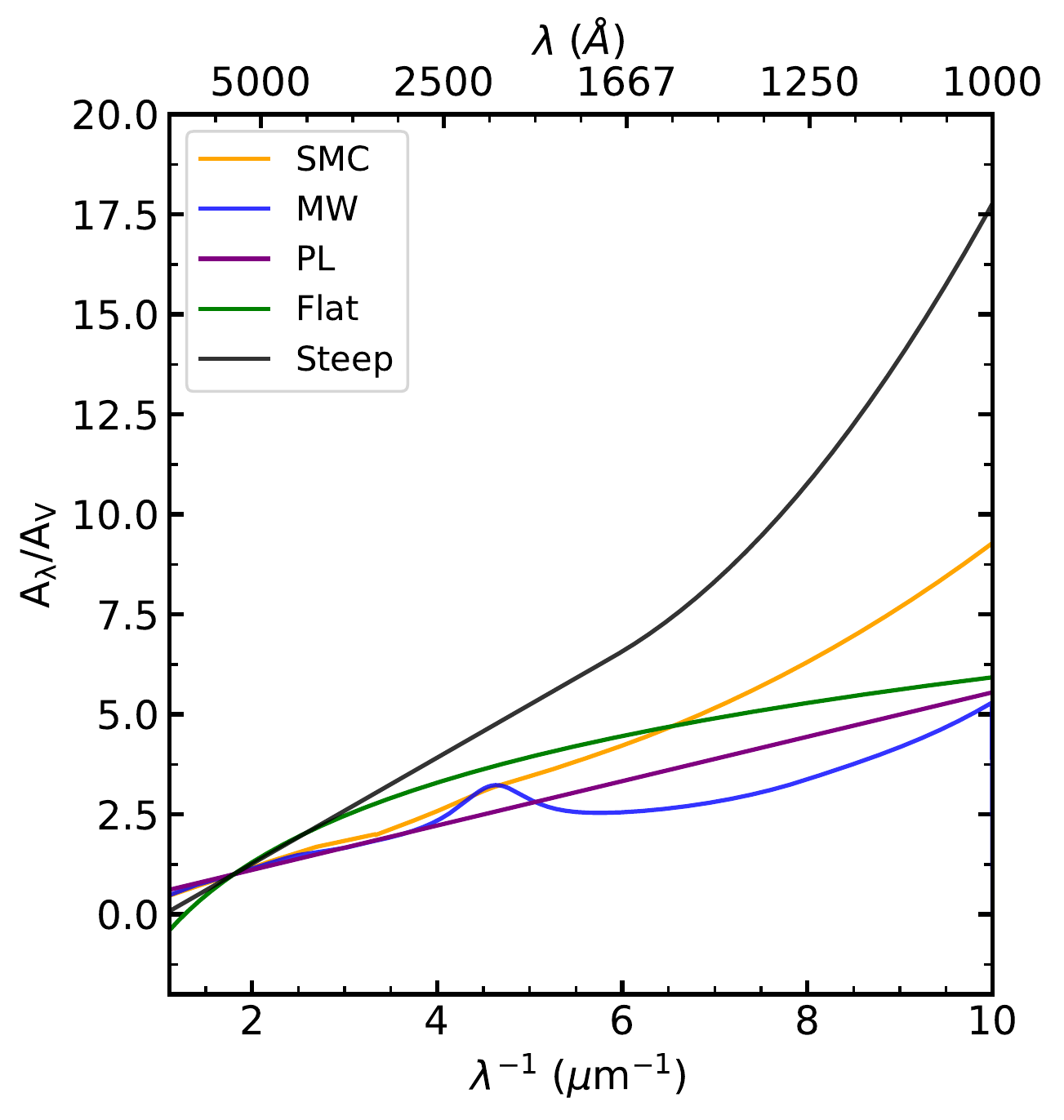}
    \caption{Comparison of the five different extinction curves used in the extinction-curve analysis; an SMC law (\citealt{richards}; yellow), a MW law (\citealt{cardelli}), a simple power-law (PL; purple), a flat ``grey'' law (\citealt{czerny}; green) and a steeper law (\citealt{zafar}; black). The different extinction curves correspond to variations in the dust grain composition and size. Only the MW curve displays the characteristic 2175\,\AA~bump produced by graphite. A comparison of how these different extinction laws affect the QSO continuum is shown in Fig.~\ref{fig:comp_av}.}
    \label{fig:extinction}
\end{figure}

\begin{figure*}
    \centering
    \includegraphics[width=6.2in]{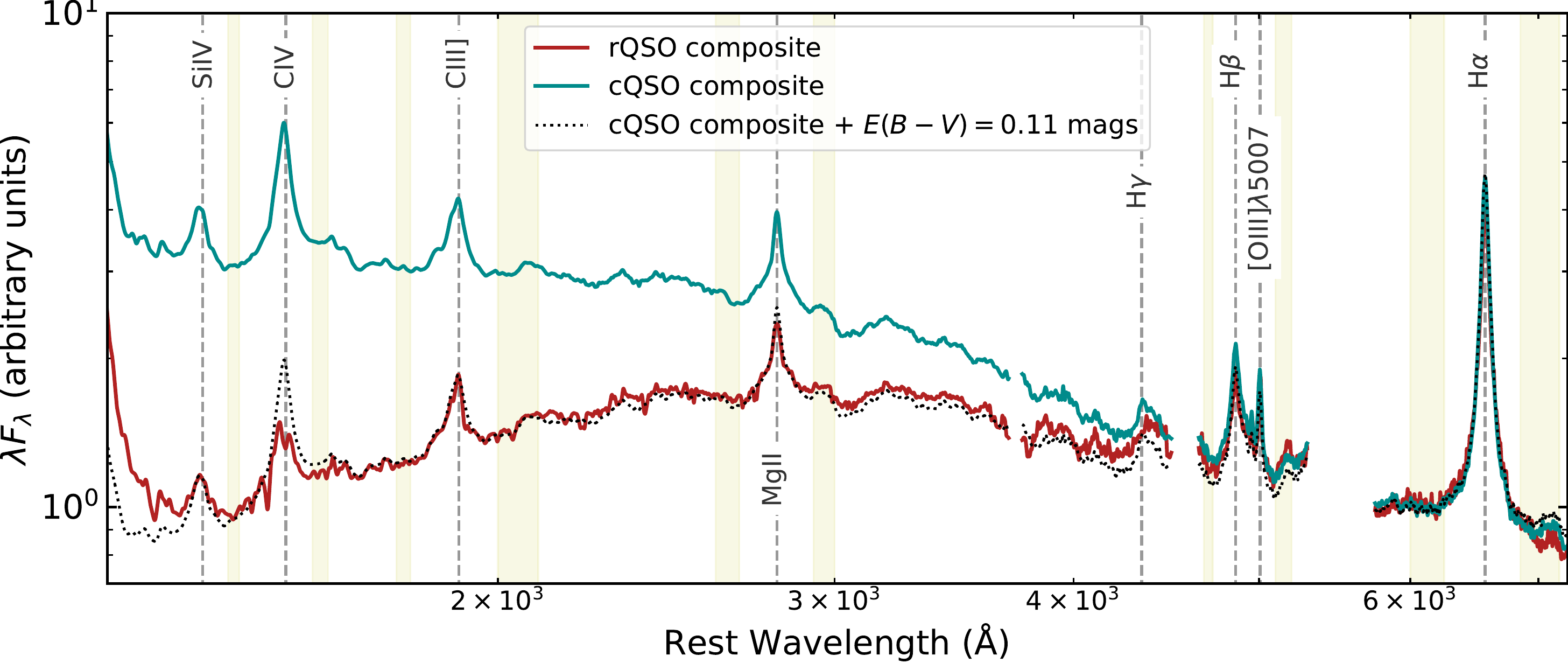}
    \caption{The geometric mean spectral composites for our rQSO and cQSO samples, normalized at 6000\,\AA. The frequency dependent deviation of the rQSO composite with respect to the cQSO composite is consistent with that expected for dust extinction. On the basis of our continuum fitting approach we find that the rQSO has an average dust extinction of $E(B-V)$\,$\sim0.11$\,mags ($A_V\sim 0.34$\,mags) which is indicated by the dotted black line; this agrees remarkably well with the rQSO composite, although there is some deviation in the \ion{C}{iv} emission-line profile. The shaded regions correspond to the windows used to define the continuum level in the emission-line fitting (see Section~\ref{sec:fitting2}) and the dashed vertical lines indicate the major emission lines. An electronic table of both composites is available through CDS.}
    \label{fig:composite}
\end{figure*}

We explored five different forms of extinction laws in this work: an SMC law (\citealt{richards}; $R_V=2.74$)\footnote{$R_V\equiv A_V/E(B-V)$: \cite{cardelli}}, a MW law (\citealt{cardelli}; $R_V=3.1$), a simple power-law (PL; $A_V\propto\uplambda^{-1}$, $R_V=4.0$ chosen to give consistent $A_V$ values compared to the other curves), a flat ``grey'' law (\citealt{czerny}; $R_V\sim2.0$) and a steeper law (\citealt{zafar}; $R_V=2.2$); a comparison of these extinction curves is displayed in Fig.~\ref{fig:extinction}. Only the MW-type curve displays the characteristic 2175\,\AA~feature expected from small carbonaceous grains present in our Galaxy. A steep extinction curve implies that the dust is predominately composed of smaller grains, which could be intrinsic or may be due to the destruction of larger grains by the intense AGN radiation field \citep{zafar}. A flatter extinction curve therefore implies the dust is composed of larger grains that either formed far our from the central region, or are produced in an outflowing wind that escapes the central region quick enough to avoid the destruction of larger grains \citep{czerny,gallerani}.

To quantify the amount of dust extinction present in each source, we constructed a composite spectrum of our cQSOs which was then fit to both the rQSOs and cQSOs using a least-squares minimization code, which varied the normalization (fixed at the NIR end) and $E(B-V)$ parameters for each of the five different extinction curves. To create the composites, the Galactic extinction-corrected spectra were first shifted to rest-frame wavelengths (using $z_{\rm VI}$; see Section~\ref{sec:fitting}) and then adjusted to a common wavelength grid. The composite was then created by taking the geometric mean across all spectra; the geometric mean was used rather than the median or arithmetic mean since it preserves the spectral shape, extinction law, and mean extinction of the composite (see Appendix in \citealt{reichard}). A comparison of our cQSO and rQSO composites is shown in Fig.~\ref{fig:composite} and a more detailed comparison of our cQSO composite to other QSO templates from the literature can be found in Appendix~\ref{sec:comp_comp}. For the extinction-curve analysis, we then masked the emission-line regions and smoothed the spectrum of the cQSO composite with a Gaussian filter ($\upsigma=3$) to produce our unreddened QSO model. It should be noted that some of the cQSOs will also contain modest amounts of extinction, which will therefore be incorporated into the cQSO composite. However, we expect the extinction towards the cQSOs to be modest; indeed, splitting the cQSO composite into the bluest and reddest 50 per~cent of sources, we estimate a spread in extinction values of $E(B-V)\sim0.013$\,mags ($A_V\sim0.04$\,mags; consistent to the results found from fitting broad-band SEDs for statistical samples in \citealt{calistro}).

The Balmer decrement provides a constraint on the amount of dust extinction along our line-of-sight towards the BLR, compared to the extinction-curve analysis which constrains the amount of dust extinction in the continuum region. However, due to the expected proximity of the accretion disc and BLR for the standard QSO model, to first order we would expect very similar levels of line-of-sight extinction. To calculate the dust extinction from the Balmer decrement, we used the equation from \cite{calzetti}:
\begin{equation}\label{eq:balmer}
     E(B-V) = \frac{1.086}{\rm k(H\,\upbeta)- \rm k(H\,\upalpha)} \rm ln\bigg(\frac{[H\,\upalpha/H\,\upbeta]_{\rm measured}}{[H\,\upalpha/H\,\upbeta]_{\rm intrinsic}}\bigg) \ ,
\end{equation}
where k($\uplambda$) is the extinction curve evaluated at $\uplambda$: in our calculation, we used a simple PL extinction curve which gives k(H\,$\upalpha$)$=3.35$ and k(H\,$\upbeta$)$=4.53$. The observed Balmer decrement [H\,$\upalpha$/H\,$\upbeta$]$_{\rm measured}$ represents the broad H\,$\upalpha$ to H\,$\upbeta$ flux ratio, obtained from the emission-line fitting (see Section~\ref{sec:fit}). The intrinsic Balmer decrement [H\,$\upalpha$/H\,$\upbeta$]$_{\rm intrinsic}$ is usually set to the theoretical ``Case B'' value of 2.88 \citep{osterbrock}. However, this value is not representative of the typical broad line Balmer decrement in unreddened QSOs; for example, the Balmer decrement of our cQSO composite is 3.40 (consistent with the average or composite Balmer decrement from \citealt{dong,glik,shen11}). In order for our Balmer reddening measurements to be consistent with our continuum-based reddening estimates (see Section~\ref{sec:extinction}), we adopted the Balmer decrement of our cQSO composite as the intrinsic ratio, taking a similar approach to Section 5.2 within \cite{glik}. From Eq.~\ref{eq:balmer}, a BLR dust extinction of $E(B-V)=0.2$\,mags, for the same measured H\,$\upalpha$ flux, would correspond to a drop in the H\,$\upbeta$ flux of around $\sim20$ per~cent.


\section{Spectral Fitting}\label{sec:fit}

To characterize the continuum and emission-line properties of our QSO sample, we fitted the \textit{X-shooter} spectra using the publicly available multi-component fitting code \textsc{PyQSOFit}\footnote{https://github.com/legolason/PyQSOFit} \citep{guo}. A detailed description of the code can be found in \cite{Guo_2019} and \cite{Shen_2019}. In this section we describe our approach to fitting the continuum emission (Section~\ref{sec:fitting}) and the emission lines (Section~\ref{sec:fitting2}).
  
\subsection{Continuum fitting: global approach} \label{sec:fitting}
For each source, we first smoothed the spectra with a 10 pixel box-car and then globally fitted both the continuum and emission lines of the entire spectrum. To fit the continuum we used a power-law, Balmer continuum (BC) and \ion{Fe}{ii} continuum components, providing our own redshifts based on the H\,$\upalpha$ broad line. Since all our sources are at a redshift $z>1.16$, \textsc{PyQSOFit} does not include a host-galaxy component; our sample is also luminous, and so we expect the effect of the host galaxy on the QSO spectra to be minimal (\citealt{calistro}; Fawcett et al. {\it in prep}; see Section~\ref{sec:ext_method}). 

The power-law continuum ($f_{\rm PL}$) is defined as
\begin{equation}
    f_{\rm PL}(\uplambda) = a_0 (\uplambda/\uplambda_0)^{a_1} \ ,
\end{equation}
where $\uplambda_0=3000$\,\AA~is the reference wavelength. The parameters $a_0$ and $a_1$ are the normalization and power-law slope, respectively.

The BC ($f_{\rm BC}$; \citealt{Dietrich_2002}) is defined as:
\begin{equation}
    f_{\rm BC}(\uplambda) = F_{\rm BE} B_{\uplambda}(T_e)\big(1-e^{-\tau_{\uplambda}}(\uplambda/\uplambda_{\rm BE})^3\big) \ ,
\end{equation}
where $F_{\rm BE}$ is the normalized flux density, $B_{\uplambda}(T_e)$ is the Planck function at the electron temperature $T_e$ and $\tau_{\uplambda}$ is the optical depth at the Balmer edge $\uplambda_{\rm BE}=3646$\,\AA. Here $F_{\rm BE}$, $T_e$ and $\tau_{\uplambda}$ are free parameters. 

The \ion{Fe}{ii} model ($f_{\rm \ion{Fe}{ii}}$) is defined as:
\begin{equation}
    f_{\rm \ion{Fe}{ii}}(\uplambda) = b_0 F_{\rm \ion{Fe}{ii}}(\uplambda,b_1,b_2) \ , 
\end{equation}
where $b_0$, $b_1$ and $b_2$ are the normalization, Gaussian FWHM used to convolve the \ion{Fe}{ii} template $F_{\rm \ion{Fe}{ii}}$, and the wavelength shift applied to the \ion{Fe}{ii} template, respectively. \textsc{PyQSOFit} provides two \ion{Fe}{ii} templates: a UV component that combines the \cite{vest} template for rest wavelengths 1000--2200\,\AA, the \cite{sal} template for 2200--3090\,\AA, and the \cite{Tsuzuki} template for 3090--3500\,\AA, and an optical component that uses the \cite{b_g} template for 3686--7484\,\AA. We chose to apply the \cite{Verner_2009} \ion{Fe}{ii} template for $\uplambda>2000$\,\AA, combined with the \cite{vest} template for $\uplambda<2000$\,\AA, to give a continuous template from 1000--12000\,\AA. We preferentially chose the \cite{Verner_2009} template for $\uplambda>2000$\,\AA~rather than the default templates provided in \textsc{PyQSOFit}, since in the \textit{X-shooter} spectra the NIR arm has a lower SNR than the UVB--VIS, and so tying the NIR \ion{Fe}{ii} template to the UVB--VIS \ion{Fe}{ii} template ensures a more physical \ion{Fe}{ii} model for the entire spectrum. 

For the QSOs with a measured dust extinction of $E(B-V)>0.03$\,mags,\footnote{This includes all of the rQSOs and 8 of the cQSOs (see Table~\ref{tab:ext_tab}).} we also include a third-order polynomial to account for the intrinsic dust extinction, defined as:
\begin{equation}
    f_{\rm poly}(\uplambda) =  \sum_{i=1}^{3}c_i(\uplambda-\uplambda_0)^i \ ,
\end{equation}
where $c_i$ are the polynomial coefficients. 

We set `rej\_abs = True', which removes any outliers falling below $3\sigma$ of the continuum for wavelengths $<3500$\,\AA~to reduce the effect of absorption lines biasing the fitting. 

From fitting the spectra, we found small disagreements between the SDSS redshift and the redshift inferred based on the observed wavelength of H\,$\upalpha$. The corrections were modest (average $\Delta z\sim 0.004$) with the most extreme correction of $\Delta z=0.0275$ applied to cQSO 1357-0307. Both the SDSS redshift ($z_{\rm SDSS}$) and amended redshift ($z_{\rm VI}$) are shown in Table~\ref{tab:sample}. Throughout the rest of the paper we use the $z_{\rm VI}$ values.

\subsection{Emission-line fitting: local continuum approach}\label{sec:fitting2}

\begin{figure*}
    \centering
    \includegraphics[width=6in]{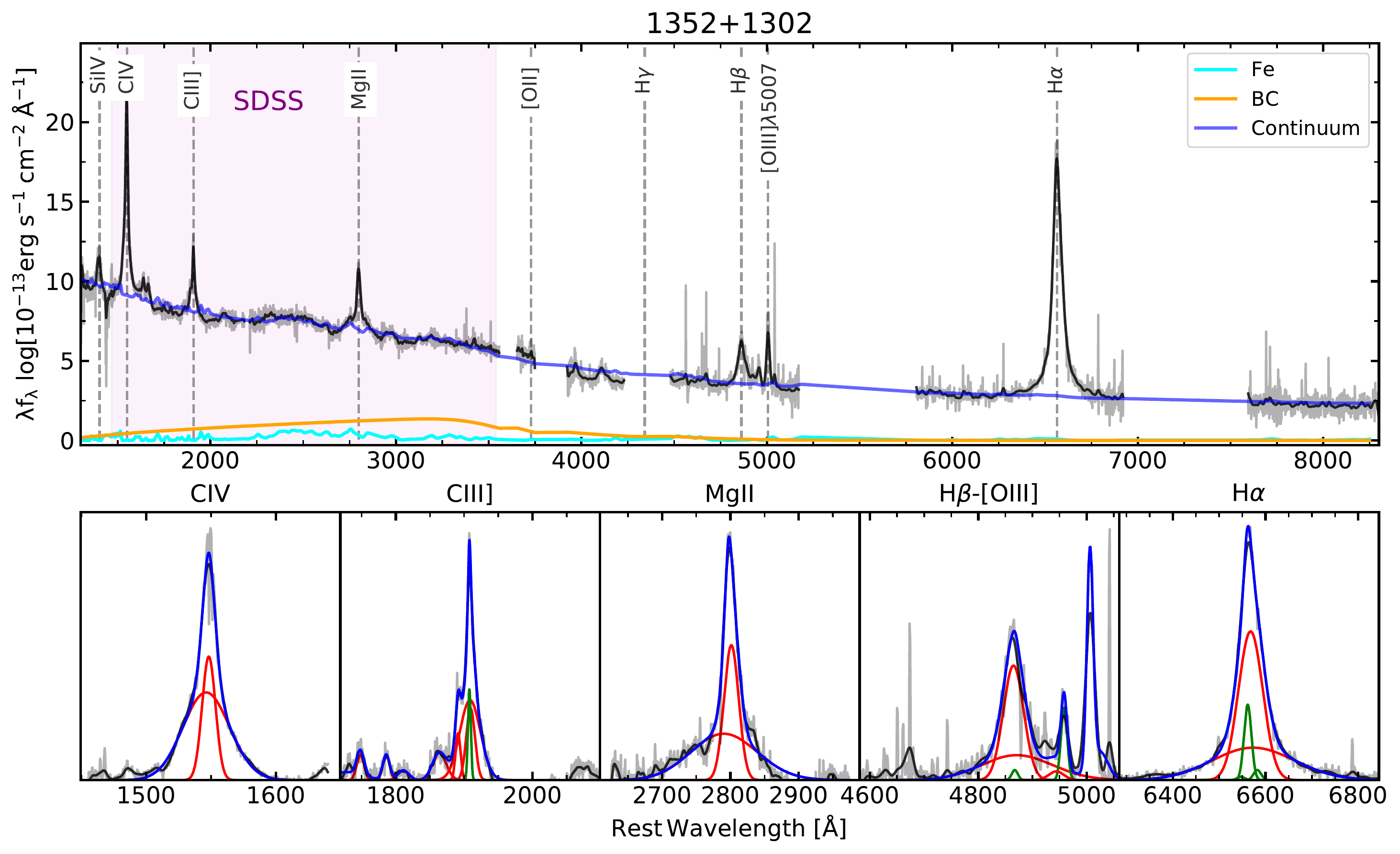}
    \caption{An example of our continuum and emission-line best-fit solutions for a single rQSO. In the top panel, the BC, \ion{Fe}{ii} and total continuum are shown in orange, cyan and blue lines, respectively. The SDSS wavelength coverage for the same source is shown in shaded magenta for comparison. In the bottom panels from left to right are the \ion{C}{iv}, \ion{C}{iii}], \ion{Mg}{ii}, H\,$\upbeta$-[\ion{O}{iii}] and H\,$\upalpha$ emission-line fits: the narrow Gaussian components are shown in green and the broad ($>1200$\,km\,s$^{-1}$) components are shown in red. The total line profile is displayed in blue.}
    \label{fig:pyqsofit}
\end{figure*}

To fit the emission lines we first subtracted from the spectra the \ion{Fe}{ii} and BC models based on the global fit, before re-fitting the emission lines using small, localised power-law continuum windows (see shaded region in Fig.~\ref{fig:composite}). The emission lines were modelled using Gaussian components: 1 Gaussian for the narrow lines and 3 Gaussians (1 narrow and 2 broad) for the broad lines. A local fitting approach for the emission lines was adopted, rather than a global approach, in order to reduce the systematic residuals from the telluric correction which can affect the performance of the global continuum fits. In M16, they compared both approaches and found that using a local continuum leads to a very small, but systematic, overestimation of the continuum emission, which will consequently lead to an underestimation of the FWHM. Depending on the emission line, we adopted additional constraints in the fitting process as described below.

For selected emission lines ([\ion{O}{iii}]$\uplambda\uplambda$4959,5007, H\,$\upbeta$, [\ion{N}{ii}]$\uplambda\uplambda$6549,6585 and [\ion{S}{ii}]$\uplambda\uplambda$6718,6732) we tied either the flux ratio, line width or both (following a similar approach to \citealt{shang}; see Table~4 therein). The two [\ion{O}{iii}]$\uplambda\uplambda$4959,5007 emission lines were first fitted using a core and a wing component for each line; the fluxes of the core components were tied to the theoretical 1:3 ratio, respectively, and the FWHM of the wing components were limited to $<2000$\,km\,s$^{-1}$ (similar to that in \citealt{calistro}). The widths for all of the other narrow emission lines were set to the fitted [\ion{O}{iii}]$\uplambda$5007 FWHM, if available, or were otherwise limited to $<1200$\,km\,s$^{-1}$. The maximum velocity offset for the narrow lines from the systemic redshift is also limited to 500\,km\,s$^{-1}$. Within the H\,$\upalpha$ complex, the flux of the [\ion{S}{ii}]$\uplambda\uplambda$6718,6732 lines were fixed to a 1:1 ratio, and the flux ratio of the [\ion{N}{ii}]$\uplambda$6549 and [\ion{N}{ii}]$\uplambda$6585 lines were set at 1:3, respectively. The H\,$\upalpha$ line was freely fit with three Gaussians to represent the broad and narrow emission-line components. The H\,$\upbeta$ emission line was then fitted using H\,$\upalpha$ as a template, fixing the peak and widths of the two broad components to be the same as that of H\,$\upalpha$. This approach was required due to the low signal-to-noise of the H\,$\upbeta$ line in $\sim28$\,per~cent of the sources. For the broad emission lines, we limited the FWHM range between $1000$ and $20000$\,km\,s$^{-1}$. We also allowed velocity offsets of up to 1000\,km\,s$^{-1}$ for all broad lines apart from \ion{C}{iv}, for which we allowed offsets of up to $4000$\,km\,s$^{-1}$ to accommodate for blueshifts associated with potential outflows \citep{rankine}. Our chosen offsets and line width limits are based on previous literature \citep{trak,Guo_2019}. As a further check, after fitting the lines we visually inspected each spectral fit to verify that our approach provided a good characterization of the emission lines and did not miss any components with broader emission or larger velocity offsets.

The uncertainties in the spectral quantities were estimated using a Monte Carlo approach within the least squares fitting code \texttt{kmpfit}, using 50 iterations. An example of our continuum and emission-line fitting is show in Fig.~\ref{fig:pyqsofit}.

\section{Accretion-disc fitting and black-hole mass measurements}\label{sec:AD}

To assess whether rQSOs and cQSOs have different accretion properties (e.g., as may be predicted if the rQSOs are an earlier black-hole growth phase), we compared the \textit{X-shooter} spectra to standard AD models. For this purpose, we adopted a geometrically thin, radiative AD model from \cite{slone}, and fit these models to the individual spectra of both the rQSOs and cQSOs, correcting for dust extinction. Three parameters completely describe these models: black-hole mass ($6.3<\rm M_{\rm BH}<10.3$\,log[M$_{\odot}$]), black-hole spin ($-1.0<a<0.998$) and mass accretion rate ($-3.0<\dot{M}<1.0$\,log\,$\rm M_{\odot}\,yr^{-1}$); these limits are based on standard QSO models and are similar to that used in C15.

To reduce the number of free parameters in the model, we can constrain M$_{\rm BH}$ using the viral relation from single-epoch spectra (as done in e.g., C15). In our work we use the FWHM of the broad H\,$\upalpha$ line to calculate M$_{\rm BH}$, since this is less affected by dust extinction than the more widely used \ion{Mg}{ii} line, consistent with previous red QSO black-hole mass estimates.\footnote{Since these are luminous QSOs we expect any contamination to the broad H\,$\upalpha$ emission line from star-formation to be negligible.} We adopted the M$_{\rm BH}$ calibration from \cite{Shen_2012}:

\begin{equation}\label{eq:bh}
    \rm log\bigg(\frac{M_{BH,vir}}{M_{\odot}}\bigg) = \mathit{a} + \mathit{b}log\bigg(\frac{\mathit{L}}{10^{44}erg\,s^{-1}}\bigg)+\mathit{c}log\bigg(\frac{FWHM}{km\,s^{-1}}\bigg) ,
\end{equation}
where $a=1.390$, $b=0.555$ and $c=1.873$ for FWHM$_{\rm H\,\upalpha}$ and $L_{5100}$. We also estimated M$_{\rm BH}$ from the \ion{Mg}{ii} emission line for completeness and to provide a consistent comparison to the M$_{\rm BH}$ values estimated by C16, using $a=1.816$, $b=0.584$ and $c=1.712$ for FWHM$_{\rm \ion{Mg}{ii}}$ and $L_{3000}$. 

To fit the AD model and determine the best-fitting parameters we used the \texttt{Python} $\upchi^{2}$ minimization code \texttt{lmfit}\footnote{https://lmfit.github.io/lmfit-py/}, adopting a least-squares algorithm. We set M$_{\rm BH}$ to the values obtained from this approximation and allowed this to vary within the errors. We ran the fitting twice: first fixing the value of $A_V$ to the non-negative best-fitting values obtained from our dust-extinction fitting (see Sections~\ref{sec:ext_method} and \ref{sec:extinction}), and then allowing this to vary as a free parameter in order to test whether a dust component is required in order to produce the best fits for the rQSOs. In both cases, the mass accretion rate and black-hole spin were left as free parameters. 
\section{Results}
Using our \textit{X-shooter} spectra for 12 rQSOs and 28 cQSOs at $1.45<z<1.65$, we have fitted the continuum and emission-line components with physically motivated models (see Sections~\ref{sec:ext_method}, \ref{sec:fit}, and \ref{sec:AD}) to constrain a range of key parameters; in the following sections we present our results.

\subsection{Quantifying the level of dust extinction in red quasars}\label{sec:ext}
\subsubsection{Continuum reddening: extinction-curve comparison} \label{sec:extinction}
\begin{table*}
\Rotatebox{90}{%
    \centering
    \begin{tabular}{cc|cc|cc|cc|cc|cc|c}
    \hline
    \hline
    & & \multicolumn{2}{c|}{SMC} & \multicolumn{2}{c|}{MW}  & \multicolumn{2}{c|}{PL} & \multicolumn{2}{c|}{Flat} & \multicolumn{2}{c|}{Steep} & H\,$\upalpha$/H\,$\upbeta$ \\ 
    \hline 
    Name & Sample & $E(B-V)$ & $\rm \upchi^{2}_r$ & $E(B-V)$ & $\rm \upchi^{2}_r$ & $E(B-V)$ & $\rm \upchi^{2}_r$ & $E(B-V)$& $\rm \upchi^{2}_{\rm r}$ & $E(B-V)$ & $\rm \upchi^{2}_r$  & $E(B-V)$\\ 
    \hline
    1020+1101\textsuperscript{$a$} & rQSO & \textbf{0.05$\pm$0.001} & \textbf{1.29} & 0.05$\pm$0.001 & 1.47 &0.02$\pm$0.00 & 1.82 & \textbf{0.03$\pm$0.001} & \textbf{1.00} & 0.06$\pm$0.0003 & 1.81 & 0.96$\pm$0.01\\  
    1049+1157 & rQSO & 0.20$\pm$0.001 & 1.96 & 0.22$\pm$0.002 & 1.31 & \textbf{0.17$\pm$0.001} & \textbf{0.92} & 0.19$\pm$0.002 & 1.73 & \textbf{0.16$\pm$0.001} & \textbf{0.93} & 1.24$\pm$0.01\\  
    1358+1145\textsuperscript{$a$} & rQSO & \textbf{0.06$\pm$0.001} & \textbf{1.99} & 0.06$\pm$0.001 & 2.35 & 0.04$\pm$0.001 & 2.30 & 0.06$\pm$0.001 & 2.39 & \textbf{0.05$\pm$0.0003} & \textbf{2.05} & 0.51$\pm$0.08\\  
    1429-0112 & rQSO & 0.19$\pm$0.001 & 1.47 & 0.19$\pm$0.001 & 1.77 & \textbf{0.18$\pm$0.001} & \textbf{1.04} & 0.17$\pm$0.001 & 2.13 & \textbf{0.17$\pm$0.001} & \textbf{1.10} & 0.36$\pm$0.01\\  
    1442+1426 & rQSO & 0.13$\pm$0.001 & 1.29 & \textbf{0.16$\pm$0.002} & \textbf{1.08} & \textbf{0.17$\pm$0.001} & \textbf{1.20} & 0.10$\pm$0.001 & 0.81 & \textbf{0.12$\pm$0.001} & \textbf{0.99} & 0.78$\pm$0.05\\  
    1523+0452 & rQSO & 0.10$\pm$0.001 & 1.33 & 0.14$\pm$0.002 & 2.17 & \textbf{0.12$\pm$0.001} & \textbf{0.95} & 0.11$\pm$0.002 & 1.90 & \textbf{0.10$\pm$0.001} & \textbf{0.90} & 0.31$\pm$0.02\\  
    1608+1207 & rQSO & \textbf{0.07$\pm$0.001} & \textbf{1.12} & 0.07$\pm$0.001 & 1.99 & 0.05$\pm$0.001 & 1.61 & 0.07$\pm$0.001 & 1.73 & 0.05$\pm$0.001 & 1.54 & -0.10$\pm$0.01\\  
    1616+0931\textsuperscript{$a$} & rQSO & 0.23$\pm$0.001 & 1.75 & 0.26$\pm$0.001 & 2.83 & \textbf{0.23$\pm$0.001} & \textbf{1.14} & 0.25$\pm$0.001 & 2.19 & \textbf{0.21$\pm$0.001} & \textbf{1.07} & 0.23$\pm$0.01\\  
    1639+1135 & rQSO & 0.07$\pm$0.001 & 2.99 & 0.03$\pm$0.002 & 7.18 & 0.05$\pm$0.001 & 2.99 & 0.06$\pm$0.001 & 4.97 & \textbf{0.06$\pm$0.001} & \textbf{2.36} & 0.71$\pm$0.01\\  
    2133+1043\textsuperscript{$a$} & rQSO & \textbf{0.09$\pm$0.001} & \textbf{1.91} & 0.09$\pm$0.001 & 3.02 & \textbf{0.08$\pm$0.001} & \textbf{1.92} & 0.08$\pm$0.001 & 2.92 & \textbf{0.08$\pm$0.001} & \textbf{1.92} & 0.46$\pm$0.04\\
    2223+1258 & rQSO & \textbf{0.08$\pm$0.001} & \textbf{2.93} & 0.07$\pm$0.001 & 4.37 & \textbf{0.07$\pm$0.001} & \textbf{2.92} & 0.06$\pm$0.001 & 4.84 & \textbf{0.06$\pm$0.001} & \textbf{2.98} & 0.003$\pm$0.01\\  
    2241-1006\textsuperscript{$a$} & rQSO & 0.03$\pm$0.001 & 10.00 & 0.02$\pm$0.001 & 11.63 & 0.02$\pm$0.001 & 9.39 & 0.02$\pm$0.001 & 11.55 & 0.02$\pm$0.001 & 9.14 & 0.01$\pm$0.01\\  
    \hline    
    0043+0114\textsuperscript{$a$} & cQSO & -0.08$\pm$0.001 & 1.72 & -0.01$\pm$0.0005 & 1.08 & -0.01$\pm$0.0003 & 0.95 & -0.02$\pm$0.0005 & 0.96 & -0.01$\pm$0.0001 & 0.96 & 0.25$\pm$0.01\\
    0152-0839 & cQSO & -0.01$\pm$0.001 & 3.03 & -0.02$\pm$0.001 & 1.86 & -0.01$\pm$0.0002 & 1.00 & -0.02$\pm$0.0005 & 1.11 & -0.02$\pm$0.0003 & 1.84 & 0.24$\pm$0.01\\
    0155-1023 & cQSO & -0.06$\pm$0.002 & 10.25 & -0.02$\pm$0.0002 & 1.78 & -0.02$\pm$0.0003 & 1.84 & -0.02$\pm$0.0004 & 1.72 & -0.02$\pm$0.0003 & 1.67 & 0.13$\pm$0.06\\
    0213-0036 & cQSO & 0.02$\pm$0.001 & 1.07 & 0.04$\pm$0.001 & 0.32 & 0.02$\pm$0.0004 & 0.35 & 0.02$\pm$0.001 & 0.31 &0.02$\pm$0.0004 & 0.37 & 0.26$\pm$0.02\\
    0223-0007\textsuperscript{$a$} & cQSO & -0.02$\pm$0.001 & 1.05 & -0.04$\pm$0.001 & 1.48 & -0.02$\pm$0.0004 & 0.96 & -0.03$\pm$0.001 & 1.03 &-0.02$\pm$0.0004 & 0.92 & 0.22$\pm$0.02\\
    0303+0027 & cQSO & -0.09$\pm$0.0003 & 4.53 & -0.02$\pm$0.0004 & 1.03 & -0.01$\pm$0.0001 & 1.31 & -0.02$\pm$0.0004 & 1.17 & -0.01$\pm$0.0002 & 1.34 & -0.27$\pm$0.004\\
    0341-0037 & cQSO & 0.02$\pm$0.001 & 1.27 & 0.03$\pm$0.001 & 1.08 & -0.03$\pm$0.001 & 0.96 & -0.01$\pm$0.001 & 0.92 & 0.02$\pm$0.0005 & 0.93 & 0.24$\pm$0.03\\
    0404-0446\textsuperscript{$a$} & cQSO & -0.06$\pm$0.002 & 8.91 & -0.03$\pm$0.003 & 2.57 & -0.01$\pm$0.001 & 4.98 & -0.02$\pm$0.0004 & 4.53 & -0.01$\pm$0.001 & 4.85 & 1.08$\pm$0.08\\
    0842+0151 & cQSO & -0.09$\pm$0.004 & 9.07 & -0.02$\pm$0.0004 & 1.32 & -0.02$\pm$0.0004 & 1.31 & -0.03$\pm$0.0005 & 1.09 & -0.02$\pm$0.0003 & 1.34 & -0.11$\pm$0.01\\
    0927+0004 & cQSO & 0.02$\pm$0.001 & 1.34 & -0.01$\pm$0.001 & 0.92 & -0.02$\pm$0.001 & 0.96 &  0.02$\pm$0.001 & 1.66 & 0.02$\pm$0.0004 & 1.96 & -0.02$\pm$0.02\\
    0934+0005 & cQSO & 0.02$\pm$0.0002 & 1.12 & 0.02$\pm$0.001 & 1.67 & 0.003$\pm$0.0004 & 1.09 & 0.001$\pm$0.001 & 1.14 & 0.004$\pm$0.0003 & 0.99 & 0.11$\pm$0.04\\
    0941+0443 & cQSO & -0.01$\pm$0.0004 & 1.89 & -0.004$\pm$0.0004 & 1.17 & -0.16$\pm$0.0003 & 1.01 & -0.01$\pm$0.0004 & 0.91 & -0.01$\pm$0.0003 & 0.95 & 0.14$\pm$0.01\\
    1001+1015 & cQSO & 0.03$\pm$0.0004 & 1.12 & -0.03$\pm$0.001 & 1.74 & 0.04$\pm$0.0003 & 1.02 & -0.03$\pm$0.001 & 1.46 & 0.003$\pm$0.0002 & 0.92 & 0.19$\pm$0.01\\
    1002+0331 & cQSO & 0.02$\pm$0.001 & 1.41 & 0.04$\pm$0.001 & 1.17 & 0.03$\pm$0.0004 & 1.17 &  0.02$\pm$0.0007 & 1.63 & 0.03$\pm$0.0003 & 1.18 & 0.02$\pm$0.04\\
    1013+0245 & cQSO & 0.11$\pm$0.003 & 1.91 & 0.13$\pm$0.002 & 1.45 & 0.09$\pm$0.002 & 1.90 & 0.12$\pm$0.002 & 0.91 & 0.11$\pm$0.002 & 1.76 & 0.33$\pm$0.04\\
    1028+1456 & cQSO & 0.008$\pm$0.001 & 1.59 & 0.08$\pm$0.001 & 1.25 & 0.06$\pm$0.001 & 1.45 & 0.008$\pm$0.002 & 1.99 & 0.06$\pm$0.001 & 1.52 & 0.04$\pm$0.02\\
    1158-0322\textsuperscript{$a$} & cQSO & 0.02$\pm$0.001 & 1.89 & 0.02$\pm$0.001 & 1.89 & 0.01$\pm$0.0005 & 1.91 & 0.005$\pm$0.001 & 2.33 & 0.01$\pm$0.0004 & 1.91 & 0.06$\pm$0.01\\
    1352+1302 & cQSO & 0.002$\pm$0.0004 & 1.66 & 0.01$\pm$0.0005 & 1.35 & 0.01$\pm$0.0004 & 1.83 & 0.003$\pm$0.001 & 1.65 & -0.004$\pm$0.0003 & 1.78 & 0.19$\pm$0.01\\
    1357-0307 & cQSO & -0.008$\pm$0.001 & 1.42 & 0.04$\pm$0.001 & 1.08 & 0.03$\pm$0.001 & 0.97 & 0.004$\pm$0.001 & 0.93 & 0.03$\pm$0.001 & 0.92 & 0.52$\pm$0.01\\
    1358+1410 & cQSO & 0.04$\pm$0.001 & 1.17 & 0.04$\pm$0.0002 & 0.91 & 0.01$\pm$0.001 & 1.18 & 0.03$\pm$0.001 & 1.19 & 0.03$\pm$0.0005 & 0.93 & 0.13$\pm$0.004\\
    1428+1001\textsuperscript{$a$} & cQSO & 0.02$\pm$0.001 & 1.64 & 0.06$\pm$0.001 & 1.63 & 0.05$\pm$0.001 & 1.87 & 0.07$\pm$0.001 & 1.70 & 0.05$\pm$0.001 & 1.80 & 0.22$\pm$0.02\\
    1502+1016 & cQSO & 0.02$\pm$0.0003 & 1.06 & 0.02$\pm$0.001 & 1.37 & 0.01$\pm$0.0004 & 0.94 & 0.03$\pm$0.001 & 1.10 & 0.01$\pm$0.001 & 0.97 & 0.17$\pm$0.02\\
    1513+1011 & cQSO & -0.01$\pm$0.0003 & 23.8 & -0.03$\pm$0.001 & 1.84 & -0.01$\pm$0.0003 & 1.42 & -0.03$\pm$0.0002 & 1.35 & -0.01$\pm$0.0004 & 1.47 & 0.48$\pm$0.02\\
    1521-0156 & cQSO & -0.07$\pm$0.001 & 19.43 & -0.06$\pm$0.0002 & 4.34 & -0.03$\pm$0.001 & 11.24 & -0.02$\pm$0.001 & 4.62 & -0.03$\pm$0.001 & 4.67 & -0.04$\pm$0.01  \\  
    1539+0534 & cQSO & -0.005$\pm$0.01 & 8.94 & -0.05$\pm$0.001 & 1.16 & -0.03$\pm$0.0005 & 1.67 & -0.06$\pm$0.001 & 1.57 & -0.04$\pm$0.0005 & 1.76 & 0.08$\pm$0.05\\
    1540+1155 & cQSO & -0.01$\pm$0.001 & 1.83 & -0.03$\pm$0.0005 & 1.08 & -0.03$\pm$0.0004 & 1.17 & -0.03$\pm$0.001 & 1.52 & -0.03$\pm$0.0004 & 1.36 & -0.13$\pm$0.05\\
    1552+0939 & cQSO & 0.08$\pm$0.001 & 1.30 & 0.09$\pm$0.001 & 1.03 & 0.07$\pm$0.001 & 1.18 & 0.10$\pm$0.001 & 1.01 & 0.04$\pm$0.001 & 1.48 & 0.29$\pm$0.04\\
    1618+1305 & cQSO & -0.07$\pm$0.001 & 1.47 & 0.003$\pm$0.001 & 1.29 & 0.005$\pm$0.001 & 1.21 & 0.009$\pm$0.001 & 1.24 & -0.004$\pm$0.001 & 1.56 & 0.12$\pm$0.05\\
    \hline
    \hline

    \end{tabular} 
}%
\caption{Table of the best-fitting dust-extinction parameters and the corresponding reduced $\upchi^{2}$ ($\upchi^{2}_{\rm r}$) obtained from fitting the SMC, MW, PL, flat and steep extinction curves to our rQSOs and cQSOs. Overall, the $E(B-V)$ values found for the five different curves are very similar. For each rQSO, the $E(B-V)$ and $\upchi^{2}_{\rm r}$ values of all extinction curves both within $0.9<\upchi^2_{\rm r}<3$ and 0.3 of the best-fitting $\upchi^{2}_{\rm r}$ value are highlighted in bold. It is worth noting that there will be additional uncertainties introduced by the spread of extinction values within our composite ($E(B-V)\sim0.013$\,mags) that would need to be taken into account in order to convert to a fully unobscured QSO (see Section~\ref{sec:ext_method}). The reddening estimated from the Balmer decrements are shown for comparison. \textsuperscript{$a$}Associated \ion{Mg}{ii} absorption.}  
\label{tab:ext_tab}
\end{table*}

Under the assumption that dust is the main cause of reddening in the rQSOs (see Section~\ref{sec:ext_method}), we quantified the amount of dust extinction along the line-of-sight by fitting the QSOs with a dust-reddened cQSO composite, adopting five different extinction curves: SMC, MW, PL, steep and flat (see Fig.~\ref{fig:extinction}). We also investigated whether there was a preference between the five different extinction curves, which would provide an insight into the nature of the dust; i.e., the size and location of the dust grains. We varied the normalization and amount of dust extinction applied to our cQSO composite using each of the five extinction-curve models, and then fit this to our QSO spectra using a least-squares minimization code. The best-fitting $E(B-V)$ constraints and the corresponding reduced $\upchi^{2}$ ($\upchi^2_{\rm r}$) values are displayed in Table~\ref{tab:ext_tab}; we note that the $E(B-V)$ values were derived with respect to the cQSO composite which will contain some amount of dust extinction, although this is generally negligible (see Section~\ref{sec:ext_method} and \cite{calistro}). A movie that demonstrates how increasing amounts of dust extinction affects the shape of the cQSO composite in comparison to the rQSO spectra is shown in Fig. 8 of the online Supplementary material.

\begin{figure}
    \centering
    \includegraphics[width=3.3in]{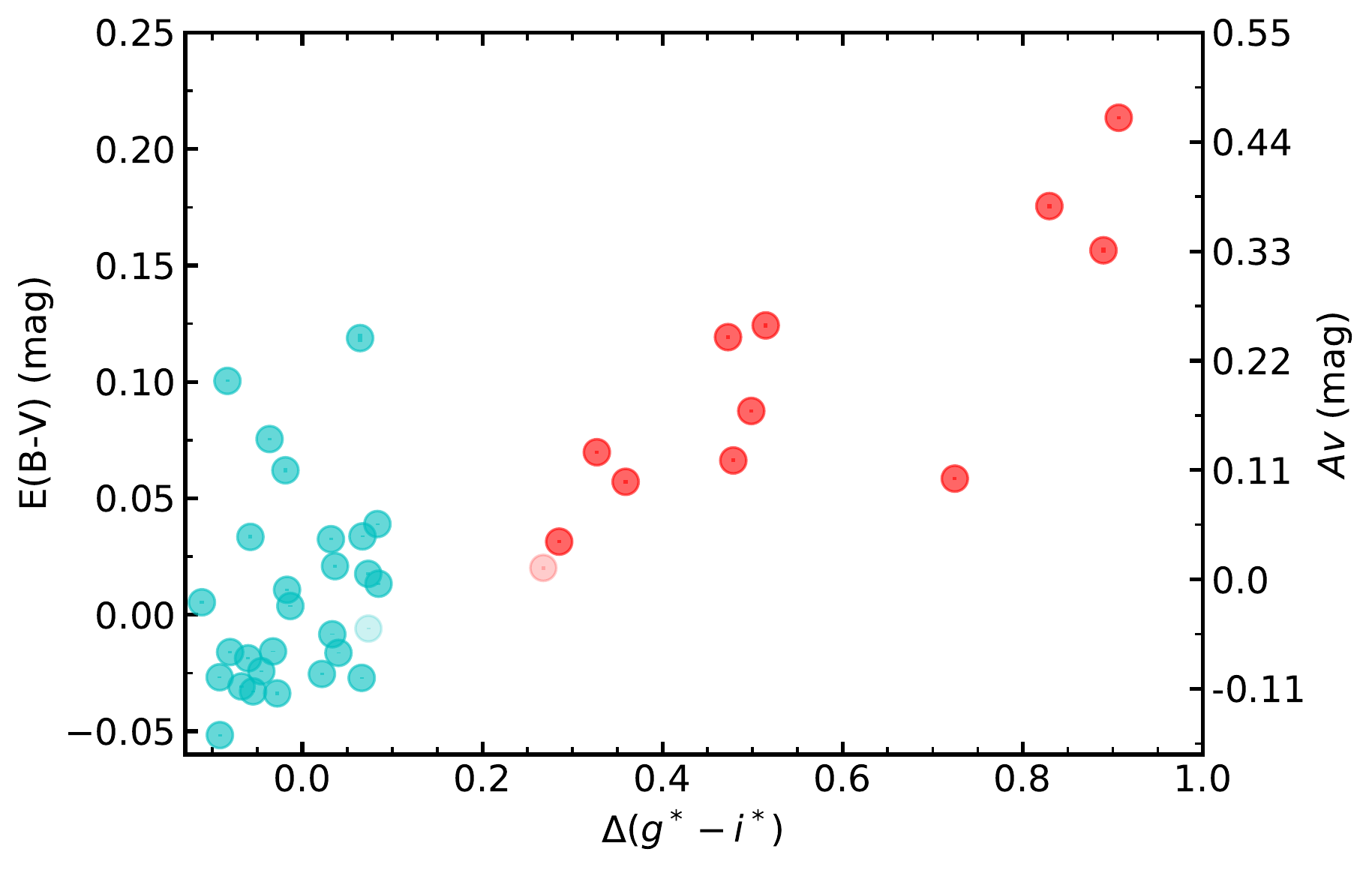}
    \caption{$\Delta(g^*-i^*)$ versus the $E(B-V)$ values obtained from the best fitting extinction curves for the rQSOs (red) and cQSOs (cyan) (see Table~\ref{tab:ext_tab}). Faded points indicate poor fits ($\upchi^2_{\rm r}>3$ or $\upchi^2_{\rm r}<0.9$). There is a clear relationship between optical colour and dust reddening.}
    \label{fig:EBV}
\end{figure}

Overall, for any given QSO, the $E(B-V)$ values obtained from the five curves are in good agreement; the standard deviation in the maximum difference between the best-fitting $E(B-V)$ values obtained from five extinction curves is $\sim0.014$\,mags. We define a fit to be good if the $\upchi^2_{\rm r}$ value is both within the range $0.9<\upchi^2_{\rm r}<3$ and within 0.3 of the $\upchi^2_{\rm r}$ value for the best-fitting extinction curve (the $E(B-V)$ and $\upchi^2_{\rm r}$ values for the good-fitting extinction curves are highlighted in bold in Table~\ref{tab:ext_tab}). By this definition, only one of the rQSOs (2241-1006) could not be well fitted using any of the curves. The poorly fitted rQSO is on the borderline of our rQSO definition and potentially has additional extinction due to an intervening absorber. The other rQSOs are well fitted by at least one of the extinction curves, which suggests that their red colours can be fully explained by dust extinction. For the rQSOs, we find extinction values in the range of $E(B-V)$\,$\sim 0.02$--$0.23$\,mags ($A_V\sim0.06$--0.7\,mags; with the additional uncertainty of $E(B-V)\sim0.013$\,mags introduced from our cQSO composite; see Section~\ref{sec:ext_method}), consistent with \cite{richards} and Figure~4 within \cite{klindt}. We find that many of the rQSOs have good fits from more than one extinction curve, preventing us from identifying a clear best-fitting extinction curve. However, our analyses do allow us to rule out certain extinction curves; we find that for all but one rQSO, it is possible to robustly rule out both the MW and flat extinction curves\footnote{It should be noted that we do not see a strong 2175\AA~bump in this source.}. This suggests that the characteristic 2175\,\AA~bump is rare in reddened QSOs, in agreement with previous studies (e.g., \citealt{zafar,temple_red}). The sharper rise of the PL, SMC, and steep extinction curves suggests that the dust is composed of smaller grains, formed within the vicinity of the SMBH where nuclear activity destroys the larger grains (\citealt{Li,zafar}; see Section~\ref{sec:dust} for a more detailed discussion). However, a larger sample of rQSOs with well-constrained $E(B-V)$ measurements is required to place robust constraints on the nature of the dust grains. Due to the lack of dust (by definition), it is not possible to distinguish between different extinction curves for the cQSOs. Excluding the poorly fitted sources, we find that there is a strong correlation between $\Delta(g^*-i^*)$ and dust extinction (see Fig.~\ref{fig:EBV}) which shows our method of selecting dust-reddened rQSOs is robust. We use the $E(B-V)$ values obtained from the best fitted extinction curve per source for further analyses. 

In our sample, 4 rQSOs and 3 cQSOs display narrow \ion{Mg}{ii} absorption features within their spectra, with $z_{abs}<z_{em}$ indicating a potential intervening non-associated absorber. For these sources there may be an additional extinction component associated with the absorbing system along the line-of-sight, and so the intrinsic QSO extinction could be slightly lower than that measured from our extinction-curve analysis; these sources are indicated within Table~\ref{tab:ext_tab}.

Fitting a dust-reddened cQSO composite to our rQSO composite (see Fig.~\ref{fig:composite}), we find that it is consistent with an average dust extinction of $E(B-V)=0.106\pm0.001$\,mags, again indicating that our rQSO sample is only moderately reddened. This is consistent with \cite{calistro}, who find a median value of $E(B-V)=0.12^{+0.21}_{-0.08}$\,mags for a sample of $\sim1000$ SDSS red QSOs, obtained from broad-band SED fitting. Interestingly, the emission-line profiles of the dust-reddened cQSO composite are similar to that in the rQSO composite, apart from \ion{C}{iv} which is more suppressed in the rQSO composite. This difference in the \ion{C}{iv} profile could be intrinsic, linked to the higher incidence of both broad and narrow absorption lines in the rQSO sample (see Section~\ref{sec:bh_dis} for a more detailed discussion), or a result of our small rQSO sample. The potential differences in the \ion{C}{iv} emission line between rQSOs and cQSOs will be explored using a much larger sample in our future DR14 spectroscopic study (Fawcett et~al. \textit{in prep}). 

\subsubsection{Broad line reddening: Balmer decrements}\label{sec:extinction_balmer}
\begin{figure*}
    \centering
    \includegraphics[width=5.6in]{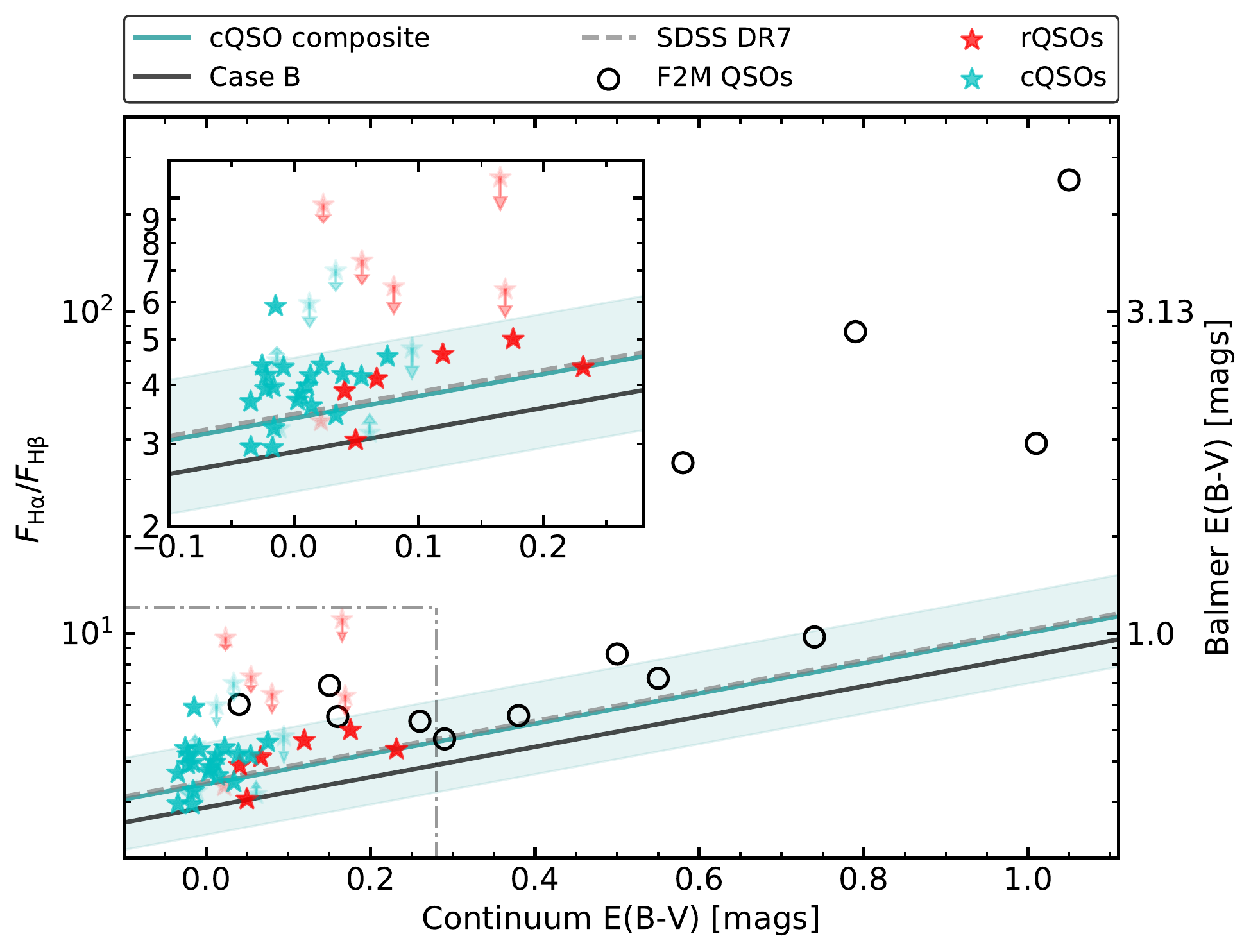}
    \caption{A comparison between the broad H\,$\upalpha$/H\,$\upbeta$ flux ratio and the best fitting dust extinction derived from the continuum for the rQSOs (red) and cQSOs (cyan). The solid black line represents the theoretical ``Case B'' (2.88), the solid cyan line represents the Balmer decrement of the cQSO composite (3.40) and the dashed grey line represents the median Balmer decrement in SDSS DR7 (3.47; \citealt{shen11}). The shaded region displays the expected scatter within the DR7 QSO population ($\sim2$; \citealt{shen11}). Faded points with arrows represent upper limits due to low SNR or undetected broad H\,$\upbeta$. The black circles indicate the heavily reddened F2M QSOs from \protect\cite{glik}. The inset shows a zoom in of the region containing the QSOs from our study.}
    \label{fig:balmer}
\end{figure*}

Using the measured broad H\,$\upalpha$ and H\,$\upbeta$ emission-line fluxes, we calculated the Balmer decrements for each rQSO and cQSO (see Section~\ref{sec:ext_method}). A comparison of the $E(B-V)$ values calculated from the continuum fitting with those calculated from the measured Balmer decrements is shown in Fig.~\ref{fig:balmer}. The heavily reddened FIRST-2MASS (F2M) QSOs from \cite{glik} are shown for comparison in addition to the expected intrinsic scatter within the DR7 QSO Balmer decrements from \cite{shen11}. 
The rQSOs display, on average, larger H\,$\upalpha$/H\,$\upbeta$ ratios, including a significantly larger fraction of upper limits that indicate the H\,$\upbeta$ line was too weak to fit (5/12: 42 per~cent of the rQSOs, compared to just 3/28: 11 per~cent of the cQSOs), consistent with a larger amounts of dust extinction along the line-of-sight towards the BLR. Excluding the upper limits, the majority of the cQSOs and rQSOs lie within the intrinsic Balmer decrement scatter as a function of $E(B-V)$. However, at the higher extinction values ($E(B-V)>0.55$\,mags) probed by the F2M sample, there is a discrepancy between the reddening derived from the continuum and that from the Balmer decrements, which could be due to poor H\,$\upbeta$ fits in noisy spectra, an intrinsically steep spectrum or a difference in the BLR--AD geometry between moderately and heavily reddened QSOs \citep{glik,kim18}. In future work (Fawcett et al. {\it in prep}) we will have the improved source statistics from our dedicated program using the Dark Energy Spectroscopic Instrument (DESI; \citealt{desi}), to push to more dust-reddened systems than those observed with SDSS and test whether there are any differences between the continuum and BLR dust reddening at higher $E(B-V)$ values.

\subsection{Continuum and emission-line properties}\label{sec:emission}
\begin{table*}
\Rotatebox{90}{%
    \centering
    \begin{tabular}{cc|cccccc|cccccc} 
    \hline
    \hline
    Name & Sample & \multicolumn{6}{c|}{FWHM [km\,s$^{-1}$]}&  \multicolumn{6}{c}{$L$ log[erg\,s$^{-1}$]}  \\
    & & \ion{C}{iv} & \ion{C}{iii}] & \ion{Mg}{ii} & [\ion{O}{iii}]  & H\,$\upbeta$ & H\,$\upalpha$ & \ion{C}{iv} & \ion{C}{iii}] & \ion{Mg}{ii} &  [\ion{O}{iii}] &  H\,$\upbeta$ &H\,$\upalpha$  \\
    \hline
1020+1101 & rQSO & 7565$\pm$98 & 5637$\pm$207 & 6399$\pm$70 & 303$\pm$142 & 6272 & 6043$\pm$63 & 44.2$\pm$0.04 & 44.0$\pm$0.01 & 44.0$\pm$0.02 & 43.0$\pm$0.1 & 43.5$\pm$0.001 & 44.5$\pm$0.002 \\  
1049+1157 & rQSO & 7423$\pm$3938 & 2606$\pm$211 & 11899$\pm$1212 & - & 7947 & 7702$\pm$160 & 44.2$\pm$0.05 & 43.6$\pm$0.11 & 44.0$\pm$0.02 & - & 43.7$\pm$0.001 & 44.5$\pm$0.003 \\
1358+1145 & rQSO & 2611$\pm$84 & 3418$\pm$498 & 2717$\pm$64 & 494$\pm$96 &  5303 & 5019$\pm$208 & 45.0$\pm$0.01 & 44.1$\pm$0.04 & 44.4$\pm$0.003 & 43.1$\pm$0.1 &  44.3$\pm$0.001 & 44.8$\pm$0.001 \\
1429-0112 & rQSO & 4665$\pm$609 & 8009$\pm$2743 & 3611$\pm$144 & 666$\pm$97 & 3207 & 3108$\pm$57 & 44.4$\pm$0.13 & 43.8$\pm$0.09 & 44.2$\pm$0.01 & 43.0$\pm$0.02 & 44.0$\pm$0.01 & 44.6$\pm$0.004 \\
1442+1426 & rQSO & 6327$\pm$133 & 3142$\pm$873 & 2933$\pm$42 & 390$\pm$487 & 2033 & 1985$\pm$39 & 44.4$\pm$0.05 & 44.3$\pm$0.04 & 44.3$\pm$0.01 & 42.4$\pm$0.1 & 43.7$\pm$0.01 & 44.4$\pm$0.004 \\
1523+0452 & rQSO & 4717$\pm$91 & 3448$\pm$553 & 2323$\pm$30 & 147$\pm$235 &  2108 & 2009$\pm$53 & 44.3$\pm$0.02 & 44.0$\pm$0.03 & 44.0$\pm$0.01 & 42.0$\pm$0.3 & 43.6$\pm$0.01 & 44.2$\pm$0.005 \\
1608+1207 & rQSO & 4569$\pm$121 & 4880$\pm$289 & 3471$\pm$183 & 164$\pm$372 & 3983 & 3886$\pm$44 & 44.3$\pm$0.02 & 43.9$\pm$0.04 & 43.9$\pm$0.02 & 41.5$\pm$0.1 & 43.9$\pm$0.003 & 44.4$\pm$0.002 \\
1616+0931 & rQSO & 7806$\pm$148 & 2258$\pm$758 & 5091$\pm$92 & 439$\pm$63 & 3727 & 3818$\pm$23 & 44.9$\pm$0.01 & 44.4$\pm$0.05 & 44.4$\pm$0.02 & 43.9$\pm$0.1 & 44.3$\pm$0.003 & 44.8$\pm$0.002 \\
1639+1135 & rQSO & 14933$\pm$579 & 6351$\pm$331 & 7033$\pm$80 & 199$\pm$174 & 6343 & 6408$\pm$55 & 44.6$\pm$0.01 & 44.3$\pm$0.02 & 44.5$\pm$0.003 & 42.8$\pm$0.1 & 44.0$\pm$0.01 & 44.8$\pm$0.001 \\
2133+1043 & rQSO & 6835$\pm$151 & 5404$\pm$1045 & 3014$\pm$74 & -  & 3414 & 3512$\pm$121 & 44.2$\pm$0.03 & 43.2$\pm$0.07 & 43.9$\pm$0.07 & - &  43.6$\pm$0.01 & 44.4$\pm$0.01 \\
2223+1258 & rQSO & 3626$\pm$394 & 2834$\pm$134 & 2901$\pm$37 & - & 3073& 2976$\pm$37 & 44.7$\pm$0.03 & 44.3$\pm$0.01 & 44.5$\pm$0.004 & - & 44.4$\pm$0.004 & 45.0$\pm$0.003 \\
2241-1006 & rQSO & 7634$\pm$71 & 7108$\pm$467 & 4707$\pm$51 & 459$\pm$461 & 5252 & 5125$\pm$79 & 44.6$\pm$0.01 & 44.2$\pm$0.05 & 44.4$\pm$0.002 & 42.0$\pm$0.1 & 44.2$\pm$0.002 & 44.7$\pm$ 0.002 \\
\hline
0043+0114 & cQSO & 6593$\pm$46 & 4490$\pm$128 & 3552$\pm$32 & - &  3082 & 3050$\pm$22 & 44.6$\pm$0.01 & 44.3$\pm$0.02 & 44.3$\pm$0.002 & - & 44.0$\pm$0.003 & 44.7$\pm$0.002 \\
0152-0839 & cQSO & 5917$\pm$31 & 4880$\pm$118 & 3621$\pm$40 & 596$\pm$124 & 3605 & 3704$\pm$40 & 44.5$\pm$0.01 & 44.1$\pm$0.005 & 44.0$\pm$0.002 & 42.4$\pm$0.1 & 43.9$\pm$0.004 & 44.5$\pm$0.002 \\
0155-1023 & cQSO & 6150$\pm$33 & 4697$\pm$201 & 3834$\pm$27 & - & 3930 & 3825$\pm$94 & 44.8$\pm$0.01 & 44.3$\pm$0.01 & 44.5$\pm$ 0.001 & - & 44.3$\pm$0.002 & 44.9$\pm$0.001 \\
0213-0036 & cQSO & 3831$\pm$78 & 3914$\pm$91 & 4048$\pm$139 & 407$\pm$257 & 3348 & 3261$\pm$58 & 44.3$\pm$0.03 & 43.7$\pm$0.02 & 43.9$\pm$0.01 & 42.3$\pm$0.1 & 43.5$\pm$0.01 & 44.1$\pm$0.004 \\
0223-0007 & cQSO & 5200$\pm$373 & 3318$\pm$70 & 3593$\pm$42 & - & 3527 & 3428$\pm$106 & 44.1$\pm$0.01 & 43.5$\pm$0.01 & 43.7$\pm$0.003 & - & 43.6$\pm$0.01 & 44.2$\pm$0.01 \\
0303+0027 & cQSO & 6191$\pm$29 & 6406$\pm$98 & 5611$\pm$58 & - & 5550 & 5459$\pm$24 & 44.8$\pm$0.01 & 44.4$\pm$0.01 & 44.4$\pm$0.003 & - & 44.3$\pm$0.001 & 44.7$\pm$0.001 \\
0341-0037 & cQSO & 3948$\pm$23 & 3168$\pm$269 & 2803$\pm$22 & 636$\pm$124 & 3019 & 3083$\pm$50 & 44.2$\pm$0.01 & 43.7$\pm$0.01 & 43.8$\pm$0.003 & 41.6$\pm$0.2 & 43.5$\pm$0.01 & 44.1$\pm$0.003 \\
0404-0446 & cQSO & 7767$\pm$584 & 3564$\pm$376 & 2187$\pm$12 & - & 1920 & 1801$\pm$77 & 44.0$\pm$0.01 & 43.4$\pm$0.02 & 43.8$\pm$ 0.004 & - & 43.1$\pm$0.02 & 43.9$\pm$0.01 \\
0842+0151 & cQSO & 6018$\pm$31 & 5523$\pm$262 & 4643$\pm$44 & - & 4326 & 4231$\pm$222 & 44.7$\pm$0.01 & 44.0$\pm$0.09 & 44.4$\pm$0.002 & - & 44.2$\pm$0.002 & 44.6$\pm$0.005 \\
0927+0004 & cQSO & 5351$\pm$43 & 7068$\pm$200 & 6561$\pm$53 & 530$\pm$408 & 6189 & 6063$\pm$74 & 44.3$\pm$0.003 & 43.6$\pm$0.01 & 43.8$\pm$0.003 & 42.6$\pm$0.02 & 43.5$\pm$0.01 & 44.0$\pm$0.01 \\
0934+0005 & cQSO & 7020$\pm$36 & 4653$\pm$483 & 3046$\pm$27 & - & 2603 & 2688$\pm$20 & 44.2$\pm$0.01 & 44.0$\pm$0.04 & 44.0$\pm$0.003 & - & 43.9$\pm$0.004 & 44.4$\pm$0.003 \\
0941+0443 & cQSO & 3937$\pm$19 & 3869$\pm$70 & 5014$\pm$129 & 521$\pm$179 & 4960 & 4834$\pm$60 & 44.7$\pm$0.003 & 44.0$\pm$0.01 & 44.1$\pm$0.002 & 42.7$\pm$0.1 & 44.1$\pm$0.001 & 44.7$\pm$0.001 \\
1001+1015 & cQSO & 5733$\pm$38 & 4595$\pm$305 & 5491$\pm$60 & 146$\pm$90 & 4939 & 4868$\pm$46 & 44.8$\pm$0.004 & 44.1$\pm$006 & 44.3$\pm$ 0.01 & 42.6$\pm$1.0 & 44.1$\pm$0.001 & 44.7$\pm$0.001 \\  
1002+0331 & cQSO & 4855$\pm$48 & 4336$\pm$236 & 2708$\pm$37 & 636$\pm$355 & 3053 & 2958$\pm$16 & 45.0$\pm$0.005 & 44.6$\pm$0.01 & 44.4$\pm$0.003 & 41.9$\pm$0.02 & 44.4$\pm$0.002 & 44.9$\pm$0.001 \\
1013+0245 & cQSO & 4479$\pm$359 & 6612$\pm$340 & 6145$\pm$219 & 252$\pm$79 & 4696 & 4364$\pm$946 & 44.5$\pm$0.004 & 43.6$\pm$0.08 & 43.8$\pm$0.01 & 41.9$\pm$0.1 & 43.1$\pm$0.01 & 43.7$\pm$0.02 \\
1028+1456 & cQSO & 4046$\pm$83 & 5898$\pm$395 & 5205$\pm$280 & - & 4377 & 4256$\pm$246 & 44.8$\pm$0.01 & 44.1$\pm$0.03 & 44.0$\pm$0.004 & - & 43.7$\pm$0.01 & 44.2$\pm$0.01 \\  
1158-0322 & cQSO & 4506$\pm$29 & 4573$\pm$253 & 3581$\pm$39 & - & 4322 & 4417$\pm$39 & 44.9$\pm$0.01 & 44.3$\pm$0.04 & 44.4$\pm$0.002 & - & 44.3$\pm$0.003 & 44.9$\pm$0.002 \\
1352+1302 & cQSO & 3183$\pm$19 & 3565$\pm$141 & 2999$\pm$37 & 806$\pm$47 & 3274 & 3190$\pm$38 & 45.0$\pm$0.01 & 44.3$\pm$0.004 & 44.4$\pm$0.002 & 43.7$\pm$0.03 & 44.3$\pm$0.003 & 44.9$\pm$0.003 \\ 
1357-0307 & cQSO & 8940$\pm$286 & 3622$\pm$121 & 2595$\pm$88 & - & 2997 & 3089$\pm$118 & 44.1$\pm$0.03 & 43.5$\pm$0.05 & 44.0$\pm$0.01 & - & 43.4$\pm$0.02 & 44.3$\pm$0.01 \\  
1358+1410 & cQSO & 3387$\pm$188 & 4280$\pm$158 & 2646$\pm$324 & - & 1908 & 1764$\pm$28 & 44.4$\pm$0.01 & 44.0$\pm$0.01 & 43.9$\pm$0.05 & - & 43.4$\pm$0.01 & 44.1$\pm$0.005 \\  
1428+1001 & cQSO & 4615$\pm$49 & 5177$\pm$794 & 5559$\pm$52 & - & 4950 & 4838$\pm$129 & 45.0$\pm$0.01 & 44.1$\pm$0.06 & 44.3$\pm$0.003 & - & 43.7$\pm$0.001 & 44.3$\pm$0.02 \\  
1502+1016 & cQSO & 5273$\pm$194 & 6165$\pm$86 & 5435$\pm$286 & 525$\pm$63 & 4672 & 4578$\pm$141 & 44.4$\pm$0.01 & 43.9$\pm$0.003 & 43.9$\pm$0.01 & 43.1$\pm$0.06 &  43.7$\pm$0.005 & 44.3$\pm$0.004 \\ 
1513+1011 & cQSO & 2628$\pm$116 & 5171$\pm$498 & 3738$\pm$119 & 529$\pm$179 & 6097 & 6015$\pm$830 & 44.6$\pm$0.01 & 45.3$\pm$0.06 & 45.0$\pm$0.003 & 43.7$\pm$0.4 & 44.6$\pm$0.004 & 45.2$\pm$0.01 \\
1521-0156 & cQSO & 3513$\pm$12 & 2907$\pm$38 & 3324$\pm$37 & 747$\pm$78 & 3901 & 3811$\pm$ 31 & 44.6$\pm$0.01 & 44.2$\pm$0.002 & 44.3$\pm$0.002 & 43.6$\pm$0.8 & 44.1$\pm$0.002 & 44.6$\pm$0.002 \\
1539+0534 & cQSO & 4132$\pm$75 & 4748$\pm$133 & 4757$\pm$59 & - & 4333 & 4422$\pm$397 & 44.5$\pm$0.01 & 44.0$\pm$0.01 & 44.1$\pm$0.003 & - & 43.6$\pm$0.01 & 44.2$\pm$0.02 \\  
1540+1155 & cQSO & 5191$\pm$82 & 4362$\pm$166 & 3427$\pm$76 & 409$\pm$235 & 4862 & 4760$\pm$278 & 44.4$\pm$0.01 & 43.9$\pm$0.01 & 44.0$\pm$0.002 & 42.2$\pm$0.1 & 43.9$\pm$0.003 & 44.3$\pm$0.01 \\
1552+0939 & cQSO & 6785$\pm$53 & 4880$\pm$289 & 3471$\pm$183 & - & 5267 & 5196$\pm$78 & 45.0$\pm$0.02 & 44.0$\pm$0.02 & 44.2$\pm$0.005 & - & 43.6$\pm$0.01 & 44.2$\pm$0.02 \\  
1618+1305 & cQSO & 6125$\pm$52 & 6386$\pm$149 & 3948$\pm$28 & - & 4185 & 4090$\pm$186 & 44.5$\pm$0.02 & 44.1$\pm$0.01 & 44.1$\pm$0.003 & - & 43.7$\pm$0.01 & 44.3$\pm$0.01 \\  

    \hline
    \hline
    \end{tabular}
    \label{tab:emission}
}%
\caption{The FWHM and extinction-corrected integrated line luminosities for the broad \ion{C}{iv}, \ion{C}{iii}], \ion{Mg}{ii}, H\,$\upbeta$ and H\,$\upalpha$ and core [\ion{O}{iii}]$\uplambda$5007 emission lines, obtained from the multi-component fitting (see Section~\ref{sec:fit}). There are no errors on the FWHM of H\,$\upbeta$ since they are tied to the FWHM of H\,$\upalpha$ in our fitting. An example of the fitting is shown in Fig.~\ref{fig:pyqsofit}. The H\,$\upbeta$--[\ion{O}{iii}] and \ion{C}{iv} emission-line profiles for the cQSOs and rQSOs are shown in Figs.~\ref{fig:con_CIV}, \ref{fig:con_OIII}, \ref{fig:red_CIV}, and \ref{fig:red_OIII}.} 
\end{table*}

From our spectral fitting (see Section~\ref{sec:fit}), we calculated the continuum luminosities, corrected for the continuum-measured dust extinction using the best-fitting $E(B-V)$ values, at rest-frame 1350\,\AA, 3000\,\AA, 5100\,\AA~and 6200\,\AA, in addition to the emission-line properties for \ion{C}{iv}, \ion{C}{iii}], \ion{Mg}{ii}, H\,$\upbeta$, [\ion{O}{iii}]$\uplambda$5007, and H\,$\upalpha$. The emission-line fitting parameters can be found in Table~\ref{tab:emission}. To assess the quality of our spectral fitting, we compared the \ion{Mg}{ii} FWHM and luminosity to that from the \cite{rakshit} catalogue, which contains spectral properties for the SDSS DR14 QSOs, using \textsc{PyQSOFit} to estimate the continuum and major emission-line properties. We find a median offset of $-1392$\,km\,s$^{-1}$ and 0.25\,dex for the FWHM and luminosity measurements of the rQSOs, respectively, and $-941$\,km\,s$^{-1}$ and 0.01\,dex for the FWHM and luminosity measurements of the cQSOs, respectively. The larger difference in the luminosity measurements for the rQSOs is a consequence of the lack of dust-extinction correction applied in the \cite{rakshit} catalogue; therefore, the luminosities for dust-reddened QSOs will be underestimated.

\begin{figure}
    \centering
    \includegraphics[width=3.3in]{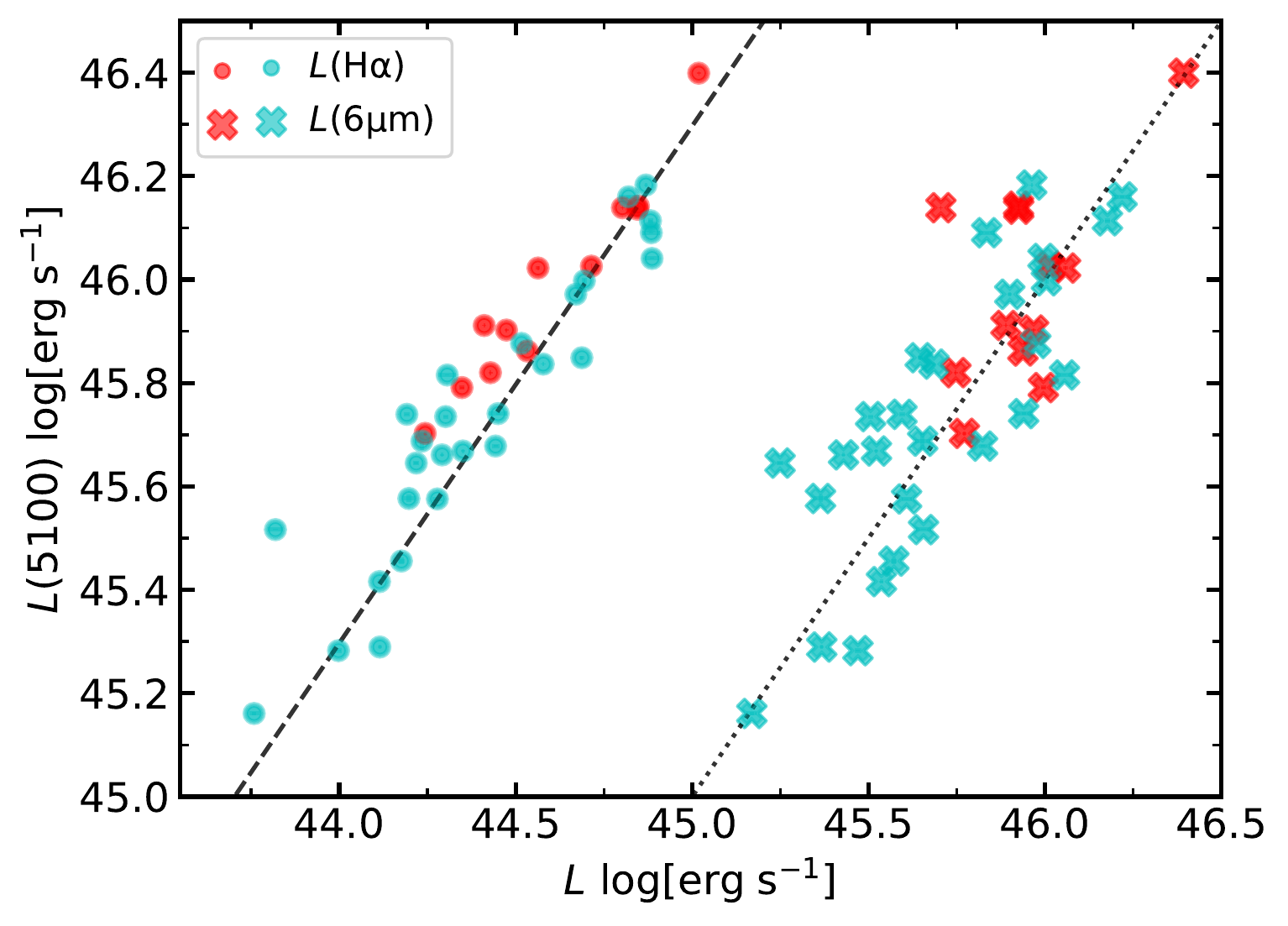}
    \caption{Extinction-corrected $L_{\rm 5100}$ versus broad H\,$\upalpha$ line luminosity (circles) and $L_{\rm 6\,\upmu m}$ (crosses) for the rQSOs (red) and cQSOs (cyan). The best fitting lines are shown as dashed and dotted lines. After correcting for extinction, the rQSOs have luminosities within the scatter of the cQSOs, for the same $L_{\rm 5100}$; however, the rQSOs are clearly biased towards higher luminosities.}
    \label{fig:lum}
\end{figure}

A comparison between the extinction-corrected 5100\,\AA~luminosity ($L_{\rm 5100}$) and the extinction-corrected H\,$\upalpha$ line luminosity and $L_{\rm 6\,\upmu m}$ is displayed in Fig.~\ref{fig:lum}. Both the H\,$\upalpha$ and $L_{\rm 6\,\upmu m}$ luminosities show a strong correlation with $L_{\rm 5100}$, with the best fitting power-law displayed as dashed and dotted lines. After correcting for extinction, the rQSOs have luminosities within the scatter of the cQSOs for the same $L_{\rm 5100}$. However, the rQSOs are clearly biased towards higher luminosities as a result of our incomplete observations (see Section~\ref{sec:obs}), which we account for in the following analyses by comparing results obtained with the full sample to those obtained using the statistically limited $L_{\rm 6\,\upmu m}$-matched sub-samples (see Section~\ref{sec:sample}).

\begin{figure*}
    \centering
    \includegraphics[width=3.2in]{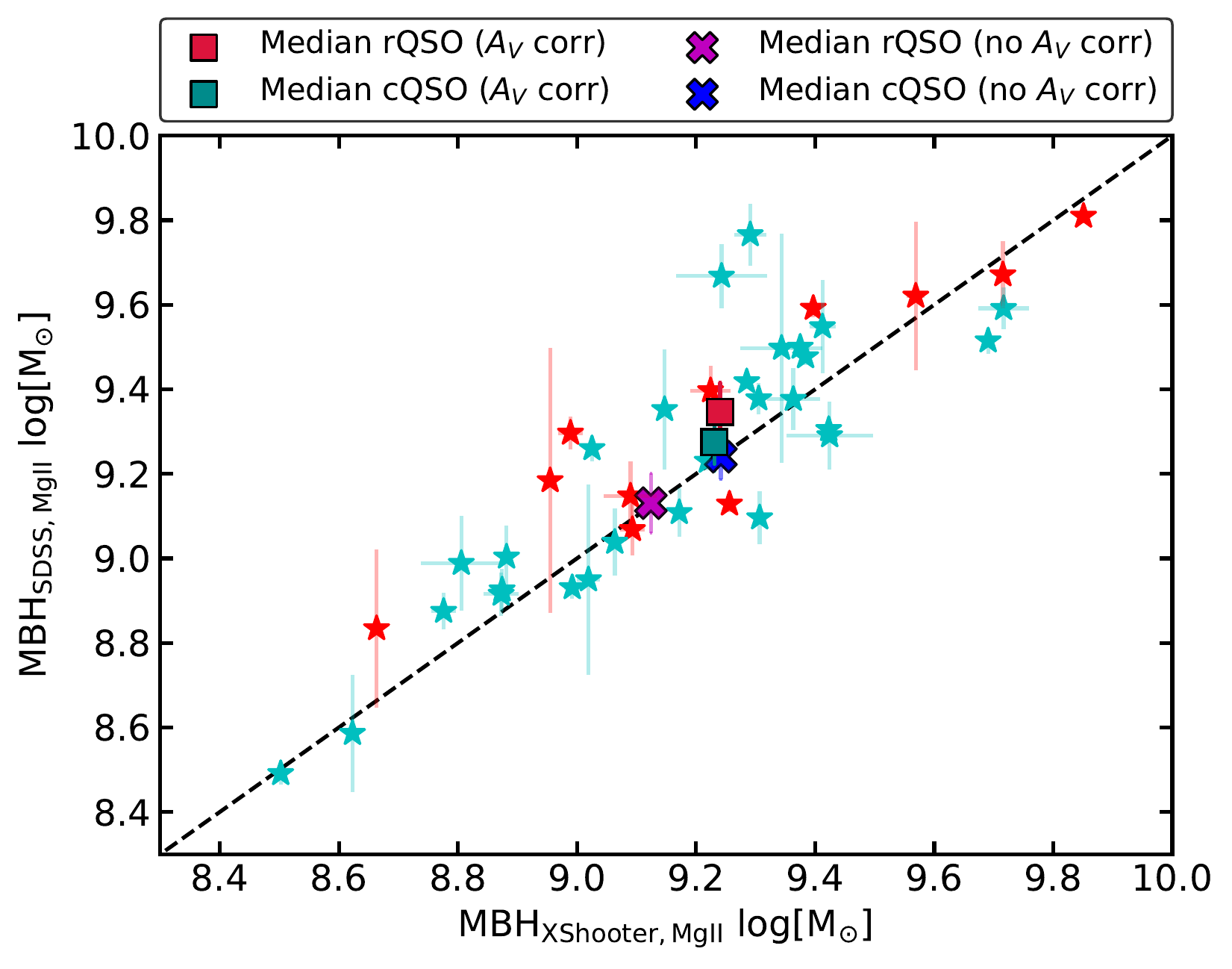}
    \includegraphics[width=3.2in]{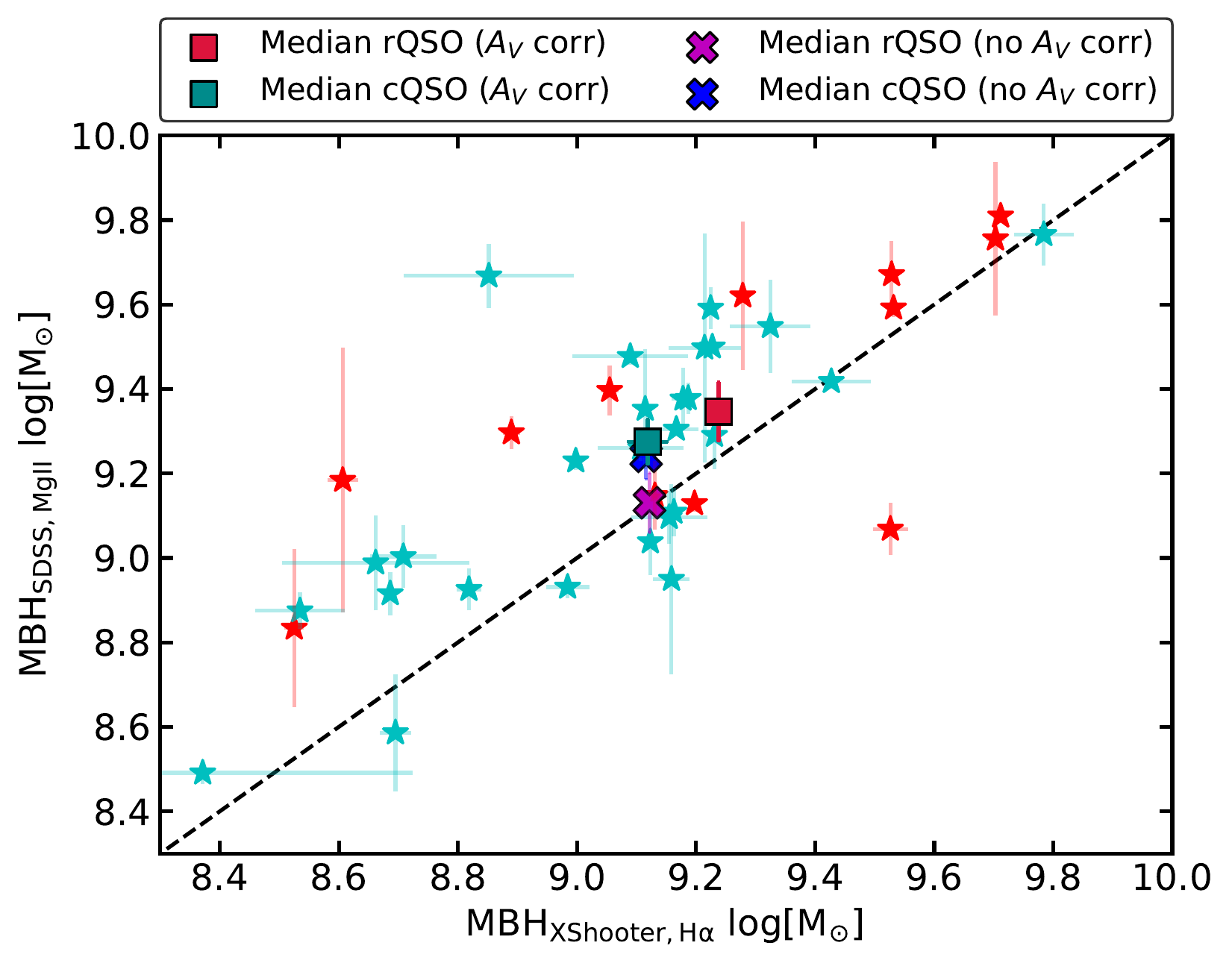}
    \caption{Black-hole mass estimates for the rQSOs (red stars) and cQSOs (cyan stars) taken from \protect\cite{rakshit} (calculated from the \ion{Mg}{ii} broad line) compared to those calculated in this work, based on the \ion{Mg}{ii} broad line (left) and H\,$\upalpha$ broad line (right) using Eq.~\ref{eq:bh}. The \textit{X-shooter} black-hole masses have been corrected for dust extinction using the best-fitting values from Section~\ref{sec:ext}; the \protect\cite{rakshit} black-hole masses have been corrected using the same correction factor applied to our \textit{X-shooter} estimates. The darker red and cyan square markers correspond to the median values where correcting for dust extinction and the magenta and blue crosses correspond to the median values without any correction for dust extinction, for the rQSOs and cQSOs, respectively. The dashed line represents the 1:1 relation, the y errors are taken from the \protect\cite{rakshit} catalogue and the x errors are estimated from MC resampling. A comparison of the extinction-corrected black-hole masses calculated from the \ion{Mg}{ii} and H\,$\upalpha$ line from the \textit{X-shooter} data is shown in Fig.~\ref{fig:mbh_mgii_ha}.}
    \label{fig:bh}
\end{figure*}

Fig.~\ref{fig:bh} shows a comparison between the M$_{\rm BH}$ from \cite{rakshit}, calculated from the \ion{Mg}{ii} line, compared to the M$_{\rm BH}$ estimated in this work, calculated from the \ion{Mg}{ii} and H\,$\upalpha$ broad line FWHM and continuum luminosities using the relations from \cite{Shen_2012} (see Section~\ref{sec:AD}). In both plots, we have calculated our M$_{\rm BH}$ values from dust-extinction corrected luminosities using the best-fit $A_V$ values (see Section~\ref{sec:ext}); this will have a greater effect on the \ion{Mg}{ii} estimation compared to H\,$\upalpha$, since shorter wavelengths are more affected by dust extinction than longer wavelengths.\footnote{The largest correction applied to a \ion{Mg}{ii}-derived M$_{\rm BH}$ due to extinction was $\approx 0.4$~dex.} We applied the same level of dust-extinction correction for the \cite{rakshit} M$_{\rm BH}$ values, to those applied for each of our sources. In Fig.~\ref{fig:bh}, the squares and crosses display the median M$_{\rm BH}$ calculated from our \textit{X-shooter} emission line \ion{Mg}{ii} (left) and H\,$\upalpha$ (right) fits, with and without a dust-extinction correction, respectively. A comparison between the \ion{Mg}{ii}- and H\,$\upalpha$-derived M$_{\rm BH}$ from our \textit{X-shooter} data is displayed in Fig.~\ref{fig:mbh_mgii_ha}.

The median extinction-corrected M$_{\rm BH}$ for the rQSOs and cQSOs are found to be $9.24\pm0.17$ and $9.12\pm0.05$\,log~M$_\odot$ calculated from the H\,$\upalpha$ line, and $9.24\pm0.15$ and $9.23\pm0.09$\,log~M$_\odot$ calculated from the \ion{Mg}{ii} line, respectively. The median M$_{\rm BH}$ of the rQSOs and cQSOs estimated from both of these estimations are consistent within $1\upsigma$ uncertainties; this is in agreement with the black-hole masses reported by \cite{calistro}, who found median values of $\sim9$\,log~M$_\odot$ for both the rQSOs and cQSOs. Further to this, after matching in $L_{\rm 6 \upmu m}$ luminosity (M$_{\rm BH}$: $9.05\pm0.26$ and $9.12\pm0.10$\,log~M$_\odot$ for the rQSOs and cQSOs, respectively) there appears to be no significant differences between the black-hole masses of rQSOs and cQSOs. It should be noted that individual estimates of M$_{\rm BH}$ calculated via the virial approach will have an intrinsic scatter due to uncertainties on the inclination angle and geometry of the BLR (see \citealt{yong} for a more detailed discussion), and also between different estimators (see \citealt{Shen_2012} for a more detailed discussion). This effect should be minimal when averaging over large samples, but due to our small samples this could be affecting the median black-hole mass estimates.

\subsection{Accretion-disc properties}\label{sec:bh_prop}
\begin{figure*}
    \centering
    \includegraphics[width=6.in]{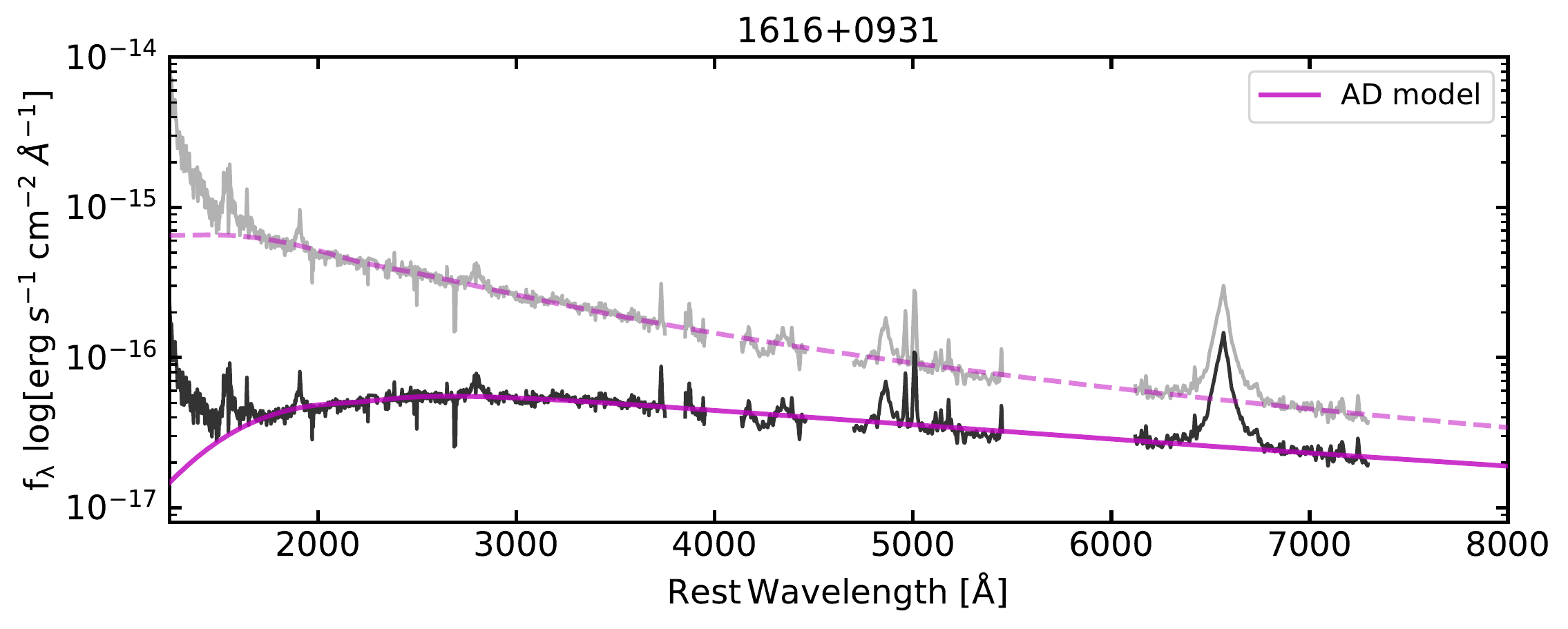}
    \caption{An example of the best-fitting AD model for one rQSO. The de-reddened spectrum is shown in grey. The solid magenta curve gives the best fitted AD model solution to the original spectrum, and the dashed magenta line gives the best-fitted model to the de-reddened spectrum. Note: the upturn in the spectrum at $\uplambda<1500$\,\AA~with respect to the AD model is due to the Ly\,$\upalpha$ emission line.}
    \label{fig:acc}
\end{figure*}

\begin{figure}
    \centering
    \includegraphics[width=3.2in]{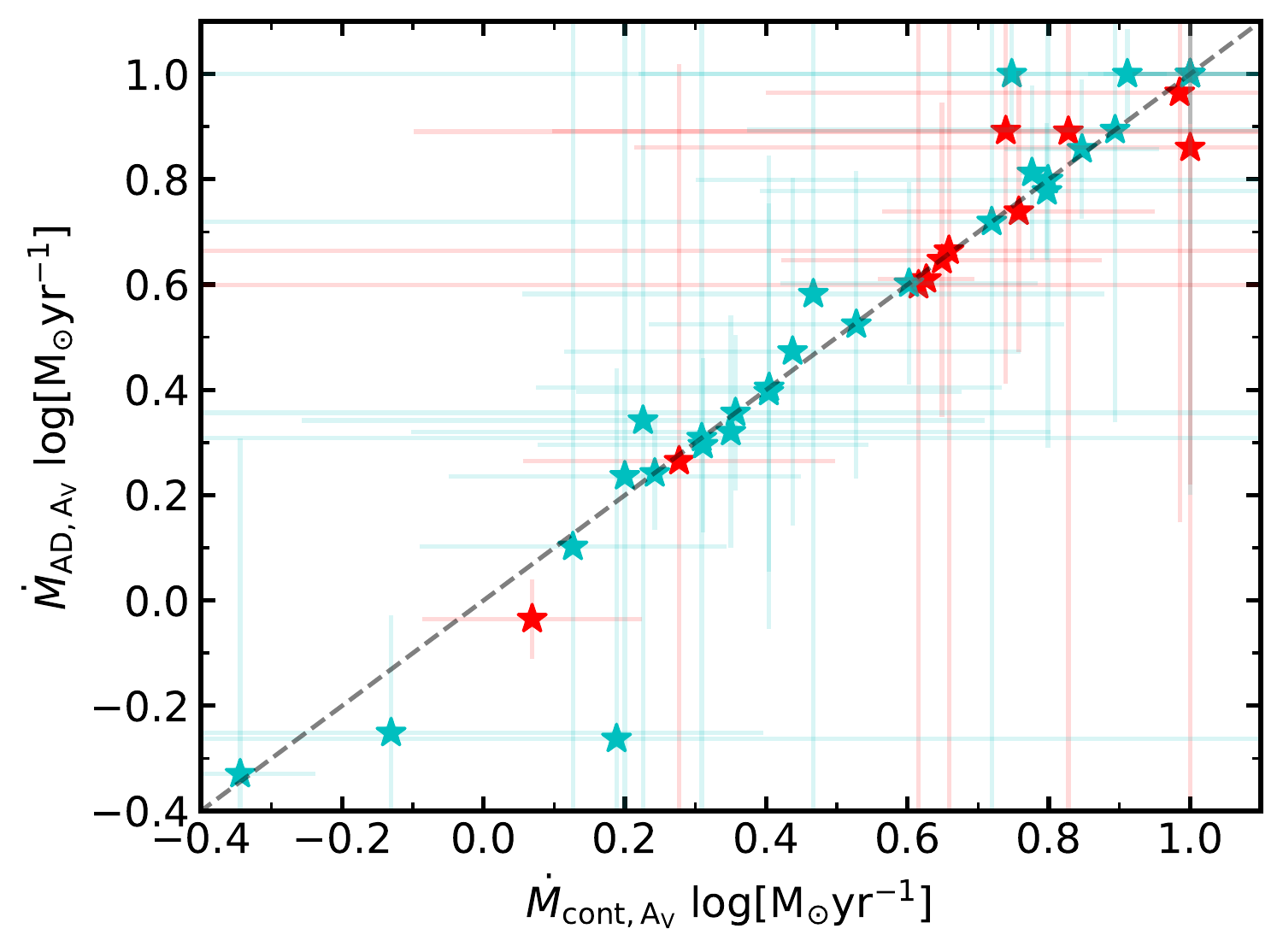}
    \caption{Best fitting mass accretion rate obtained from the AD model without fixing $A_V$ versus that obtained with fixing $A_V$ to the best-fitting extinction values from our extinction-curve analysis (see Section \ref{sec:ext}). Overall, we find similar mass accretion rate values obtained from both fitting methods.}
    \label{fig:AD_noav}
\end{figure}

\begin{table*}
\centering
    \caption{Table of the best-fitting AD parameters and virial black-hole mass estimates for our rQSOs and cQSOs (see Section~\ref{sec:AD}). Columns from left to right display the: (1) shortened source name utilized in this paper, (2) QSO sub-sample (rQSO or cQSO), (3) virial black-hole mass (M$_{\rm BH}$) estimate using the broad H\,$\upalpha$ line (see Section~\ref{sec:AD}), (4) black-hole spin ($a$), (5) mass-accretion rate ($\dot{M}$), (6) bolometric luminosity ($L_{\rm bol}$), (7) Eddington ratio ($\uplambda_{\rm edd}$), defined as $L_{\rm Edd}/L_{\rm bol}$ and (8) the reduced $\upchi^2$ of the fit ($\upchi^{2}_{\rm r}$). The errors correspond to $1\upsigma$ standard errors.} 
    \label{tab:accretion}
    \begin{tabular}{cccccccc} 
    \hline
    \hline
    Name & Sample &  M$_{\rm BH}$ & $a$ & $\dot{M}$  & $L_{\rm bol}$ & log $\uplambda_{\rm edd}$ & $\upchi_{\rm r}^{2}$\\ 
    & & log\,[M$_{\odot}$] & & log\,[M$_{\odot}$yr$^{-1}$]  & log\,[erg\,s$^{-1}$] & &  \\
    \hline
1020+1101 & rQSO & 9.53$\pm$0.01 & 0.86$\pm$0.19 & 0.28$\pm$0.22 & 46.5$\pm$0.22 & -1.08$\pm$0.31 & 1.76 \\  
1049+1157 & rQSO & 9.70$\pm$0.02 & 0.82$\pm$0.13 & 0.07$\pm$0.16 & 46.3$\pm$0.16 & -1.46$\pm$0.16 & 1.47 \\  
1358+1145 & rQSO & 9.51$\pm$0.03 & 0.60$\pm$3.71 & 0.62$\pm$1.54 & 46.9$\pm$1.54 & -0.73$\pm$2.20 & 3.66  \\  
1429-0112 & rQSO & 9.05$\pm$0.01 & -1.00$\pm$6.63 & 0.83$\pm$0.93 & 46.2$\pm$0.93 & -0.93$\pm$0.93 & 2.02 \\  
1442+1426 & rQSO & 8.56$\pm$0.02 & -0.64$\pm$4.46 & 1.00$\pm$0.79 & 46.5$\pm$0.79 & -0.21$\pm$1.11 & 1.16 \\  
1523+0452 & rQSO & 8.52$\pm$0.02 & -1.00$\pm$4.09 & 0.73$\pm$0.64 & 46.1$\pm$0.64 & -0.50$\pm$1.69 & 4.31 \\  
1608+1207 & rQSO & 9.12$\pm$0.01 & -1.00$\pm$1.61 & 0.65$\pm$0.23 & 46.1$\pm$0.23 & -1.18$\pm$0.23 & 2.01 \\  
1616+0931 & rQSO & 9.29$\pm$0.005 & 0.20$\pm$10.18 & 0.66$\pm$3.39 & 46.4$\pm$3.38 & -0.96$\pm$5.05 & 1.54 \\  
1639+1135 & rQSO & 9.71$\pm$0.01 & 0.63$\pm$0.13 & 0.63$\pm$0.07 & 46.9$\pm$0.07 & -0.92$\pm$0.12 & 3.68 \\  
2133+1043 & rQSO & 9.09$\pm$0.03 & -1.00$\pm$12.38 & 0.76$\pm$0.19 & 46.2$\pm$0.19 & -1.04$\pm$0.27 & 4.00 \\  
2223+1258 & rQSO & 9.23$\pm$0.01 & 0.35$\pm$0.22 & 1.00$\pm$0.08 & 46.8$\pm$0.08 & -0.50$\pm$0.12 & 8.88 \\  
2241-1006 & rQSO & 9.47$\pm$0.01 & -0.50$\pm$2.97 & 0.99$\pm$0.59 & 46.5$\pm$0.59 & -1.09$\pm$0.81 & 11.90 \\  
\hline
0043+0114 & cQSO & 9.02$\pm$0.01 & -0.31$\pm$0.59 & 1.00$\pm$0.12 & 46.5$\pm$0.12 & -0.59$\pm$0.15 & 2.03 \\
0152-0839 & cQSO & 9.11$\pm$0.01 & 0.43$\pm$0.52 & 0.60$\pm$0.18 & 46.5$\pm$0.18 & -0.68$\pm$0.27 & 1.91 \\
0155-1023 & cQSO & 9.30$\pm$0.02 & 0.70$\pm$0.19 & 0.85$\pm$0.11 & 47.1$\pm$0.11 & -0.28$\pm$0.16 & 2.79 \\
0213-0036 & cQSO & 8.69$\pm$0.01 & -0.91$\pm$3.11 & 0.35$\pm$0.45 & 45.8$\pm$0.45 & -1.02$\pm$0.45 & 1.29 \\
0223-0007 & cQSO & 8.82$\pm$0.02 & 0.10$\pm$2.91 & 0.31$\pm$0.80 & 46.0$\pm$0.80 & -0.88$\pm$1.23 & 2.02 \\
0303+0027 & cQSO & 9.41$\pm$0.01 & 0.69$\pm$0.83 & 0.80$\pm$0.50 & 47.1$\pm$0.50 & -0.45$\pm$0.51 & 3.65 \\
0341-0037 & cQSO & 8.71$\pm$0.01 & -1.00$\pm$2.18 & 0.44$\pm$0.32 & 45.8$\pm$0.32 & -0.98$\pm$0.70 & 1.67 \\
0404-0446 & cQSO & 8.33$\pm$0.03 & 0.00$\pm$2.11 & 0.75$\pm$0.53 & 46.4$\pm$0.53 & -0.09$\pm$0.76 & 36.00 \\
0842+0151 & cQSO & 9.28$\pm$0.04 & 0.48$\pm$3.93 & 0.72$\pm$1.46 & 46.7$\pm$1.46 & -0.67$\pm$1.62 & 3.94 \\
0927+0004 & cQSO & 9.19$\pm$0.01 & 0.93$\pm$0.03 & -0.34$\pm$0.11 & 45.9$\pm$0.11 & -1.38$\pm$0.18 & 1.98 \\
0934+0005 & cQSO & 8.78$\pm$0.01 & -1.00$\pm$0.11 & 0.78$\pm$0.02 & 46.2$\pm$0.02 & -0.70$\pm$0.02 & 6.43 \\
0941+0443 & cQSO & 9.32$\pm$0.01 & 0.76$\pm$0.42 & 0.40$\pm$0.27 & 46.7$\pm$0.27 & -0.75$\pm$0.40 & 2.93 \\
1001+1015 & cQSO & 9.23$\pm$0.01 & 0.73$\pm$0.36 & 0.31$\pm$0.23 & 46.6$\pm$0.23 & -0.75$\pm$0.25 & 2.25 \\  
1002+0331 & cQSO & 9.02$\pm$0.004 & -0.41$\pm$11.09 & 1.00$\pm$2.15 & 46.5$\pm$2.14 & -0.62$\pm$2.15 & 2.19 \\
1013+0245 & cQSO & 8.85$\pm$0.60 & -0.10$\pm$8.41 & 0.19$\pm$2.04 & 45.8$\pm$2.04 & -0.98$\pm$2.48 & 1.84 \\
1028+1456 & cQSO & 9.10$\pm$0.05 & 0.90$\pm$0.47 & 0.23$\pm$0.48 & 46.5$\pm$0.48 & -0.69$\pm$0.71 & 1.47 \\  
1158-0322 & cQSO & 9.39$\pm$0.01 & 0.51$\pm$0.76 & 0.80$\pm$0.41 & 47.1$\pm$0.41 & -0.43$\pm$0.42 & 4.16 \\
1352+1302 & cQSO & 9.16$\pm$0.01 & 0.18$\pm$0.27 & 1.00$\pm$0.08 & 46.7$\pm$0.08 & -0.56$\pm$0.08 & 4.00 \\  
1357-0307 & cQSO & 8.93$\pm$0.03 & 0.32$\pm$3.93 & 0.53$\pm$0.29 & 46.4$\pm$0.29 & -0.64$\pm$0.42 & 1.96 \\  
1358+1410 & cQSO & 8.44$\pm$0.01 & -1.00$\pm$0.33 & 0.91$\pm$0.06 & 46.3$\pm$0.06 & -0.23$\pm$0.06 & 6.79 \\  
1428+1001 & cQSO & 9.22$\pm$0.02 & 0.97$\pm$0.04 & 0.13$\pm$0.22 & 46.4$\pm$0.22 & -0.91$\pm$0.22 & 1.73 \\  
1502+1016 & cQSO & 9.17$\pm$0.02 & 0.81$\pm$0.20 & 0.20$\pm$0.25 & 46.5$\pm$0.25 & -0.83$\pm$0.26 & 1.68 \\  
1513+1011 & cQSO & 9.86$\pm$0.11 & 0.97$\pm$0.002 & 0.24$\pm$0.005 & 46.5$\pm$0.005 & -1.56$\pm$0.01 & 3.70 \\  
1521-0156 & cQSO & 9.12$\pm$0.01 & -0.73$\pm$3.17 & 0.89$\pm$0.52 & 46.3$\pm$0.52 & -0.89$\pm$0.78 & 7.61 \\  
1539+0534 & cQSO & 9.15$\pm$0.07 & 0.83$\pm$0.31 & 0.40$\pm$0.33 & 46.7$\pm$0.33 & -0.52$\pm$0.43 & 4.79 \\  
1540+1155 & cQSO & 9.24$\pm$0.05 & 0.76$\pm$1.71 & 0.36$\pm$1.35 & 46.6$\pm$1.35 & -0.68$\pm$1.36 & 2.40 \\  
1552+0939 & cQSO & 9.22$\pm$0.01 & 1.00$\pm$0.12 & -0.13$\pm$0.53 & 46.1$\pm$0.53 & -1.18$\pm$0.56 & 1.44 \\  
1618+1305 & cQSO & 9.03$\pm$0.04 & 0.10$\pm$1.52 & 0.47$\pm$0.41 & 46.2$\pm$0.41 & -0.98$\pm$0.58 & 3.74 \\  
    \hline
    \hline
    \end{tabular}
\end{table*}

Following Section~\ref{sec:AD}, we fitted a thin AD model to the spectra, fixing $A_V$ to the best-fit values from our extinction-curve analysis (see Section~\ref{sec:extinction}), in order to compute the mass accretion rate, black-hole spin and Eddington ratio parameters (see Fig.\,\ref{fig:acc} for an example of the best-fitting AD model to a rQSO). For 83 per~cent (10/12) of the rQSOs and 86 per~cent (24/28) of the cQSOs, the thin AD model provided a good fit to the QSO continuum ($\upchi^2_{\rm r}<5$); the best fitting parameters and $\upchi^2_{\rm r}$ values are displayed in Table~\ref{tab:accretion}. Since both the majority of the rQSOs and cQSOs are well fitted with a thin AD model, this suggests that there are no significant differences between the accretion discs of cQSOs and rQSOs, once the effects of dust extinction are taken into account. We also refitted the thin AD models to the spectra, this time leaving $A_V$ as a free parameter, and found consistent levels of dust extinction for the majority of the rQSOs. For a few moderately dust-reddened sources we found AD solutions not requiring dust, but overall we obtained consistent AD parameters to those found adopting a fixed $A_V$ (see Fig.~\ref{fig:AD_noav} for a comparison of the $\dot{M}$, with and without fixing $A_V$ in the AD fitting). This also suggests that dust-reddening is a consistent explanation for the red colours in the rQSOs, even if differences in accretion processes are considered. It is worth noting that we did not include an additional disc-wind component in our model, which has been found to affect the shape of QSO spectra in a similar way to dust-reddening (C15); more complex AD models will be tested in a future, larger \textit{X-shooter} study. We find good agreement to C16 between our best fitting AD parameters for the C15 cQSOs (see Appendix~\ref{sec:cap_comp} for a more detailed discussion).

\begin{figure*}
    \centering
    \includegraphics[width=5.6in]{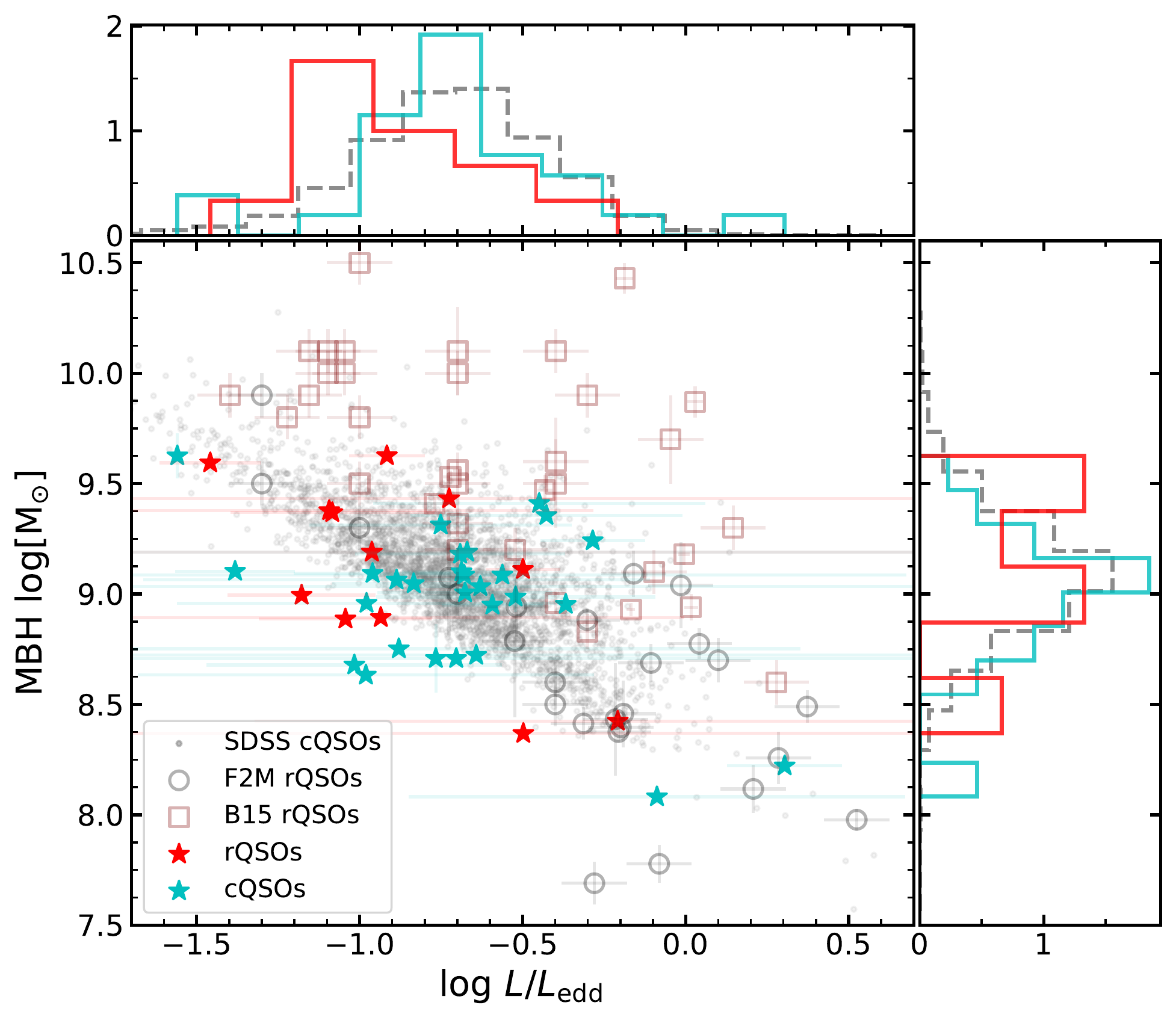}
    \caption{Black-hole mass versus Eddington ratio for our rQSOs (red stars) and cQSOs (cyan stars) explored in this paper. The black-hole masses are determined from a virial relation utilizing the broad H\,$\upalpha$ and the Eddington ratios are derived from fitting a thin AD model to our rQSO and cQSO spectra. Overplotted are the F2M QSOs (black circles) using the values from \protect\cite{Urrutia_2012,Kim_2015,kim18}, and the heavily reddened QSOs from \protect\cite{ban12,ban} (maroon squares); both of these samples favour higher Eddington ratios, albeit with very different selections. Our parent SDSS DR7 uniformly-selected cQSOs are plotted in grey. The corresponding black-hole mass and Eddington ratio distributions for our rQSOs, cQSOs and the SDSS DR7 sample are shown in the adjacent histograms as a solid red, solid cyan and dashed grey lines, respectively. The uncertainties for our rQSO and cQSO samples are calculated using MCMC. The error bars for the F2M QSOs are taken from \protect\cite{Urrutia_2012} and correspond to 0.1\,dex estimates. \protect\cite{ban} do not provide uncertainties for the Eddington ratios and so we assume a 0.1\,dex estimate.}
    \label{fig:edd}
\end{figure*}

\begin{table}
    \centering
    \caption{Table of the median AD properties for our rQSO, cQSO, $L_{\rm 6\,\upmu m}$-matched, radio-detected (rad det) and radio-undetected (rad undet) samples. See Table~\ref{tab:accretion} for a description of the column names.}
    \begin{tabular}{ccccc}
        \hline
        \hline
         & M$_{\rm BH}$ & $\uplambda_{\rm Edd}$ & $a$ & $\dot{M}$ \\
         & log\,[M$_{\odot}$] & log & & log\,[M$_{\odot}$yr$^{-1}$] \\
        \hline
        rQSO & 9.24$\pm$0.17 & -0.95$\pm$0.10 & -0.15$\pm$0.47 & 0.70$\pm$0.07 \\
        cQSO & 9.12$\pm$0.05 & -0.68$\pm$0.04 & 0.46$\pm$0.23 & 0.45$\pm$0.14 \\
        rQSO$_{L_{\rm 6\,\upmu m}}$ & 9.05$\pm$0.26 & -0.96$\pm$0.19 & -0.64$\pm$0.61 & 0.74$\pm$0.17 \\
        cQSO$_{L_{\rm 6\,\upmu m}}$ & 9.12$\pm$0.10 & -0.67$\pm$0.03 & -0.45$\pm$0.47 & 0.81$\pm$0.09 \\
        rQSO$_{\rm rad\,det}$ & 9.40$\pm$0.21 & -1.00$\pm$0.07 & 0.42$\pm$0.55 & 0.64$\pm$0.19 \\
        cQSO$_{\rm rad\,det}$ & 9.17$\pm$0.03 & -0.60$\pm$0.23 & 0.75$\pm$0.16 & 0.37$\pm$0.06 \\
        rQSO$_{\rm rad\,undet}$ & 9.16$\pm$0.28 & -0.61$\pm$0.21 & -0.57$\pm$0.38 & 0.86$\pm$0.13 \\
        cQSO$_{\rm rad\,undet}$ & 8.99$\pm$0.13 & -0.69$\pm$0.06 & 0.05$\pm$0.29 & 0.73$\pm$0.12 \\
        \hline
        \hline
    \end{tabular}
    \label{tab:AD_average}
\end{table}

We find no significant differences in the accretion rate or black-hole spin for our rQSOs compared to the cQSOs (see Table~\ref{tab:AD_average} for the median AD properties). The best-fitting black-hole spin values span the whole range ($-1$ to 0.998), with an average value of $-0.15\pm0.47$ and $0.46\pm0.23$ for the rQSOs and cQSOs, respectively. If differences in the post-merger spin evolution of black-holes were responsible for the radio excess in rQSOs \citep{garofalo}, then we might expect systematic differences between the black-hole spin distributions of the two samples. However, in the luminosity-matched samples we find the two black-hole spin distributions to be consistent; applying the two-sided Kolmogorov-Smirnov (K-S) test this is statistically not significant ($p$-value\,$=0.88$). The median mass accretion rates are $0.70\pm0.07$ and $0.45\pm0.14$\,log\,[M$_{\odot}$yr$^{-1}$] for the rQSOs and cQSOs, respectively; these differences are statistically significant ($p$-value\,$=0.09$), but after matching the two samples in $L_{\rm 6\,\upmu m}$ the distributions are consistent ($p$-value\,$=0.42$), with average values of $0.74\pm0.17$ and $0.81\pm0.09$\,log\,[M$_{\odot}$yr$^{-1}$] for the rQSOs and cQSOs, respectively. Splitting out the radio-detected QSOs within our samples we explored whether there were any differences in accretion properties between the radio-detected and undetected sources (see Table~\ref{tab:AD_average}). Significant differences would support previous studies that suggest radio-loud quasars have higher black-hole spin values as per the ``spin paradigm'', whereby relativistic jets can be formed from the rotational energy of the spinning SMBH \citep{wilson,schulze}. However, we do not find any significant differences, although our sample is too small to draw robust conclusions.

Fig.~\ref{fig:edd} shows the distribution of M$_{\rm BH}$ as a function of Eddington ratio. There are small, tentative ($p$-value\,$=0.04$) differences in the Eddington ratios, with rQSOs displaying on average lower Eddington ratios compared to cQSOs (median values of $-0.95\pm0.10$ and $-0.68\pm0.04$ for the rQSOs and cQSOs, respectively); tentative differences remain after matching in $L_{\rm 6\upmu m}$ ($p$-value\,$=0.03$), although the sample size is now very small (7 rQSOs and 7 cQSOs). This is contrary to what was found in previous red QSO studies (e.g., \citealt{Urrutia_2012,Kim_2015,ban,kim18}; plotted on Fig.~\ref{fig:edd}) who claim that red QSOs have higher Eddington ratios compared to typical QSOs which could be evidence for a phase of rapid black-hole growth. However, the radio properties of our samples are not representative of the overall QSO population; for example, we have many more radio-loud cQSOs compared to rQSOs. It should also be noted that no difference in M$_{\rm BH}$ or Eddington ratio (inferred from the FWHM of broad emission lines) between SDSS selected rQSOs and cQSOs was found in our previous studies \citep{klindt,calistro}, using the \cite{Shen_2012} and \cite{rakshit} samples, respectively. Consequently, a larger \textit{X-shooter} sample with fewer biases is required to robustly test whether there are significant differences in the Eddington ratio between rQSOs and cQSOs.
\section{Discussion}

We have used the \textit{X-shooter} spectra of a sample of 40 red and blue/typical QSOs at $1.45<z<1.65$, in order to constrain their line and continuum-emission, dust-extinction, and AD properties. From these data we can assess whether the enhanced radio emission we have previously found in red QSOs are due to differences in the accreting engine or outflow properties.

From fitting a dust-reddened cQSO composite to our QSO samples (see Section~\ref{sec:extinction}), we find that the red colours in 11/12 ($\sim92$\,per~cent) of our rQSOs can be fully explained by dust, and rule out both the MW and flat extinction curves for all but one of the rQSOs. The amount of dust reddening in our rQSO sample is modest, with $E(B-V)$ values ranging from $\sim0.02$--$0.23$\,mags ($A_V\sim0.06$--0.7\,mags; see Table~\ref{tab:ext_tab}). We also find a strong correlation between $\Delta (g^*-i^*)$ and $E(B-V)$ which suggests our method of selecting dust-reddened QSOs is robust (see Fig.~\ref{fig:EBV}). On the basis of our emission-line fitting approach, we also measured the broad H\,$\upalpha$/H\,$\upbeta$ Balmer decrements and find on average good agreement within the intrinsic scatter for the rQSOs and cQSOs as a function of $E(B-V)$. We also find higher Balmer decrements, including upper limits, for the rQSOs compared to the cQSOs which indicates higher levels of dust along the line-of-sight towards the BLR (see Section~\ref{sec:extinction_balmer}).

Using the black-hole masses calculated from the broad H\,$\upalpha$ emission (Section~\ref{sec:emission}) and the dust-extinction measurements from our extinction-curve fitting (Section~\ref{sec:extinction}), we fitted a simple thin AD model to our rQSOs and cQSOs to test whether thin disc models adequately explain the nature of the accretion-disc emission in these QSOs (see Section \ref{sec:bh_prop}). We find good fits for 83 per~cent (10/12) of the rQSOs and 86 per~cent (24/28) of the cQSOs, with no significant differences in the black-hole spin, mass accretion rate or black-hole mass parameters when accounting for luminosity. We find tentative evidence that rQSOs have lower Eddington ratios compared to cQSOs (see Table~\ref{tab:AD_average}); this difference remains after matching in $L_{\rm 6\,\upmu m}$, although we are extremely limited in source statistics (only 7 rQSOs and 7 cQSOs are matched in $L_{\rm 6\,\upmu m}$). Leaving $A_V$ as a free parameter we also recover consistent levels of dust extinction to our continuum measurements, and find similar accretion properties to those obtained using a fixed $A_V$ (see Fig.~\ref{fig:AD_noav}). This also suggests that the red colours in rQSOs are fully consistent with modest levels of dust obscuration of an otherwise normal QSO. 

On the basis of our analyses using \textit{X-shooter} spectra the only significant differences between rQSOs and cQSOs appears to be the presence of larger amounts of dust along the line-of-sight towards the rQSOs. One scenario that can connect this dust to the fundamental differences in the radio properties between rQSOs and cQSOs is winds and outflows interacting with a dust and gas rich ISM/circumnuclear environment. Therefore, in the following sub-sections we focus on constraining the location and composition of the dust and search for differences in the presence and properties of outflows between our rQSO and cQSO samples.

\subsection{The nature of dust in red QSOs}\label{sec:dust}
The shape of a dust-extinction law, determined by the total-to-selective extinction $R_V$ (where $R_V\equiv A_V/E(B-V)$; \citealt{cardelli}), is primarily due to the composition of the dust and distribution of grain sizes. In general, a lower value of $R_V$ corresponds to a steeper extinction curve and smaller dust grains, and vice versa. Consequently, we can potentially learn about the physical properties of the dust by identifying the best-fitting dust-extinction laws, which are defined by different values of $R_V$, for each of the QSOs. Typically, studies have used an SMC-like extinction curve ($R_V=2.74$) to explain dust reddening in QSOs, although the effectiveness of this for describing the dust in all QSOs has been debated. For example, \cite{czerny} construct a QSO extinction curve based on the blue and red composite spectra from \cite{richards} and find a flatter ``grey'' curve that deviates from an SMC-like curve at shorter wavelengths, which has also been found in other studies \citep{gaskell,gaskell_benker,mehdipour,temple_red}. They found their derived grey extinction curve is best described by dust containing no graphite, with a fairly large grain size of $>0.016\,\upmu$m, and suggested that this lack of graphite might support a scenario where the dust is formed in an outflowing wind (e.g., \citealt{elvis}). On the other hand, \cite{zafar} analysed high extinction QSOs for which the SMC law could not provide a good fit, deriving an average extinction curve that is steeper than the SMC law, with a best-fit value of $R_V=2.2\pm 0.2$. They suggest that the steepness of this curve implies a lack of large dust grains, which may have been destroyed by the activity of the AGN.

Across the overall rQSO sample, we do not find a significant preference for one extinction curve over another (see Section~\ref{sec:extinction} and Fig.~\ref{fig:EBV}); however, we rule out the flat and MW extinction curve for all but one rQSO. This could imply that rQSOs have dust composed of smaller grains, described by a steeper curve, suggesting a higher level of nuclear activity or that the dust is formed within the vicinity of the central nucleus \citep{zafar}. This is consistent with previous work which have found the characteristic 2175\,\AA~bump (expected from small carbonaceous grains present in our Galaxy) absent in rQSOs \citep{richards,czerny,zafar,temple_red}. This result is also in agreement with \cite{willott}, who claim that flatter extinction curves are driven by redshift selection effects incorporated in composites that are constructed from large QSO samples. Therefore, distinguishing between different extinction curves requires the analysis of individual spectra or composites constructed over a narrow redshift range.

We can also potentially gain insight into the dust composition from empirically determining the value of $R_V$ in each spectrum. To do this, we normalize the spectra relative to the observed-frame flux of the QSOs in the \textit{WISE} \textit{W1} band (3.4\,$\upmu$m), and then calculate $A_V$ and $E(B-V)$ in order to determine $R_V\equiv A_V/E(B-V)$. The \textit{W1} band is likely dominated by thermal emission from dust grains at $z=1.5$ (e.g., \citealt{stern12,assef}), and therefore normalizing the spectra relative to this band can allow us to estimate the intrinsic $A_V$ and $E(B-V)$ directly from the spectra. From our analysis we do not find a significant difference in the $R_V$ value between the rQSOs and cQSOs, although we do find tentative evidence for relatively low $R_V$ values ($\sim2$) for the rQSOs. These low $R_V$ values are equivalent to a steeper dust-extinction curve, consistent with the results from our extinction-curve analysis. However, we are restricted to moderate levels of $A_V$ due to the shallowness of SDSS, as well as a small number of sources. Our future study analysing the dust-extinction properties of DESI rQSOs, which will push to much more extinguished systems than those identified in the SDSS, will robustly constrain the nature of dust in rQSOs.

\subsection{Accretion-driven winds in red QSOs?}\label{sec:bh_dis}
Overall, we find no significant differences in the accretion properties between rQSOs and cQSOs, and therefore we consider other mechanisms that could be driving the differences in the radio emission. One possibility is that rQSOs have more prominent winds and outflows than cQSOs, which could lead to enhanced radio emission via interactions and shocks with the ISM gas. We can search for the signatures of outflows by analysing the emission-line profiles: strong line asymmetries in the \ion{C}{iv} broad line or very high velocity dispersions in the forbidden lines (e.g., [\ion{O}{iii}]$\uplambda$5007) are clear indicators of powerful ionized outflows \citep{carniani,zakamska_2016,rankine,calistro,jarvis_21}. 

In \cite{calistro}, they found a higher incidence of high velocity \ion{C}{iv} blueshifts ($>1000$\,km\,s$^{-1}$) in rQSOs, compared to typical QSOs, for a statistical sample of $\sim1800$ QSOs using the DR14 catalogue \citep{rakshit}. From measuring the \ion{C}{iv} blueshifts of our sample, we find tentative evidence for larger blueshifts in the rQSOs compared to the cQSOs; 50 per~cent (6/12) of rQSOs display a \ion{C}{iv} blueshift with $v>1000$\,km\,s$^{-1}$, compared to 29 per~cent (8/28) of the cQSOs. However, after matching in $L_{\rm 6\,\upmu m}$, this difference is not significant ($p$-value\,$=0.43$), and analysing a larger luminosity-matched sample obtained from \cite{rankine}, we also do not find a difference (Fawcett et~al. \textit{in prep}). 

The [\ion{O}{iii}]$\uplambda$5007 emission line traces outflows on larger scales (outside of the BLR), compared to UV emission-line tracers. Fitting the [\ion{O}{iii}]$\uplambda$5007 emission line with a core (narrow) and wing (broad) component, \cite{calistro} found evidence for larger FWHMs in the wing components of a sample of SDSS red QSOs compared to typical QSOs, at $z<1$ with $L_{\rm 6\,\upmu m}>44.5$\,ergs$^{-1}$, suggesting that outflows are more prevalent within the red QSO population. In our \textit{X-shooter} sample, the [\ion{O}{iii}]$\uplambda$5007 emission line is only detected in 82 per~cent and 46 per~cent of the rQSOs and cQSOs, respectively, and so any analyses using this line will be limited. Overall, we find evidence for broad [\ion{O}{iii}]$\uplambda$5007 profiles in both the red and control sample; however, we do not find any significant differences in the FWHM of either the [\ion{O}{iii}]$\uplambda$5007 core or wing components ($p$-values\,$=0.96$ and 0.97 for the [\ion{O}{iii}]$\uplambda$5007 core and wing FWHM, respectively). We find tentative evidence for larger blueshifts in the [\ion{O}{iii}]$\uplambda$5007 wing components of rQSOs; 42 per~cent (5/12) of rQSOs display an [\ion{O}{iii}]$\uplambda$5007 wing with a blueshift of $v>400$\,km\,s$^{-1}$ compared to 21 per~cent (6/28) of the cQSOs. However, matching in $L_{\rm 6\,\upmu m}$ this difference is not significant ($p$-value\,$=0.32$), although there are too few sources for this to be conclusive. The lack of any significant differences in the [\ion{O}{iii}]$\uplambda$5007 line could be due to a number of factors such as the small number of sources or low number of [\ion{O}{iii}]$\uplambda$5007 detections in our sample. The [\ion{O}{iii}]$\uplambda$5007 and \ion{C}{iv} profiles and fits for the cQSOs and rQSOs are shown in Figs.~\ref{fig:con_CIV}, \ref{fig:con_OIII}, \ref{fig:red_CIV}, and \ref{fig:red_OIII}. A future larger \textit{X-shooter} sample is required to robustly tie down any differences in the outflow properties between rQSOs and cQSOs. 

In Alexander et~al. (\textit{in prep}) we find a link between rQSOs and Low-ionization Broad Absorption Line QSOs (LoBALs; BALQSOs that display additional low-ionization species such as \ion{Mg}{ii} and \ion{Al}{iii}, e.g., \citealt{trump}); LoBALs tend to have redder optical colours on average and have enhanced radio emission compared to nonBALs (also identified in other studies, e.g., \citealt{morabito}). However, after removing the BALQSOs from the parent rQSO sample there is still a radio enhancement compared to the typical QSOs, suggesting that the presence of dust is more crucial to the fundamental differences in the radio properties of rQSOs than the presence of a BAL wind (Alexander et~al. \textit{in prep}). This, together with the lack of evidence for stronger winds in rQSOs, could suggest that the radio emission in rQSOs is more closely related to circumnuclear/ISM opacity rather than outflow power. For example, the radio emission could be produced via shocks (causing synchrotron emission) between the wind/outflow and the circumnuclear/ISM gas and dust (e.g., \citealt{liu,zakamska_2016}). 
\section{Conclusions}
We have studied a sample of rQSOs and cQSOs at $1.45<z<1.65$ using wide-band \textit{X-shooter} optical--NIR spectra in order to measure the dust-extinction, emission and accretion-disc properties. From our analyses we find that:

\begin{itemize}
    \item \textbf{The colours of red QSOs are fully consistent with dust extinction (see Fig.~\ref{fig:EBV} and \ref{fig:balmer}):} From our extinction-curve fitting analysis, we find that the optical colours in the majority (11/12) of our rQSOs are consistent with dust, with moderate amounts of extinction $E(B-V)\sim0.02$--$0.23$\,mags ($A_V\sim0.06$--$0.7$\,mags). Exploring the nature of the dust in rQSOs, we rule out the flat and MW extinction curves and find similarly good fits for both the steep and PL extinction curves. Analysing the Balmer decrements we find that the rQSOs have larger Balmer decrements, including the upper limits, compared to the cQSOs; this is evidence for more dust along the line-of-sight to the BLR in the rQSOs. See Sections~\ref{sec:extinction}, \ref{sec:extinction_balmer} and \ref{sec:dust}.
    \item \textbf{There are no significant differences in the accretion properties of red and typical (control) QSOs after correcting for dust extinction (see Table~\ref{tab:AD_average} and Fig.~\ref{fig:edd}):} We find that a simple thin accretion disc can describe the accretion engine of 83 per~cent of the rQSOs and 86 per~cent of the cQSOs. Refitting the AD models with $A_V$ as a free parameter, we recover consistent accretion parameters and find similar levels of dust extinction to those found in the extinction-curve analysis, for the majority of the QSOs. After correcting for any bias in luminosity between the two samples, we find no significant differences in the accretion properties of rQSOs such as black-hole spin, mass accretion rate or black-hole mass. We tentatively find lower Eddington ratios in the rQSOs compared to the cQSOs, but a larger unbiased sample is required to robustly confirm or refute this result. See Sections~\ref{sec:emission} and \ref{sec:bh_prop}. 
    \item \textbf{There are no significant differences in the outflow properties of red and typical QSOs:} From fitting the emission lines, we find a larger fraction of rQSOs display strong blueshifts in both the \ion{C}{iv} ($v>1000$\,km\,s$^{-1}$) and [\ion{O}{iii}]$\uplambda$5007 wing ($v>400$\,km\,s$^{-1}$) profiles, although these differences are not significant. We also find no significant differences between rQSOs and cQSOs in terms of the FWHM of either the core or wing [\ion{O}{iii}]$\uplambda$5007 component, although we are limited by the small number of [\ion{O}{iii}]$\uplambda$5007 detections. Our future large SDSS DR14 study will more robustly test these results. See Section~\ref{sec:bh_dis}.
\end{itemize}

Our results suggest that the main difference between red and typical QSOs is the presence of dust rather than significant differences in the accretion or outflow properties. On the basis of these results, a potential self-consistent scenario that links the enhanced radio emission to the dust in red QSOs is a more dust and gas-rich environment, in which the radio emission is due to shocks produced by either an outflow or a jet interacting with a higher opacity ISM/circumnuclear environment in the red QSOs.

This paper demonstrates the capability and quality of \textit{X-shooter} spectra in determining the nature of red QSOs, and future larger \textit{X-shooter} samples will robustly tie down and refine many of these results. Our future work will also utilize the Dark Energy Spectroscopic Instrument (DESI; \citealt{desi}) optical spectra (observed wavelength of 360--980\,nm) of $\sim2$--3 million quasars, which will push to much fainter and more obscured systems ($A_V$ up to $\sim$2--3\,mags) than those targeted in the SDSS. This sample, combined with NIR spectroscopic follow-up, can potentially provide us with both with the impressive source statistics and the broader range in $A_V$ required to robustly constrain many of the results presented in this paper. These data will also provide the potential to measure differences in the dust composition with increasing $A_V$ and to, ultimately, conclusively determine the connection between the presence of dust and the enhanced compact radio emission in red QSOs.
\section{Acknowledgements}
We would like to thank the anonymous referee for their constructive comments. We acknowledge a quota studentship through grant code ST/S505365/1 funded by the Science and Technology Facility Council (VAF), the Faculty of Science Durham Doctoral Scholarship (LK), the Science and Technology Facilities Council (DMA, DJR, through grant codes ST/P000541/1 and ST/T000244/1). EL acknowledges the support of grant ID: 45780 Fondazione Cassa di Risparmio Firenze. This work was supported by the Medical Research Council (LKM through grant code MR/T042842/1).
We want to thank D. M. Capellupo for providing us with their reduced \textit{X-shooter} spectra, ESO for their help and communication and S. Rakshit for spectral fitting advice. 

Funding for the SDSS and SDSS-II has been provided by the Alfred P. Sloan Foundation, the Participating Institutions, the National Science Foundation, the U.S. Department of Energy, the National Aeronautics and Space Administration, the Japanese Monbukagakusho, the Max Planck Society, and the Higher Education Funding Council for England. The SDSS Web Site is http://www.sdss.org/.
The SDSS is managed by the Astrophysical Research Consortium for the Participating Institutions. The Participating Institutions are the American Museum of Natural History, Astrophysical Institute Potsdam, University of Basel, University of Cambridge, Case Western Reserve University, University of Chicago, Drexel University, Fermilab, the Institute for Advanced Study, the Japan Participation Group, Johns Hopkins University, the Joint Institute for Nuclear Astrophysics, the Kavli Institute for Particle Astrophysics and Cosmology, the Korean Scientist Group, the Chinese Academy of Sciences (LAMOST), Los Alamos National Laboratory, the Max-Planck-Institute for Astronomy (MPIA), the Max-Planck-Institute for Astrophysics (MPA), New Mexico State University, Ohio State University, University of Pittsburgh, University of Portsmouth, Princeton University, the United States Naval Observatory, and the University of Washington.

This publication makes use of data products from the Wide-field Infrared Survey Explorer, which is a joint project of the University of California, Los Angeles, and the Jet Propulsion Laboratory/California Institute of Technology, funded by the National Aeronautics and Space Administration.

This work is based on observations collected at the European Southern Observatory under ESO programmes 0101.B-0739(A) and 088.B-1034.
\section{Data Availability Statement}
The data underlying this article will be shared on reasonable request
to the corresponding author.
The rQSO and cQSO composites used in this paper are available at CDS via anonymous ftp to cdsarc.u-strasbg.fr (130.79.128.5) or via \\ \url{https://cdsarc.unistra.fr/viz-bin/cat/J/MNRAS} and \\ \url{https://github.com/VFawcett/XshooterComposites}.
\bibliography{bib}
\bibliographystyle{mnras}
\appendix

\section{2229-0832: a synchrotron-dominated red QSO?}\label{sec:intrinsic}
The QSO 2229-0832 is the most radio bright QSO in our sample, as well as being the only \textit{Fermi}-detected source (see Section~\ref{sec:sample}). From fitting a dust-reddened cQSO composite to this source, we find that none of the dust-extinction curves provide a good fit; this could suggest that other factors are dominating the red colours, such as synchrotron emission. In a previous study we explored whether synchrotron emission can explain the optical colours in a sample of SDSS DR7 rQSOs \citep{klindt}. To do this, we fitted a synchrotron emission model to the radio--optical data of 3FGL J0045.2-3704 to generate a synchrotron-dominated template (see Appendix~A in \citealt{klindt}, and \citealt{klindt17} for more details). This template was then scaled from the 1.4\,GHz flux of the QSO sample to estimate the expected $i$-band flux from a synchrotron component, comparing to that from the DR7 catalogue. Only $\sim8$ per~cent of the radio-detected rQSOs were found to have optical colours that were potentially contaminated by synchrotron emission. Following this approach, we find that the predicted $i$-band magnitude estimated from a synchrotron component is comparable ($<0.36$\,mag difference) to the SDSS $i$-band magnitude for 2229-0832, suggesting that a synchrotron component may be dominant in this source. For the other sources in our sample, the predicted $i$-band magnitude estimated from a synchrotron component are inconsistent ($>0.5$\,mags difference) with the SDSS $i$-band magnitude.

One clear signature of synchrotron emission is the characteristic double-bump structure in the SED, usually found in Flat Spectrum Radio Quasars (FSRQs) and BL Lacertae Objects (BL Lacs). The lowest energy bump is due to synchrotron emission in the radio--X-ray and the higher energy bump is due to a synchrotron self Compton mechanism in the X-ray--gamma-rays. Therefore finding a gamma-ray counterpart in addition to a double-bumped SED in a QSO is a good indicator of a dominant synchrotron component. The QSO 2229-0832 displays two clear bumps in the SED which is further evidence that the red colours of this source are likely dominated by synchrotron emission rather than dust.

\section{Black-hole mass comparison}
\begin{figure}
    \centering
    \includegraphics[width=3.3in]{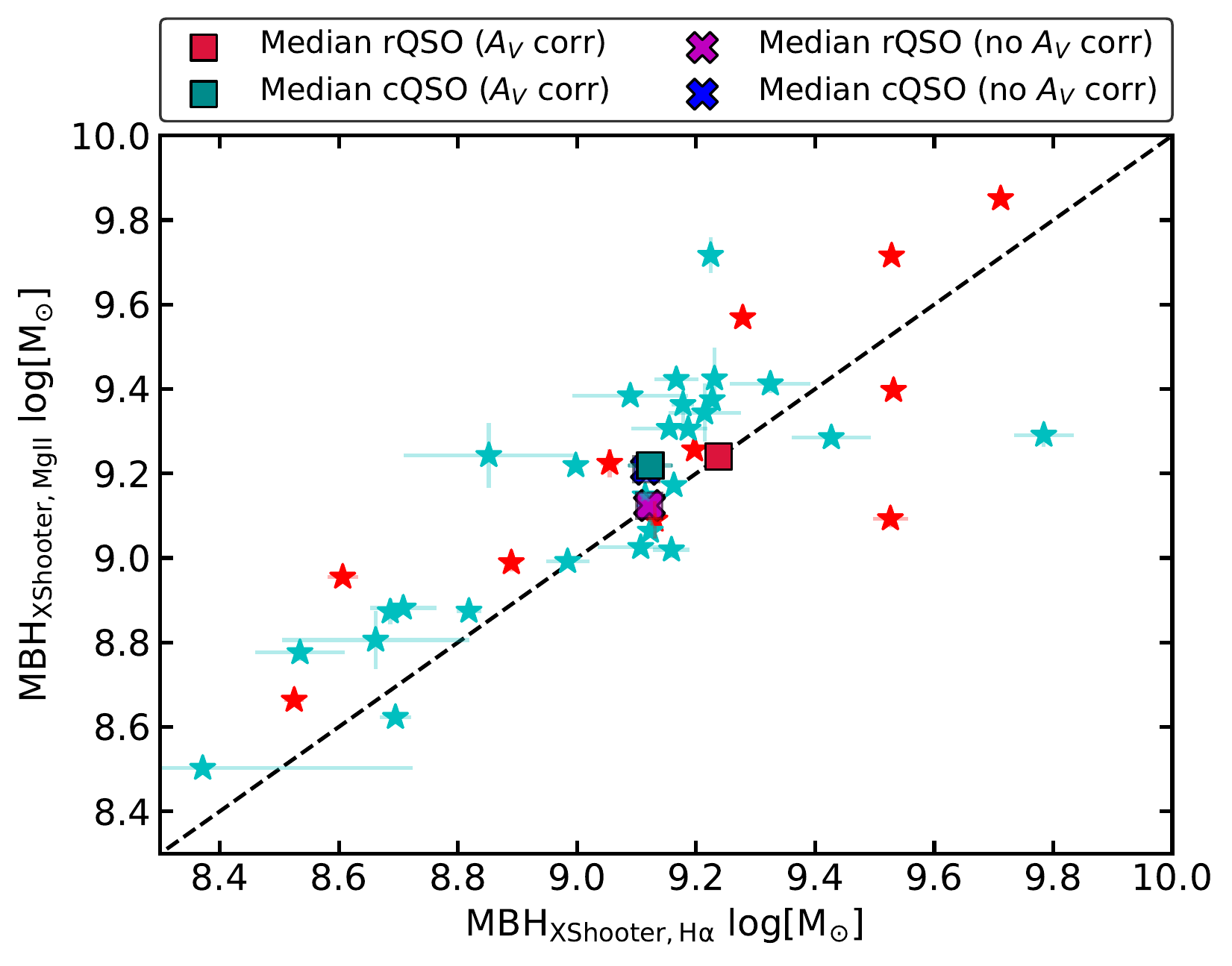}
    \caption{The extinction-corrected black-hole masses calculated from the \ion{Mg}{ii} line versus the H\,$\upalpha$ line for our \textit{X-shooter} spectra. See Fig.~\ref{fig:bh} for a description of the data, markers and colours plotted.}
    \label{fig:mbh_mgii_ha}
\end{figure}

Fig.~\ref{fig:mbh_mgii_ha} displays the extinction-corrected black-hole masses calculated using the broad \ion{Mg}{ii} versus H\,$\upalpha$ from our \textit{X-shooter} spectra. The squares show the median values for the rQSOs and cQSOs (corrected for dust extinction); the median rQSO black-hole mass estimated from the two emission lines is in good agreement, but the median cQSO H\,$\upalpha$ black-hole mass is slightly lower than that from the \ion{Mg}{ii} emission line. This difference could be due to small number statistics or the intrinsic scatter between different black-hole mass estimators \citep{Shen_2012}.

\section{Comparison to C15}\label{sec:cap_comp}
In this section we compare some of the emission-line and black-hole properties for the 15 C15 cQSOs calculated in this work to those from C16 and M16.

\begin{figure*}
    \centering
    \includegraphics[width=2.3in]{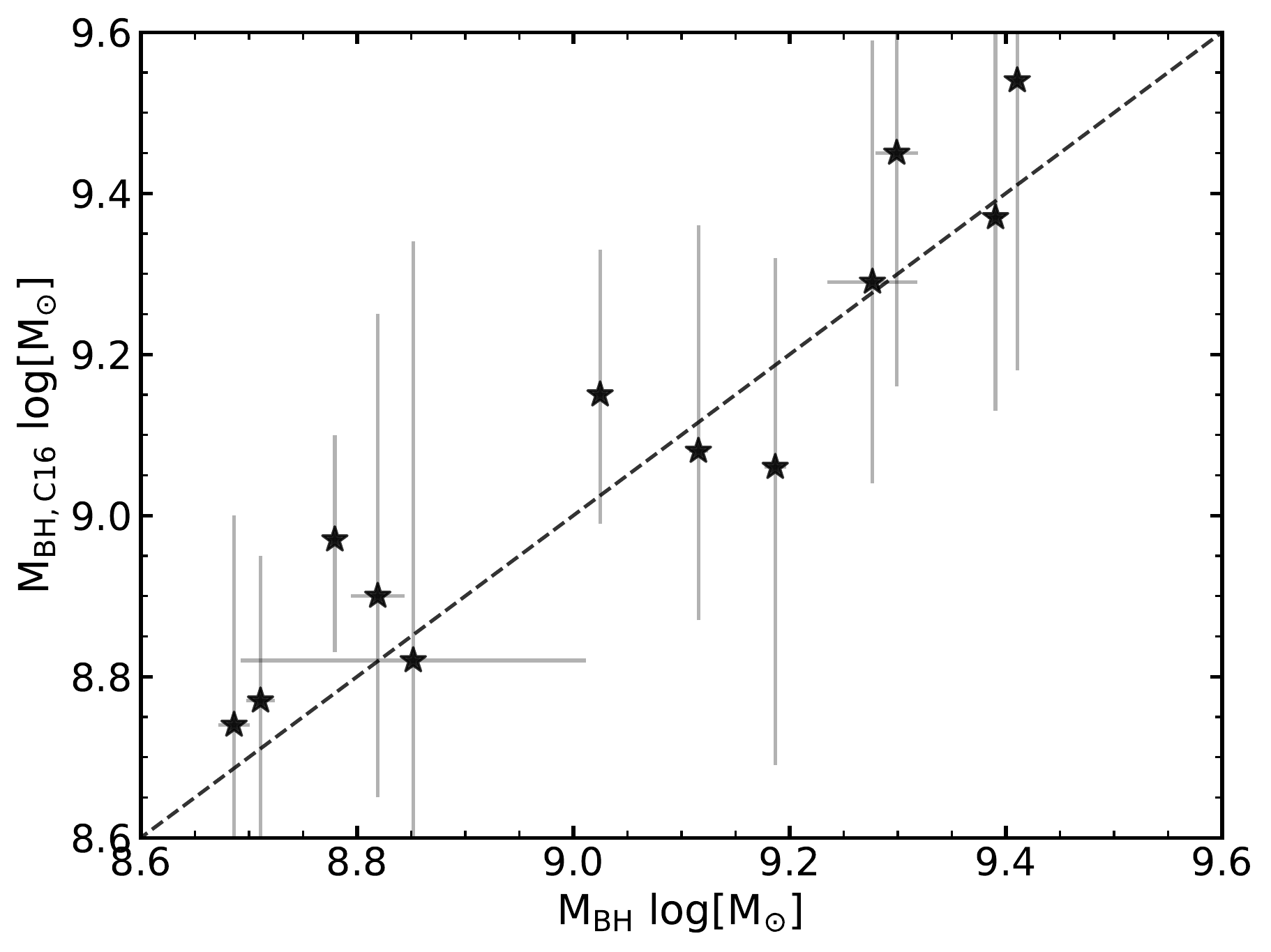}
    \includegraphics[width=2.3in]{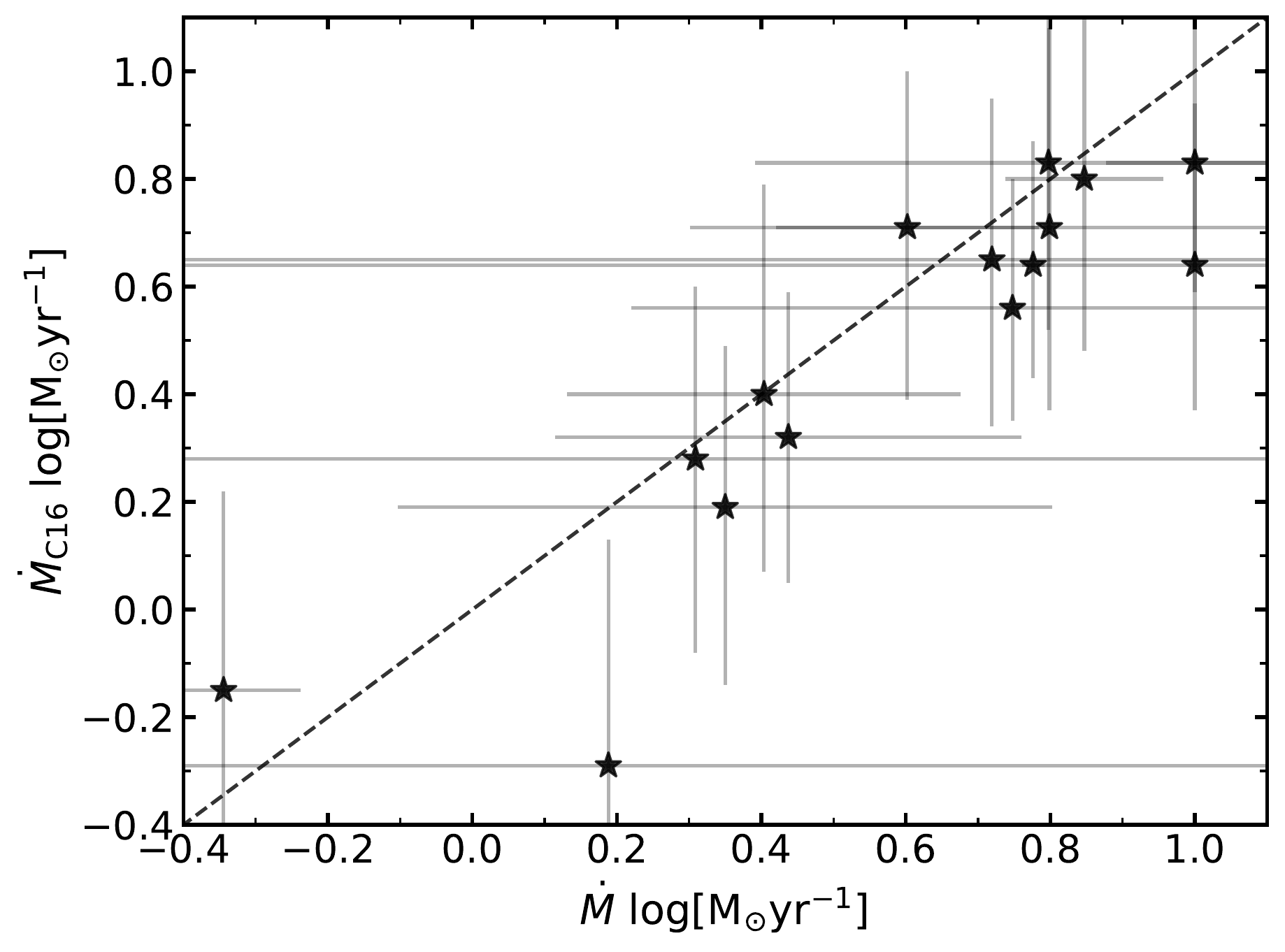}
    \includegraphics[width=2.3in]{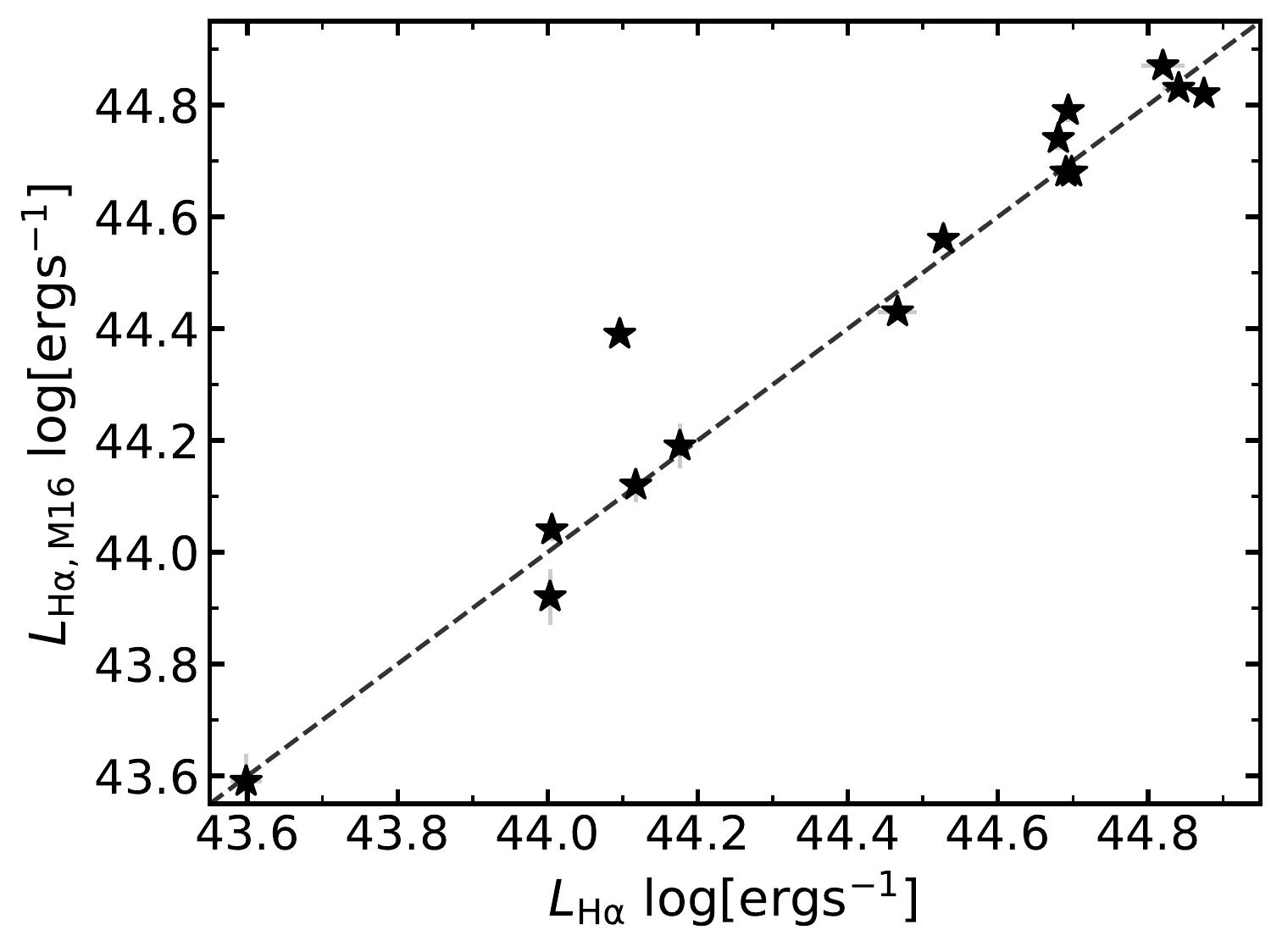}
    \caption{M$_{\rm BH}$ (left), mass accretion rate (middle), and H$\upalpha$ emission line luminosities (right) obtained from this work for the C15 cQSOs, compared to the equivalent parameters obtained for the same sample by C16 and M16.}
    \label{fig:cap_bh}
\end{figure*}

In Fig.~\ref{fig:cap_bh} (left) we compare M$_{\rm BH}$ calculated from the H\,$\upalpha$ broad line (using relations from \citealt{Shen_2012}; see Section~\ref{sec:AD}) to those obtained from AD fitting in C16. Overall, we find good agreement between the two methods. Fig.~\ref{fig:cap_bh} (middle) shows a comparison of the mass accretion rate obtained from our accretion-disc fitting (see Section~\ref{sec:AD}), compared to that in C16. Fig.~\ref{fig:cap_bh} (right) shows a comparison of the total H$\upalpha$ emission line luminosities (not corrected for dust extinction) based on our emission line fitting (see Section~\ref{sec:fit}), compared to that from M16.

\section{Control composite comparison}\label{sec:comp_comp}
\begin{figure*}
    \centering
    \includegraphics[width=6in]{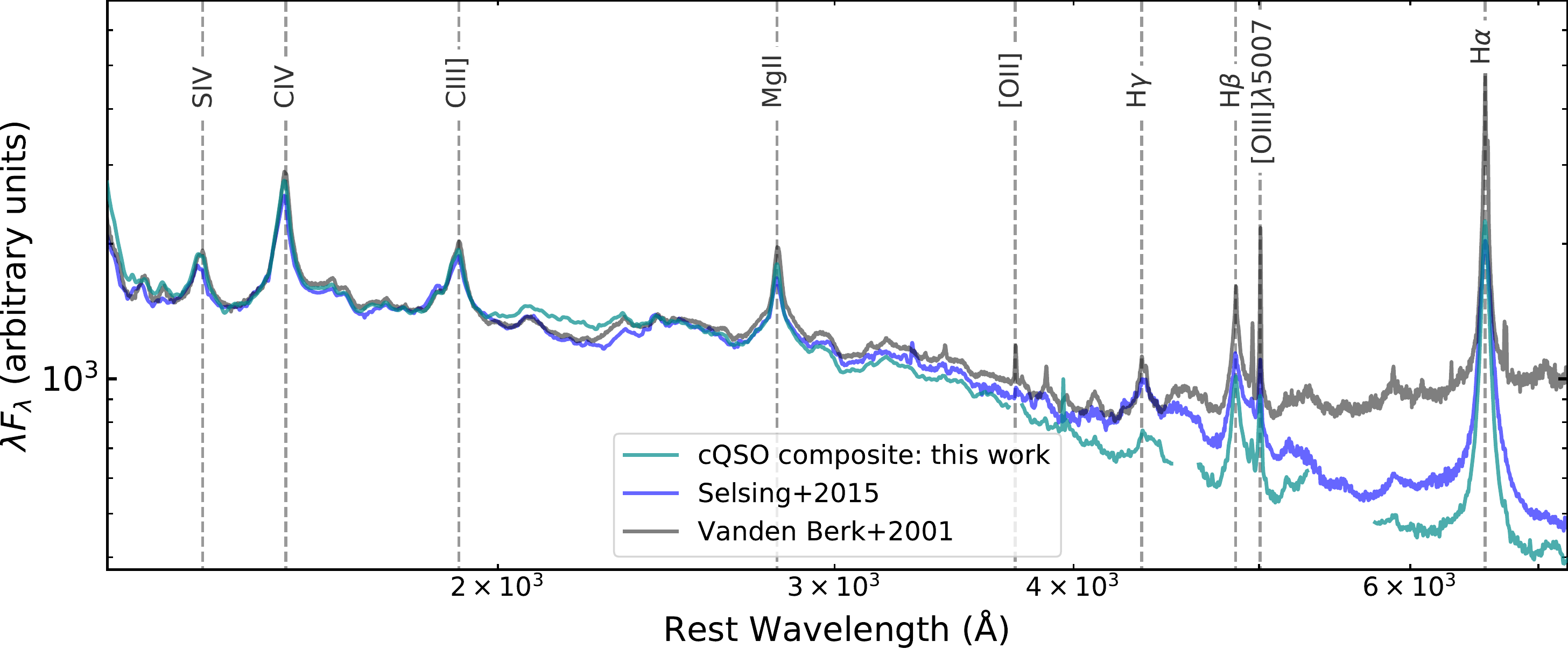}
    \caption{Comparison of our cQSO composite (cyan; see Section~\ref{sec:ext_method}) to that in V01 (black) and S16 (blue). All three composite are similar until $\sim4000$\,\AA; the V01 composite flattens out due to host-galaxy contamination by low redshift QSOs. The S16 composite is slightly redder than our composite, which could be due to a difference in colour in the underlying sample, a difference in luminosities, greater host-galaxy contribution or just small number variations. }
    \label{fig:cQSO_comp}
\end{figure*}
Our measured continuum extinction values are calculated with respect to our cQSO composite, under the assumption that this is a good representation of the typical QSO population. To test this assumption, we compared our cQSO composite to existing QSO templates: \cite{vanden} (hereafter V01) and \cite{selsing} (hereafter S16). 

V01 is the most widely used QSO template in the literature. It is composed of over 2200 SDSS spectra that span a redshift range of $0.044<z<4.789$. Due to the large range in redshifts of the QSOs that comprise this composite and the narrow wavelength range covered by SDSS, a variety of different types of systems will be contributing to the composite at different wavelengths. For example, the lowest redshift QSOs, which will be more host-galaxy dominated, will be predominantly contributing towards the reddest end of the composite, whereas the highest redshift QSOs, and therefore the more luminous by selection biases, will be contributing predominantly towards the bluest end. S16, on the other hand, use only seven QSOs in a redshift range of $1<z<2$, observed by \textit{X-shooter}, to construct their composite. This has the advantage of the broader wavelength range of \textit{X-shooter}, combined with a narrower range in redshift, which will result in similar types of systems contributing at all wavelengths. However, the seven QSOs are chosen to be very bright ($r<17$\,mags), and therefore may not be representative of the overall QSO population. Our \textit{X-shooter} sample consists of 28 and 12 luminous ($16<r<18$\,mags) cQSOs and rQSOs.

\begin{figure*}
    \includegraphics[width=6in]{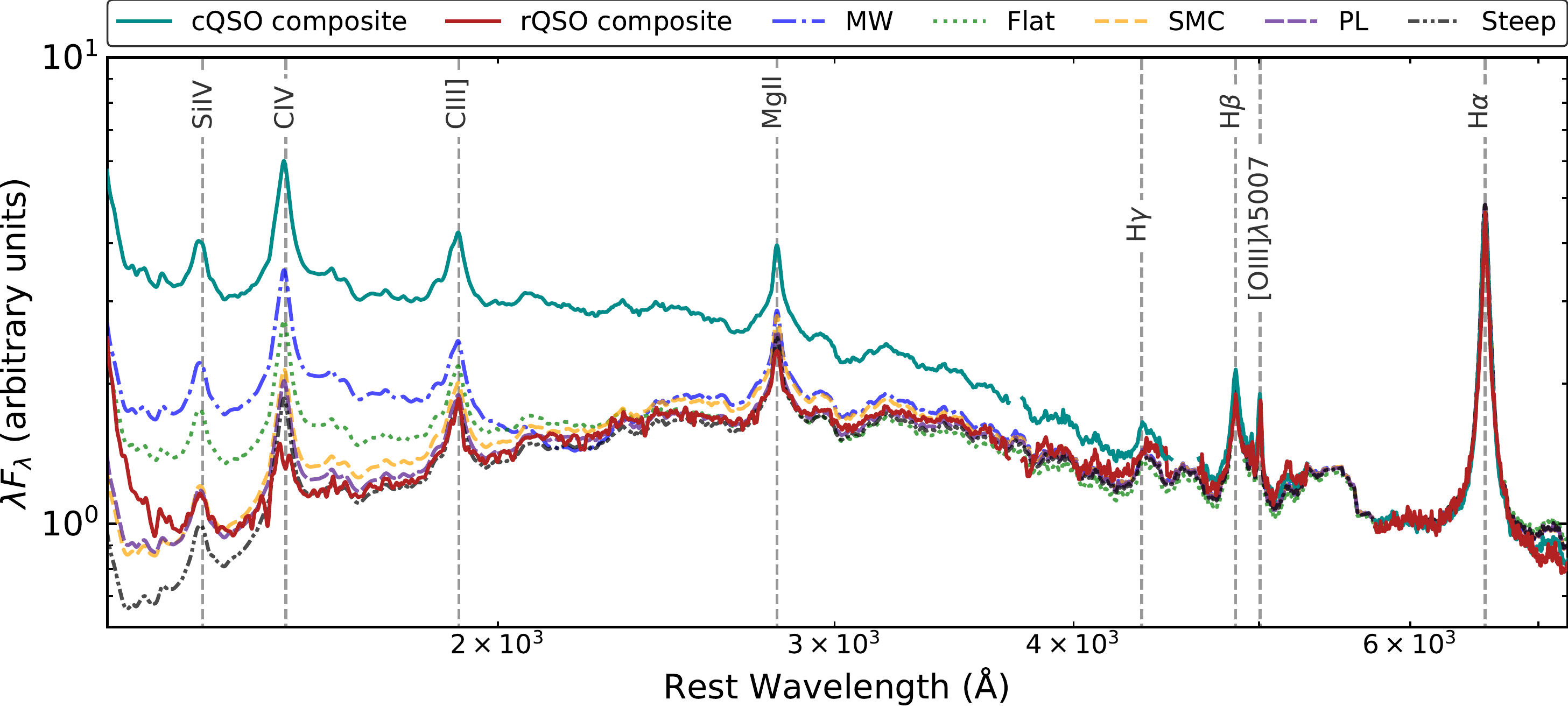}
    \caption{Comparison of how the five different extinction laws used in this study affect the continuum shape of the cQSO composite (solid cyan curve) with $E(B-V)=0.11$\,mags. The rQSO composite is also displayed (solid red curve); reddening the cQSO composite with a simple PL gives the best fit to the rQSO composite. The dashed grey lines correspond to the major emission lines.}
    \label{fig:comp_av}
\end{figure*}

Fig.~\ref{fig:cQSO_comp} displays the comparison between our cQSO composite, and those from V01 and S16, all normalized to 1450\,\AA. Below $\sim4000$\,\AA, the three composites are fairly similar. Above 4000\,\AA, the V01 template flattens out compared to both S16 and our composites; this is expected due to the strong host-galaxy contamination from the low redshift QSOs included in the V01 template. The S16 composite appears to be redder relative to our cQSO composite. Comparing the $g^*-i^*$ colours of the seven \textit{X-shooter} QSOs included in the S16 composite to our red and control QSO colour selections, we find that six have colours consistent with our cQSOs and one is slightly redder (although not red enough to be selected as an rQSO); this could be driving the differences in the composite. Equally, the low source statistics in both of the composites, the difference in luminosities between the samples, or the level of host-galaxy contamination could also be affecting the shape.

\section{Spectra, [\ion{O}{iii}] and \ion{C}{iv} profiles}
\begin{figure*}
    \centering
    \includegraphics[width=6.in]{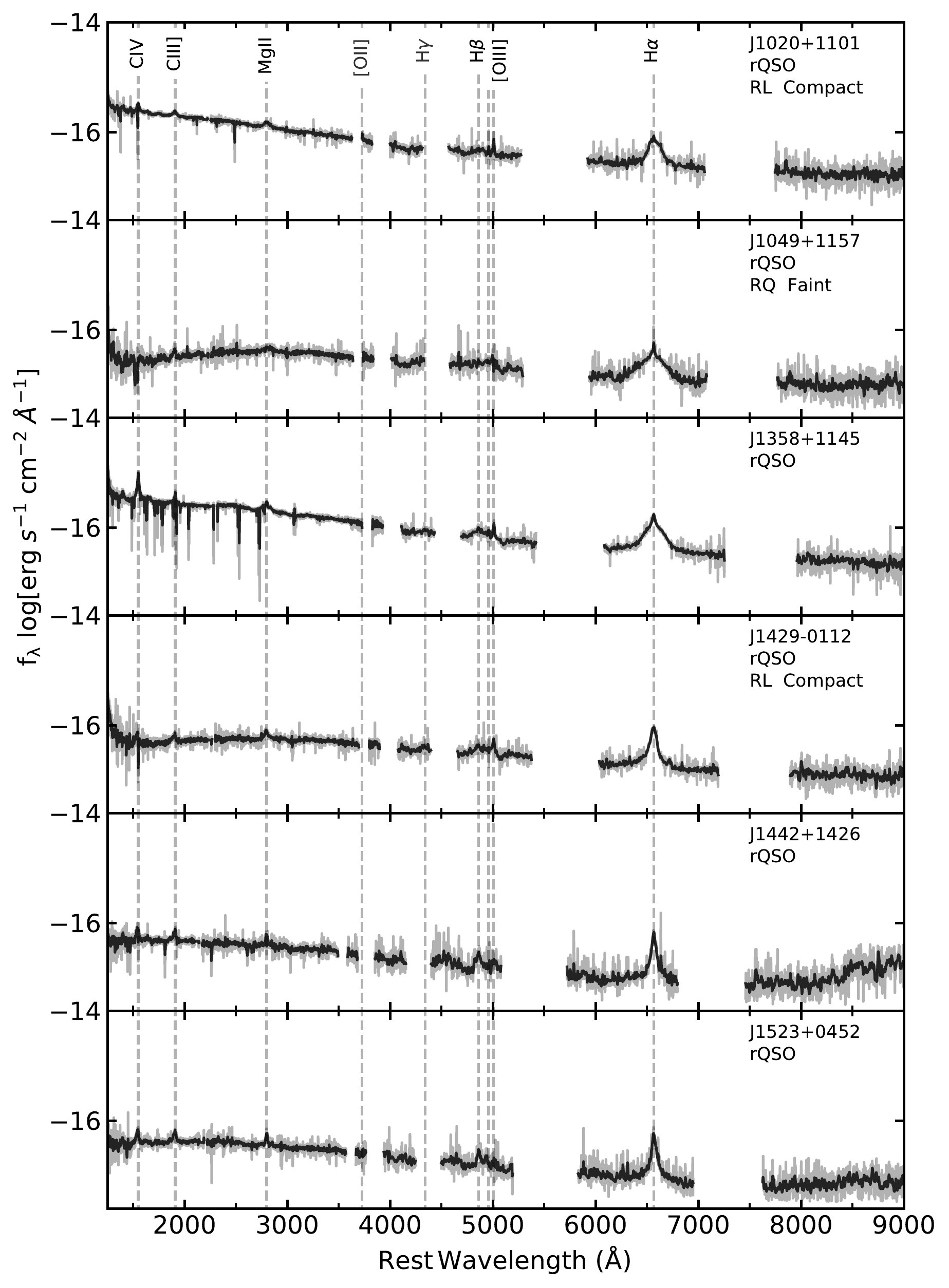}
\end{figure*}
\begin{figure*}
    \includegraphics[width=6in]{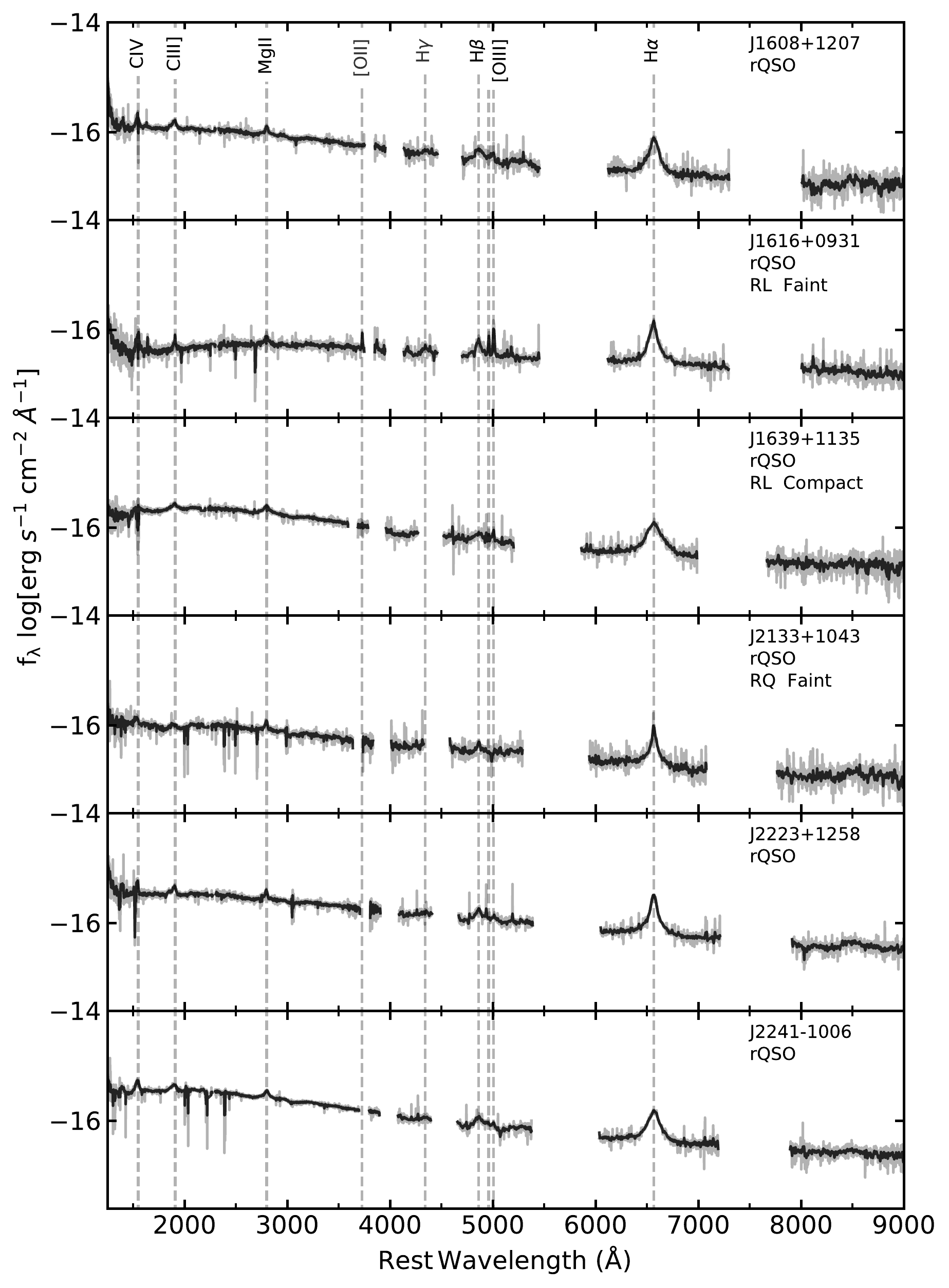}
    \caption{Finalized spectra for rQSO sample. The name, sample, whether a source is radio-loud (RL) or radio-quiet (RQ) and the radio morphology is displayed at the top right of each spectrum. The dashed grey lines correspond to the major emission lines, labelled in the top panel.}
    \label{fig:spec_rQSO}
\end{figure*}
\begin{figure*}
    \includegraphics[width=6in]{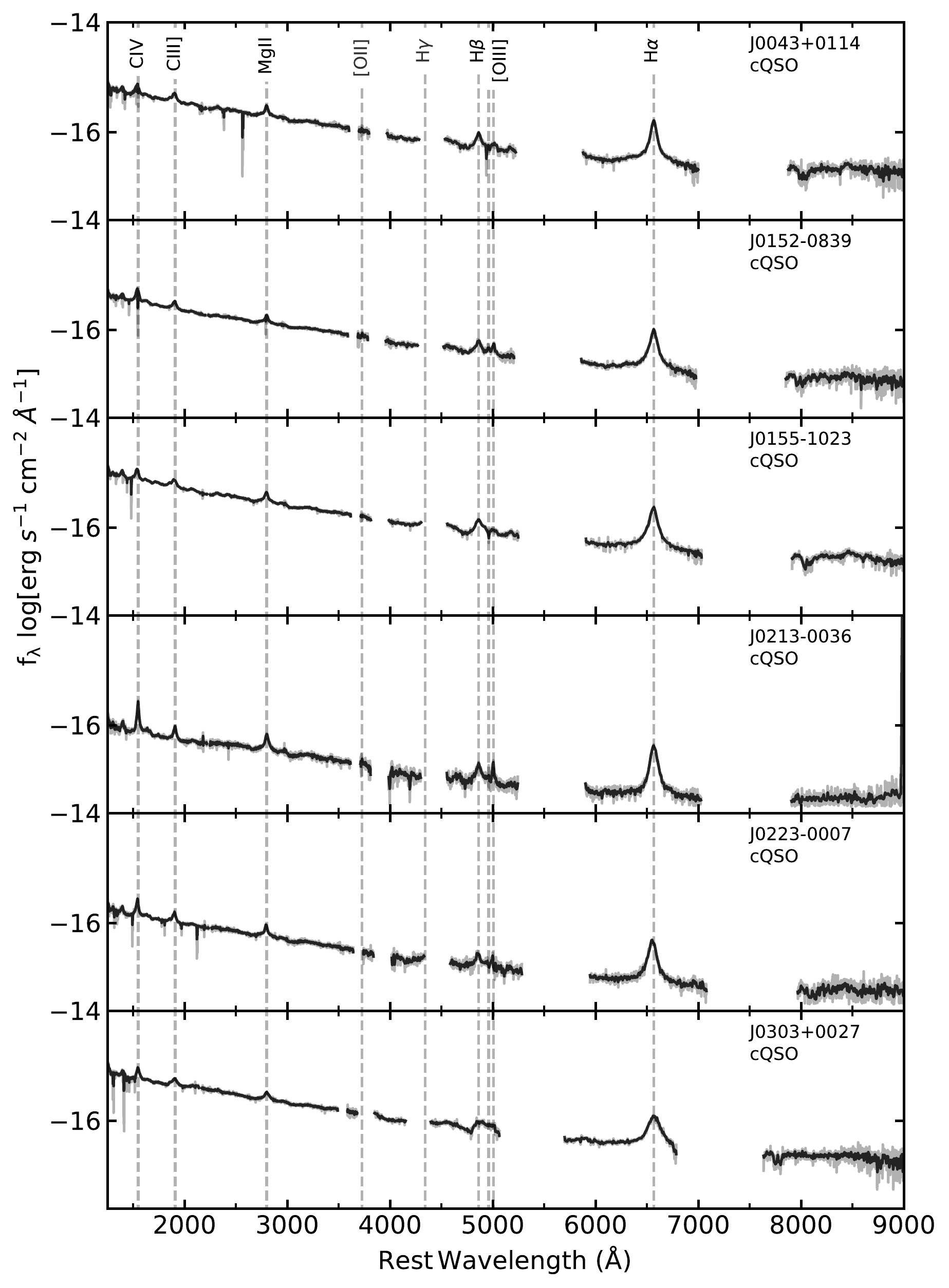}
\end{figure*}
\begin{figure*}
    \includegraphics[width=6in]{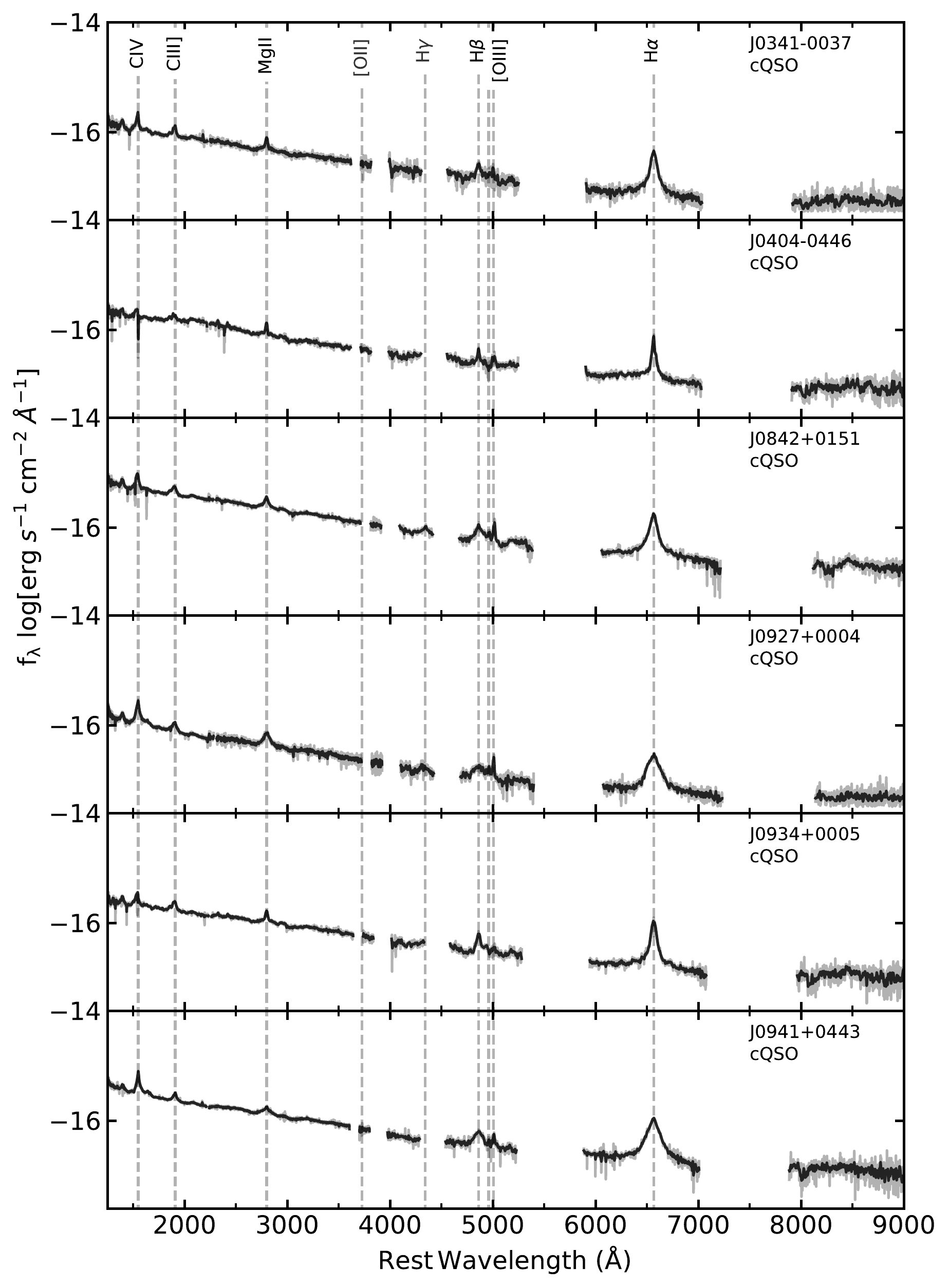}
\end{figure*}
\begin{figure*}
    \includegraphics[width=6in]{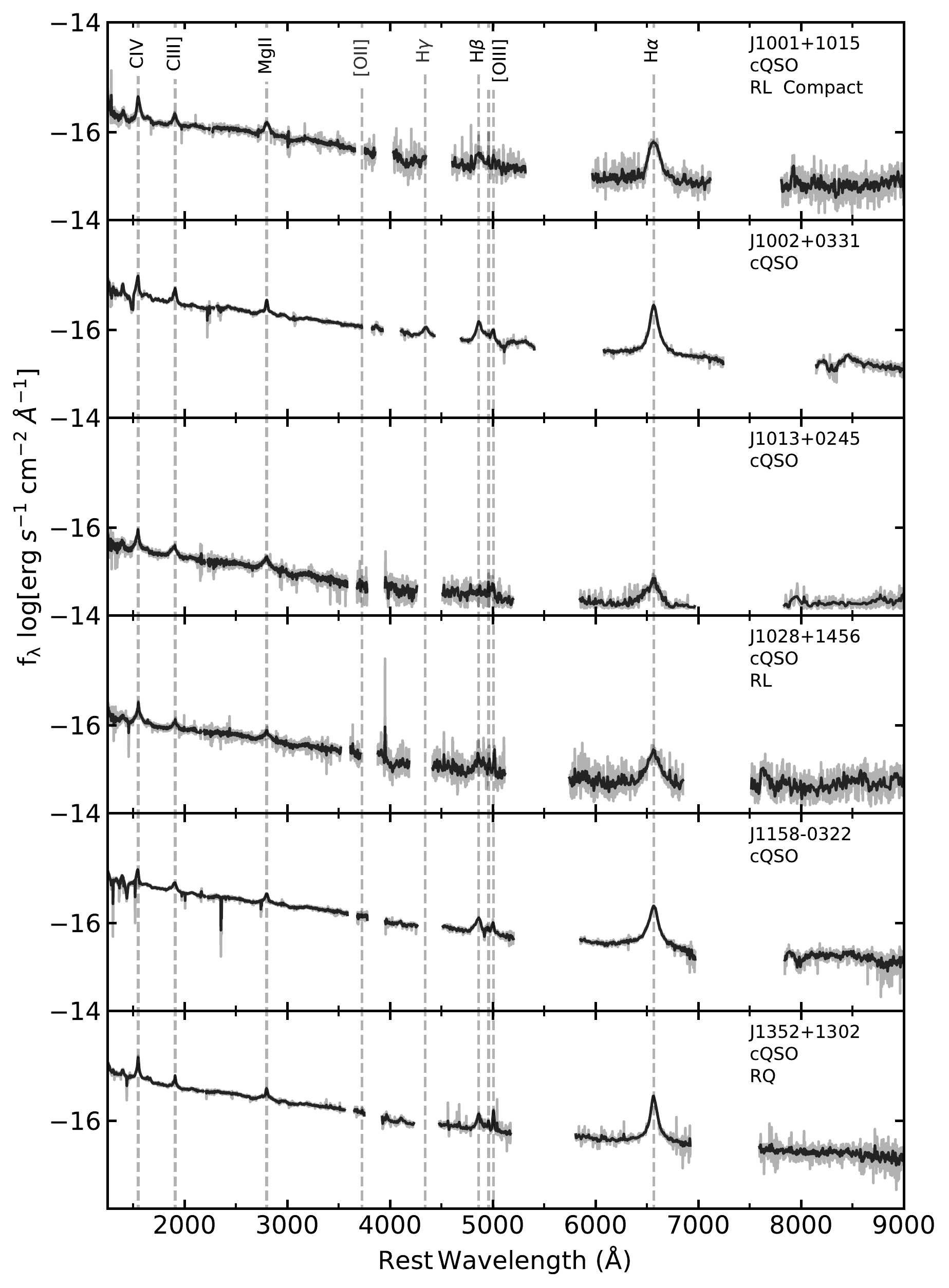}
\end{figure*}
\begin{figure*}
    \includegraphics[width=6in]{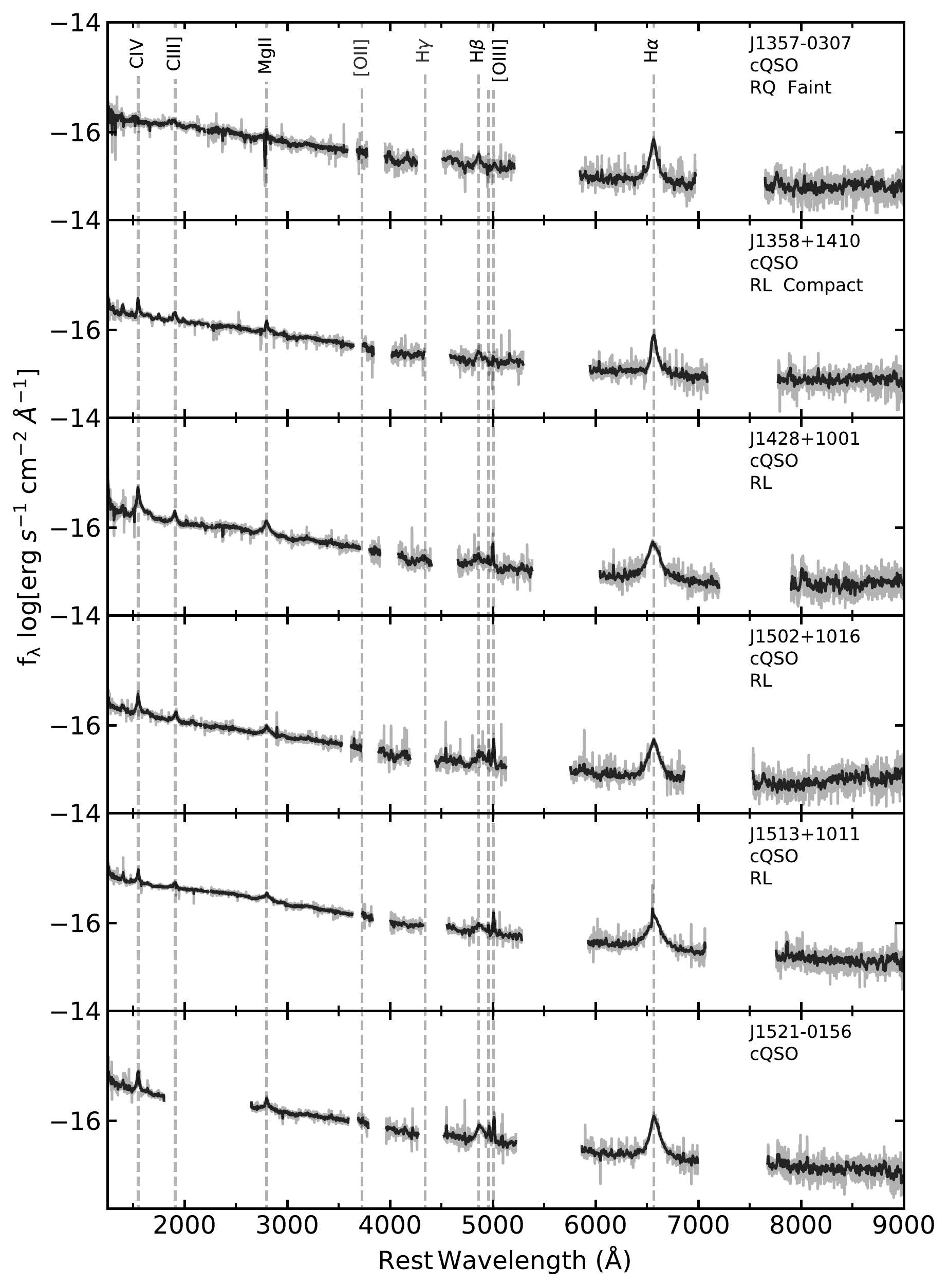}
\end{figure*}
\begin{figure*}
    \includegraphics[width=6in]{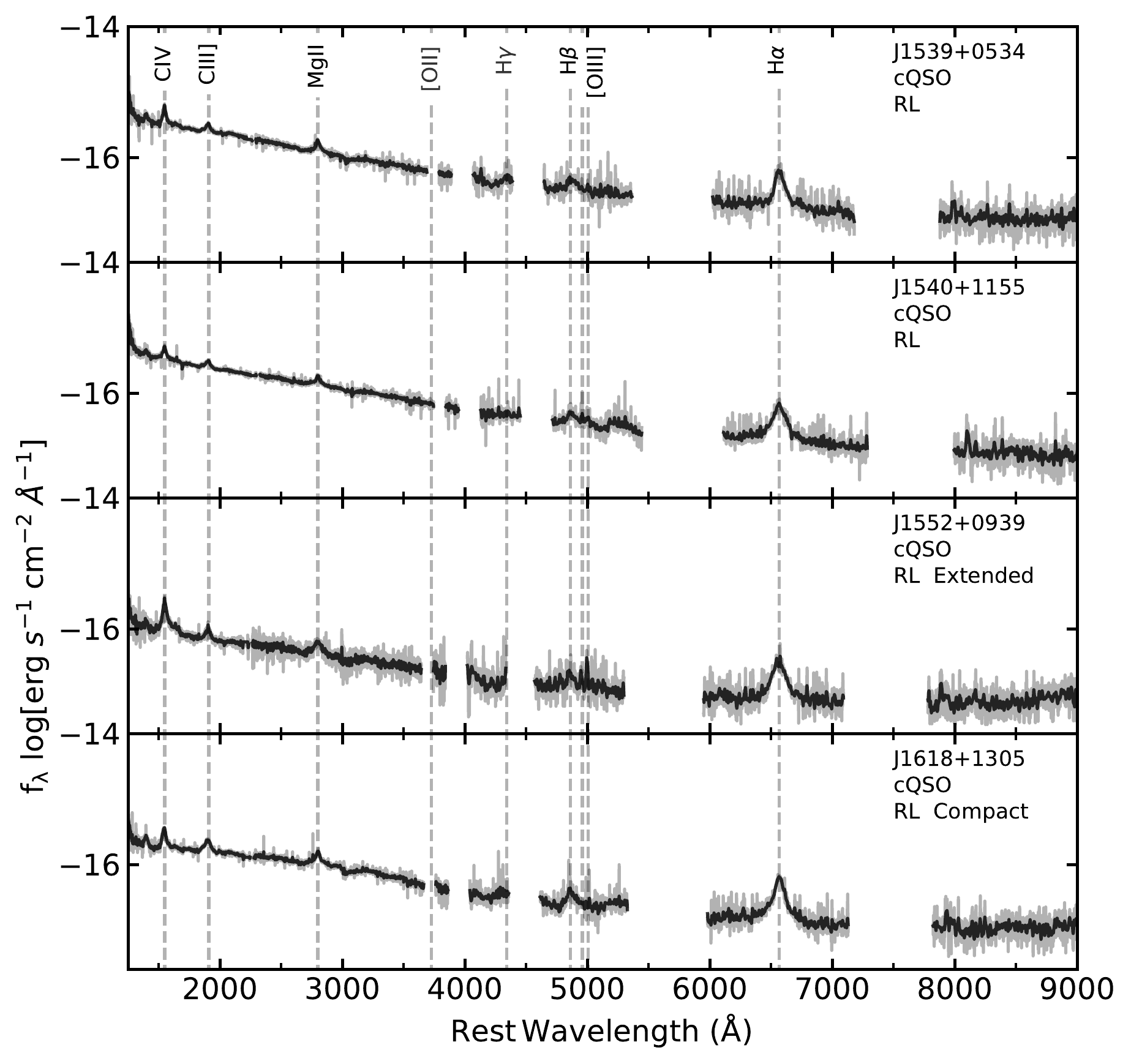}
    \caption{Finalized spectra for cQSO sample, the including C15 sources. The name, sample, whether a source is radio-loud (RL) or radio-quiet (RQ) and the radio morphology is displayed at the top right of each spectrum. The dashed grey lines correspond to the major emission lines, labelled in the top panel.}
    \label{fig:all_spec}
\end{figure*}

\begin{figure*}
    \includegraphics[width=6in]{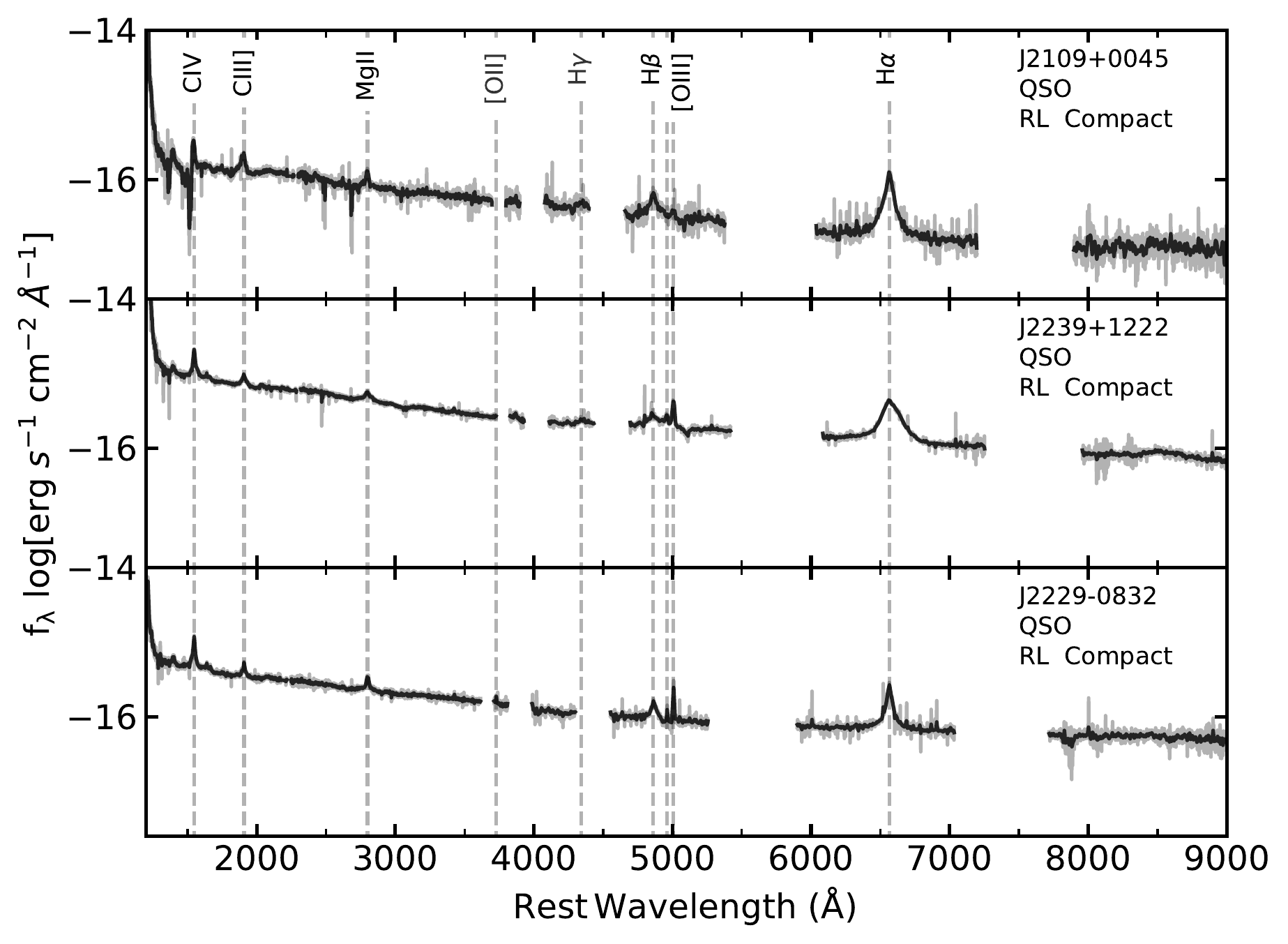}
    \caption{Finalized spectra for the QSOs excluded from our analyses (see Section~\ref{sec:sample}); 2229-0832 is the \textit{Fermi}-detected source and 2109+0045 and 2239+1222 had an incorrect level of Galactic extinction originally applied that moves them out of our rQSO threshold. The name, sample, whether a source is radio-loud (RL) or radio-quiet (RQ) and the radio morphology is displayed at the top right of each spectrum. The dashed grey lines correspond to the major emission lines, labelled in the top panel.}
    \label{fig:ex_spec}
\end{figure*}

Figs.~\ref{fig:spec_rQSO}, \ref{fig:all_spec}, and \ref{fig:ex_spec} display the spectra for the rQSOs, cQSOs and three QSOs not included in our final sample (see Section~\ref{sec:sample}), respectively.

\begin{figure*}
    \includegraphics[width=6in]{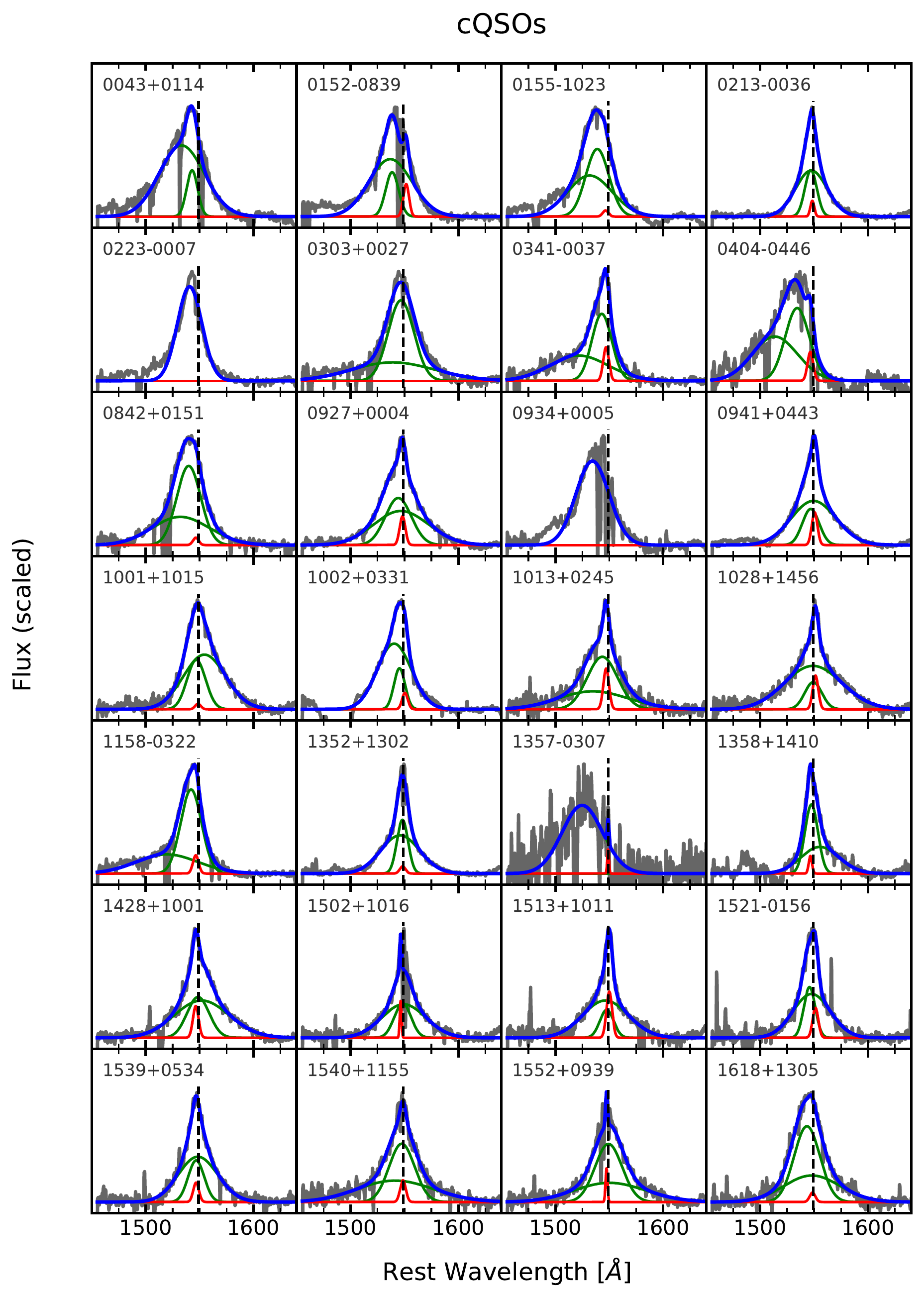}
    \caption{\ion{C}{iv} profiles for the cQSOs. The y-axis is scaled to 1.4 times the maximum flux. The red and green lines indicate the narrow and broad Gaussian fits, respectively, and the blue line shows the overall fit profile. The source name and rest wavelength of \ion{C}{iv} is also displayed.}
    \label{fig:con_CIV}
\end{figure*}

\begin{figure*}
    \includegraphics[width=6in]{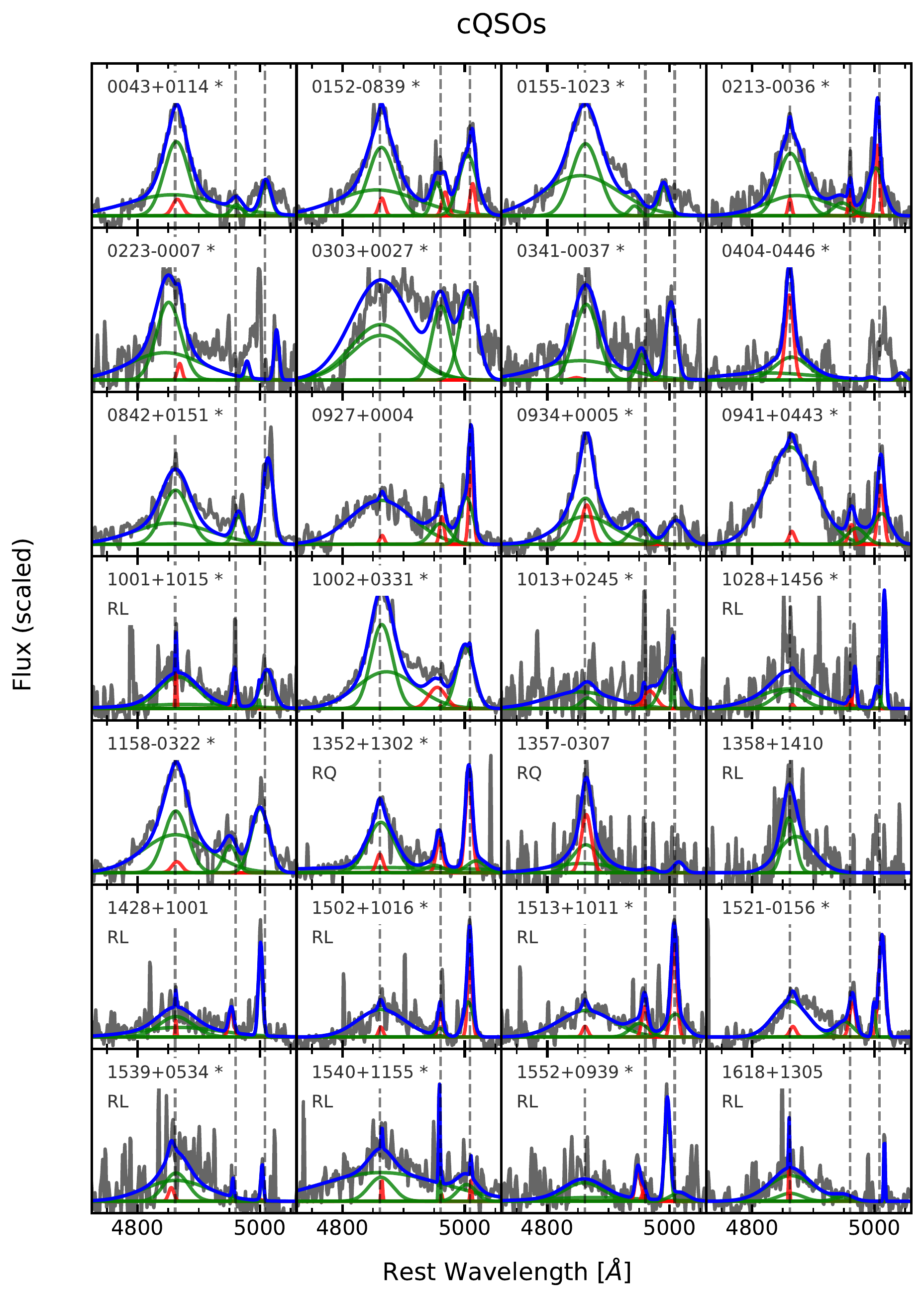}
    \caption{H\,$\upbeta$--[\ion{O}{iii}] profiles for the cQSOs. The y-axis is scaled to 1.3 times the maximum flux. The red and green lines indicate the narrow core and broad wing fits, respectively, and the blue line shows the overall fit profile. The source name and whether it is radio-quiet (RQ) or radio-loud (RL) is also displayed. Significant [\ion{O}{iii}] detections are indicated by a star next to the source name.}
    \label{fig:con_OIII}
\end{figure*}

\begin{figure*}
    \includegraphics[width=5.5in]{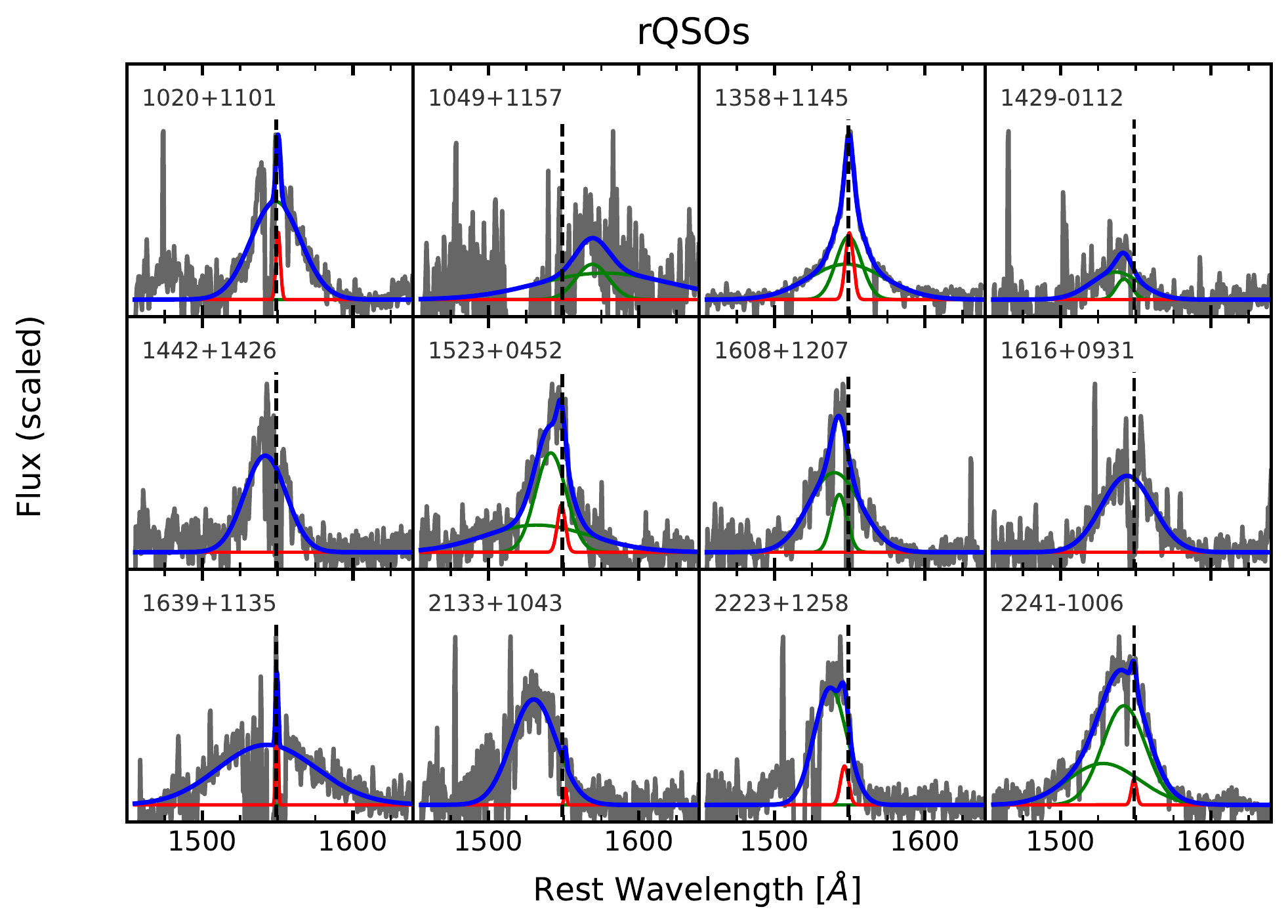}
    \caption{\ion{C}{iv} profiles for the rQSOs. The y-axis is scaled to 1.4 times the maximum flux. The red and green lines indicate the narrow and broad Gaussian fits, respectively, and the blue line shows the overall fit profile. The source name and rest wavelength of \ion{C}{iv} is also displayed. The rQSO 1049+1157 shows a strong BAL feature blueward of the \ion{C}{iv} peak which will have an effect on the quality of the fit; after removing this source from our sample, we still do not find any differences in the \ion{C}{iv} blueshift between the rQSOs and cQSOs (see Section~\ref{sec:bh_dis}).}
    \label{fig:red_CIV}
\end{figure*}

\begin{figure*}
    \includegraphics[width=5.5in]{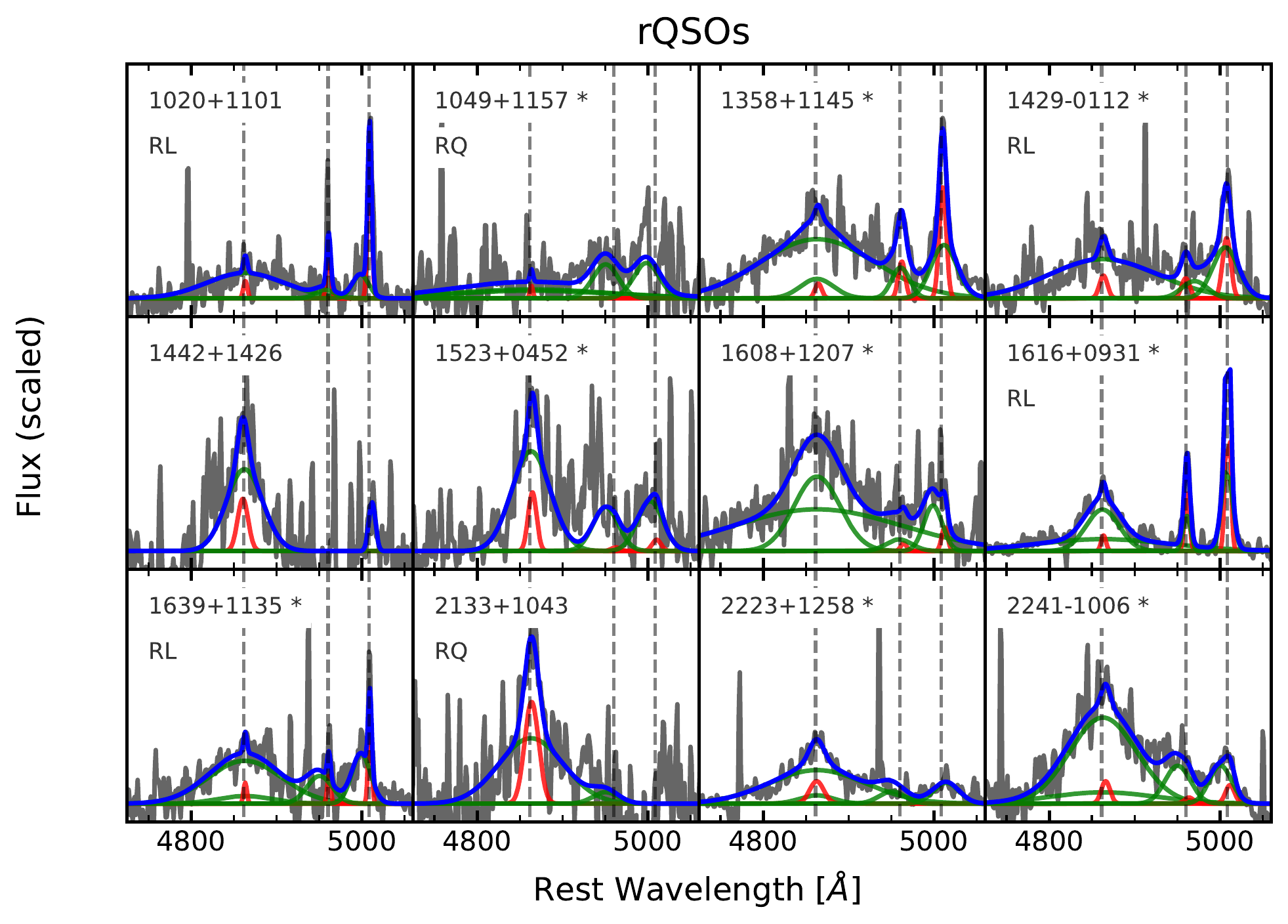}
    \caption{H\,$\upbeta$--[\ion{O}{iii}] profiles for the rQSOs. The y-axis is scaled to 1.3 times the maximum flux. The red and green lines indicate the narrow core and broad wing fits, respectively, and the blue line shows the overall fit profile. The source name and whether it is radio-quiet (RQ) or radio-loud (RL) is also displayed. Significant [\ion{O}{iii}] detections are indicated by a star next to the source name.}
    \label{fig:red_OIII}
\end{figure*}

Figs.~\ref{fig:con_CIV} and \ref{fig:con_OIII} display the \ion{C}{iv} and H\,$\upbeta$--[\ion{O}{iii}] thumbnails for the cQSOs, respectively, and Figs.~\ref{fig:red_CIV} and \ref{fig:red_OIII} display the \ion{C}{iv} and H\,$\upbeta$--[\ion{O}{iii}] thumbnails for the rQSOs, respectively. The red and green lines indicate the narrow and broad Gaussian fits, respectively, and the blue line indicates the overall fitted profile. There are a number of rQSOs and cQSOs that display both broad [\ion{O}{iii}] and extreme \ion{C}{iv} blueshifts; however, there is are no significant differences between the rQSOs and cQSOs. This suggests that, although both rQSOs and cQSOs can host powerful outflows, it is not clear that they are more prevalent in the rQSO population.

\label{lastpage}
\bsp
\end{document}